\documentclass[ejs]{imsart}

\RequirePackage{amsthm,amsmath,amsfonts,amssymb}
\RequirePackage[authoryear]{natbib}%% uncomment this for author-year citations
\RequirePackage[colorlinks,citecolor=blue,urlcolor=blue]{hyperref}
\RequirePackage{graphicx}
\usepackage{graphicx}
\usepackage{subfigure}
\usepackage{comment}
\usepackage{color}
\makeatletter\providecommand\current@color{0 g 0 G}\let\default@color\current@color\makeatother

\arxiv{2405.03042}
\startlocaldefs
\theoremstyle{plain}

\newtheorem{theorem}{Theorem}[section]

\usepackage{amsmath}
\usepackage{mathtools}
\usepackage{subfigure}
\usepackage{bm}
\usepackage{bbm}
\usepackage{algorithm2e}
\usepackage{booktabs}
\usepackage{caption}
\usepackage{enumerate}

\newtheorem{definition}{Definition}

\newtheorem{Lemma}{Lemma}

\newtheorem{assumption}{Assumption}

\def\algo{\texttt{PSIMF}}
\newcommand{\diag}{{\rm diag}}

\theoremstyle{remark}

\endlocaldefs

\begin{document}
\begin{frontmatter}
\title{Functional Post-Clustering Selective Inference with Applications to EHR Data Analysis}
%\title{A sample article title with some additional note\thanksref{t1}}
\runtitle{Functional Post-Clustering Selective Inference}
%\thankstext{T1}{A sample additional note to the title.}

\begin{aug}
%%%%%%%%%%%%%%%%%%%%%%%%%%%%%%%%%%%%%%%%%%%%%%%
%% Only one address is permitted per author. %%
%% Only division, organization and e-mail is %%
%% included in the address.                  %%
%% Additional information can be included in %%
\author[A]{\fnms{Zihan}~\snm{Zhu}\ead[label=e1]{zhzhu1@wharton.upenn.edu}},
\author[B]{\fnms{Xin}~\snm{Gai}\ead[label=e2]{xin.gai@vanderbilt.edu}}
\and
\author[C]{\fnms{Anru R.}~\snm{Zhang}\ead[label=e3]{anru.zhang@duke.edu}}
%%%%%%%%%%%%%%%%%%%%%%%%%%%%%%%%%%%%%%%%%%%%%%
%% Addresses                                %%
%%%%%%%%%%%%%%%%%%%%%%%%%%%%%%%%%%%%%%%%%%%%%%
\address[A]{Department of Statistics \& Data Science, Wharton School, University of Pennsylvania, Philadelphia, PA 19104, USA\printead[presep={,\ }]{e1}}

\address[B]{Department of Biostatistics, Vanderbilt University, Nashville, TN 37203, USA\printead[presep={,\ }]{e2}}

\address[C]{Department of Biostatistics \& Bioinformatics and Department of Computer Science, Duke University, Durham, NC 27705, USA\printead[presep={,\ }]{e3}}

\runauthor{Z. Zhu, X. Gai, A. R. Zhang}
\end{aug}

\begin{abstract}
In electronic health records (EHR) analysis, clustering patients according to patterns in their data is crucial for uncovering new subtypes of diseases. Existing medical literature often relies on classical hypothesis testing methods to test for differences in means between these clusters. Due to selection bias induced by clustering algorithms, the implementation of these classical methods on post-clustering data often leads to an inflated type-I error. In this paper, we introduce a new statistical approach that adjusts for this bias when analyzing data collected over time. Our method extends classical selective inference methods for cross-sectional data to longitudinal data. We provide theoretical guarantees for our approach with upper bounds on the selective type-I and type-II errors. We apply the method to simulated data and real-world Acute Kidney Injury (AKI) EHR datasets, thereby illustrating the advantages of our approach.
\end{abstract}



\end{frontmatter}
%%%%%%%%%%%%%%%%%%%%%%%%%%%%%%%%%%%%%%%%%%%%%%
%% Please use \tableofcontents for articles %%
%% with 50 pages and more                   %%
%%%%%%%%%%%%%%%%%%%%%%%%%%%%%%%%%%%%%%%%%%%%%%
%\tableofcontents

\begin{sloppypar}
%%%%%%%%%%%%%%%
\section{Introduction}\label{sec:intro}
%%%%%%%%%%%%%%%

Testing for a difference in means between groups of functional data is fundamental to answering research questions across various scientific areas \citep{fan1998test,cuevas2004anova,zhang2007statistical,zhang2014analysis}. Recently, there has been an increasing demand for post-clustering inference of functional data, namely, testing the difference between groups discovered by clustering algorithms. In particular, the electronic health records (EHR) system contains a rich source of longitudinal observational data, covering patient demographics, vital signs, and biochemical markers, making these data ideal for identifying subphenotypes of patients. With the increasing prevalence of EHR data, longitudinal data clustering methods used to evaluate patient subphenotypes have become more commonly applied in clinical research, especially in the analysis of vital signs, laboratory values, interventions, etc \citep{MANZINI2022104218,ramaswamy2021ckd,lou2021learning,chen2022learning,zeldow2021functional}. Post-clustering inference for functional data is a challenging problem and existing testing methods are often not applicable. The main challenge of this problem is the selection bias, which would lead to inflated false discoveries if uncorrected, induced by clustering algorithms. In more detail, the clustering forces separation regardless of the underlying truth, making the $p$-value spuriously small. In practice, empirical observations reveal that applying classical methods often leads to spuriously small $p$-values \citep{hall2007two,zhang2007statistical,horvath2012inference,qiu2021two}. This is an instance of a broader phenomenon termed {\it data snooping} \citep{ioannidis2005most}, referring to the misuse of data analysis to find patterns in data that can be presented as statistically significant, thus leading to potentially false conclusions.

The selective inference framework is commonly employed as a remedy for selection bias. However, the focus of selective inference has primarily been on data with discrete observations. Due to the nature of EHR data, there is an urgent demand for a novel selective inference framework that accommodates continuous functional datasets with unaligned observations.

To address this challenge, in this paper, we develop a valid test for the difference in means between two clusters estimated from the functional data, named {\it \underline{P}ost-clustering \underline{S}elective \underline{I}nference for \underline{M}ulti-feature \underline{F}unctional Data} (\algo). To handle the continuity of functional datasets, which often contain large timesteps and cannot be treated as discrete data, our method finds the low-rank spectral representation for the continuous data based on kernel ridge regression.  {Throughout the paper, the clustering step is treated as a two-cluster procedure, so the inferential target is a selected two-cluster contrast rather than a general multi-cluster comparison.} To address the selection bias in the inference procedure, we propose a selective inference framework leveraging the clustering information. Next, we discuss the EHR phenotyping problem before
introducing more details of our procedure.

%%%%%%%%%%%%%%%%%
\subsection{Application: Phenotyping Based on Electronic Health Records}\label{sec:phenotyping-EHR}
%%%%%%%%%%%%%%%%%

Phenotyping refers to the process of identifying specific clinical characteristics or patterns of patients. The application of longitudinal clustering methods to electronic health records (EHR) data has proven to be a powerful tool for phenotyping, offering novel insights into patient heterogeneity and disease progression. Numerous studies have similarly utilized longitudinal clustering methods with EHR data to identify various patient subtypes and advance clinical research. For instance, researchers studied type 2 diabetes mellitus (T2DM) patients by analyzing their data on various biochemical markers \citep{MANZINI2022104218}. These markers included glycated hemoglobin (HbA1c), body mass index (BMI), and diastolic and systolic blood pressures, among others. By applying longitudinal deep learning clustering methods on EHR, \cite{MANZINI2022104218} identified seven different subtypes of T2DM. In addition, a hybrid semimechanistic modeling methodology was introduced to analyze the progression of chronic kidney disease (CKD) \citep{ramaswamy2021ckd}. When applied to the EHR data of CKD patients, the model effectively identified five distinct patient subpopulations. Through this pioneering method, the emphasis was placed on harnessing longitudinal data to understand disease progression phenotypes, thereby aiming to streamline individualized treatment strategies for each subgroup. 

%%%%%%%%%%%%%%%%%
\subsection{Main Contributions}\label{sec:contributions}
%%%%%%%%%%%%%%%%%

Our work introduces a post-clustering selective inference framework for functional data and provides theoretical guarantees to control selective errors under the Gaussian distributional assumption. Our framework comprises three parts:
\begin{enumerate}
	\item We utilize low-dimensional embedding to transform high-dimensional functional data into low-dimensional tensors (i.e., three-way arrays) while simultaneously imputing missing values. This embedding is a linear transformation that preserves normality, resulting in a random tensor where each slice follows a matrix normal distribution.
	
	\item We propose an estimator to evaluate the unknown covariance matrices of the matrix normal distribution and use the estimated covariance matrices to perform a whitening transformation.
	
	\item We define the selective $p$-value based on the tensor obtained through low-dimensional embedding and the whitening transformation. Inspired by previous work, our selective $p$-value leverages clustering information to reduce selection bias and control the selective type-I error. Furthermore, we prove that the proposed $p$-value is the conditional probability of a scaled Chi-square distribution truncated to a subset. We also introduce a Monte Carlo approximation to estimate the proposed selective $p$-value.
\end{enumerate}
Our work presents two major novelties compared to previous works \citep{gao2024selective,chen2023selective,yun2023selective,hivert2022post}. First, our selective inference framework addresses functional data with missing values and multiple features, while previous works often focus on vector inputs. We impute missing values through low-dimensional embedding, specifically using basis expansion regression. This linear transformation preserves both null and alternative hypotheses, transforming records of a feature into a low-dimensional vector. The resulting data has a tensor structure induced by the multiple features. Consequently, we extend selective inference for matrix inputs \citep{gao2024selective} into the tensor case and define the selective 
$p$-value.

Second, we employ the sample covariance estimator for the whitening transformation. Unlike previous works that often assume covariance matrices are scaled identity matrices, this assumption may not hold in our scenarios with multi-feature functional observations. Therefore, estimators for the scaled parameter, such as the mean estimator \citep{gao2024selective}, may fail in functional settings. To address this issue, we demonstrate that the problem essentially boils down to estimating the covariance of a truncated normal distribution, and we develop a sample covariance estimator accordingly. We prove that the sample covariance estimator is consistent under the null hypothesis and our selective inference framework then controls the selective type-I error. Furthermore, we show that the statistical power converges to 1, and the proposed selective inference framework is asymptotically powerful.

The merit of the proposed procedure is illustrated in a real data example on Acute Kidney Injury (AKI) EHR data. AKI is a potentially life-threatening condition that impacts approximately 20\% of hospitalized patients in the United States \citep{wang2012acute}. Given this prevalence, early warning of patient outcomes becomes crucial as it can significantly improve prognosis \citep{macleod2009ncepod}. Identifying new subphenotypes often serves as the foundation for such early warnings. The most direct, insightful, and currently available indicator for AKI is the temporal trajectory of creatinine. We apply our proposed method to the inference after longitudinal clustering of AKI based on creatinine. 
In conducting this, we utilized EHR data from the \textit{MIMIC-IV} database \citep{johnson2020mimic,johnson2023mimic,goldberger2000physiobank}. Our approach yields results that are both meaningful and credible.

The code of the proposed \algo{} is available online (\url{https://github.com/Telvc/PMISF}).

%%%%%%%%%%%%%%%%%
\subsection{Related work}\label{sec:related-work}
%%%%%%%%%%%%%%%%%

\paragraph{Selective Inference.} In classic statistical inference, hypotheses are assumed to be predetermined before observing the dataset. However, in a broad range of supervised and unsupervised learning tasks, such as regression and clustering, the hypotheses are often data-driven. Consequently, the model selection step introduces selection bias, rendering classical inference methods inadequate. To address this issue, \citet{berk2013valid}, \citet{fithian2014optimal}, and \citet{lee2016exact} developed the selective inference framework, a process for making statistical inferences that account for the selection effect. Building on the work of \citet{lee2016exact}, selective inference has been extensively applied to the high-dimensional linear models \citep{tibshirani2016exact, yang2016selective, loftus2015selective, charkhi2018asymptotic, taylor2018post, hyun2021post, jewell2022testing}. In recent years, \citet{gao2024selective} proposed an elegant selective inference framework for conducting hypothesis tests on post-clustering datasets, inspiring a series of subsequent studies on this topic \citep{chen2023selective, zhang2019valid, hivert2022post, yun2023selective,chen2023testing}. While most existing work focuses on post-clustering inference for discrete data, this paper aims to develop a selective inference framework for multi-feature functional data.

\paragraph{Functional Clustering.} In this paper, we investigate post-clustering inference for functional data. Functional clustering involves categorizing curves, functions, or shapes based on their patterns or structures. This method has been explored in functional data analysis literature due to its practical applications. For example, \cite{abraham2003unsupervised, serban2005cats, kayano2010functional, coffey2014clustering, giacofci2013wavelet} developed two-stage clustering methods that leverage functional basis expansion. These methods reduce the dimensionality of functional data through basis expansion regression before implementing clustering techniques for low-dimensional vectors. In contrast, \cite{peng2008distance, chiou2007functional} proposed methods that select the basis using functional principal components (FPC), avoiding the need for a prespecified set of basis functions. Additionally, other research directions include functional clustering approaches such as leveraging the FPC subspace-projection \citep{chiou2012dynamical, chiou2008correlation} and model-based clustering \citep{banfield1993model, james2003clustering, jacques2014model, heinzl2014clustering}.

{\paragraph{Sparse functional data.} For sparsely and irregularly observed trajectories, population mean and covariance functions can be estimated by pooling observations across subjects rather than requiring dense records for every individual. \citet{yao2005functional} developed the PACE framework for sparse longitudinal data, and \citet{zhang2016sparse} established unified asymptotic results for pooled local-linear mean and covariance estimators from sparse to dense regimes, including asymptotic normality. These approaches estimate population-level functional quantities. By contrast, the ridge map in Section~\ref{s2.1} is a subject-level linear embedding used to construct the inputs to clustering and selective inference.}

\paragraph{Modeling of Matrix Distribution.} In this paper, we model multi-feature functional data using the matrix normal distribution. To implement the whitening transformation in our proposed selective inference framework, we propose to estimate the block covariance matrix, determined by the Kronecker product of two covariance matrices. Estimating the block covariance matrix has been extensively explored in the literature \citep{dawid1981some, dutilleul1999mle, yin2012model, tsiligkaridis2013covariance, zhou2014gemini, hoff2015multilinear, ding2018matrix, hoff2023core}.

%%%%%%%%%%%%%%%%%
\subsection{Notation and Preliminaries}\label{sec:notation}
%%%%%%%%%%%%%%%%%

In this paper, we denote $\mathcal{MN}(\mu, \Sigma_1, \Sigma_2)$ as the matrix normal distribution with mean $\mu$ and covariance matrices $\Sigma_1, \Sigma_2$. Specifically, if $Z$ is a random matrix with i.i.d. standard Gaussian entries, then $\mu + \Sigma_1^{1/2} Z \Sigma_2^{1/2} \sim \mathcal{MN}(\mu, \Sigma_1, \Sigma_2)$. For any positive integer $m$, $\mathbb{S}_+^m$ denotes the collection of all $m$-by-$m$ symmetric positive semi-definite matrices. For any matrix $A\in\mathbb{R}^{m\times n}$, we denote $\lVert A\rVert_F$ as its Frobenius norm, and $\text{vec}(A)\in\mathbb{R}^{mn}$ denotes the vectorization of $A$, which is defined as follows:
\begin{equation*}
	\begin{split}
		& \text{vec}\left(\begin{bmatrix}
	A_{11} & \cdots & A_{1m}\\
	\vdots & \ddots & \vdots\\
	A_{n1} & \cdots & A_{nm}
\end{bmatrix}\right) \\
= & \begin{bmatrix}
	A_{11} & \cdots & A_{1m} & A_{21} & \cdots & A_{2m} & \cdots & A_{n1} & \cdots & A_{nm}
\end{bmatrix}^\top \in \mathbb{R}^{mn}.
\end{split}
\end{equation*}
For any Hilbert space $\mathcal{H}$, we denote $\lVert\cdot\rVert_{\mathcal{H}}$ as the associated norm. For any vector function $\mu = (\mu_1,\mu_2,\cdots,\mu_n):\mathcal{D}\rightarrow\mathbb{R}^n$, where $\mathcal{D}\subseteq\mathbb{R}$ is the domain, we denote $\lVert \mu\rVert_{\infty} = \sup_{x\in\mathcal{D}}\max\{|\mu_1(x)|,|\mu_2(x)|,\cdots,|\mu_n(x)|\}$. For any functions $f, g$, we define $f\odot g$ as their Cartesian product. Namely, for any $(f\odot g)(x,y) = f(x)\cdot g(y)$, where $x, y$ are in the domains of $f, g$ respectively. For any two matrices $A\in \mathbb{R}^{m\times n}, B\in \mathbb{R}^{p\times q}$, define 
$$A\otimes B = \begin{bmatrix}
	a_{11} B & \cdots & a_{1n}B\\
	\vdots & \ddots & \vdots\\
	a_{m1} B & \cdots & a_{mn}B
\end{bmatrix} \in \mathbb{R}^{(pm) \times (qn)}$$
as their Kronecker product. For any positive integer $n$ and any $n$-mode tensor $\mathcal{A}\in\mathbb{R}^{i_1\times\cdots\times i_n}$, define $\mathcal{A}[j,:,\cdots,:]$ as its $j$th mode-1 slice, $\mathcal{A}[:,j,:,\ldots,:]$ as its $j$th mode-2 slice, etc. For any matrix $U\in\mathbb{R}^{i_m\times K}$, where $i_1,\ldots,i_n, K$ are positive integers, define $U\times_m \mathcal{A}\in\mathbb{R}^{i_1\times\cdots i_{m-1}\times K\times i_{m+1}\times i_n}$ as their mode-$m$ tensor product. The cardinality of a set $A$ is denoted by $|A|$.

%%%%%%%%%%%%%%%%%
\subsection{Paper Organization}\label{sec:roadmap}
%%%%%%%%%%%%%%%%%

This paper is organized as follows. In Section \ref{s2}, we introduce the problem formulation of post-clustering inference for functional data. Section \ref{s3} introduces our proposed method \algo. Section \ref{s4} proves how our method achieves bounded selective type-I and type-II errors. Lastly, Section \ref{s5} presents our numerical experiments on synthetic data to validate our theory and on real-world Acute Kidney Injury (AKI) EHR data.

%%%%%%%%%%%%%%%%%
\section{Problem Formulation}\label{s2}
%%%%%%%%%%%%%%%%%

This section introduces the problem of post-clustering inference for functional data, illustrated in the context of the EHR data analysis. EHR contains records of diverse features for different patients, where each record of a feature and a patient forms a trajectory of functional data. We consider the EHR data from $n$ patients and $m$ features. We observe $\mathcal{W} = (W_{ij})_{i\in[n], j\in [m]}$ for each subject $i\in[n]$ and feature $j\in [m]$, where $W_{ij}$ is the observed data of the $i$th subject and $j$th feature within a certain period recording their physical features. Let $\Omega = (\Omega_{ij})_{i\in[n], j\in [m]}$ be the corresponding time points of the record $\mathcal{W}$, where $\Omega_{ij}\coloneqq (t_{ijk})_{k\in[r_{ij}]}\in\mathbb{R}^{r_{ij}}$ is the record of time points for the $j$th feature of the $i$th subject, $r_{ij}$ is the number of time points for this record, and $W_{ij} \coloneqq (W_{ij}(t_{ijk}))\in\mathbb{R}^{r_{ij}}$ is the record for the $j$th feature of the $i$th patient. For all $i\in[n], j\in[m]$ and $k\in[r_{ij}]$, denote the time point of observations as $\{t_{ijk}\} \subseteq [0,T]$. In summary, the data for each subject $i$ comprises $m$ features, with each feature $j$ represented by a vector $W_{ij}$ corresponding to the time points $\Omega_{ij}$. Our observations are thus $W_{ij}(t)$, for $i\in [n]$, $j\in[m]$, and $t \in \Omega_{ij}$. Real EHR data often contain missing values, resulting in $\Omega_{ij}$ frequently having low cardinality. Given these data, our objective is to uncover the phenotypes of the subjects, i.e., the potential clusters among these subjects.

%%%%%%%%%%%%%%%%%
\subsection{Model Setup}\label{sec:setup}
%%%%%%%%%%%%%%%%%

Next, we introduce the model of functional post-clustering inference. We assume the measurements of each feature $W_{ij}$ along time follow a Gaussian process. This implies that the feature records on a set of time points follow a multivariate normal distribution. Given the similarity between subjects and for the sake of analytical simplicity, we assume that these Gaussian processes $W_{ij}$ share a common covariance function across all subjects $i$. Furthermore, as each subject $i$ contains multiple features, the record $W_i = (W_{ij})_{j\in [m]}$ for the subject $i$ could be viewed as a multivariate Gaussian process, which is formally defined as follows.
\begin{definition}[Multivariate Gaussian process]
	We denote $f \sim \mathcal{MGP}(\mu,R)$ and say $f \coloneqq (f_1, f_2, \cdots, f_m)$ is a multivariate Gaussian process on $[0,T]$ with the vector-valued mean function $\mu\coloneqq (\mu_1, \cdots, \mu_m): [0, T]\rightarrow \mathbb{R}^m$ and covariance function $R\coloneqq (R_{j_1j_2})_{j_1,j_2\in[m]}: [0,T]\times[0,T]\rightarrow \mathbb{R}^{m\times m}$, if the following holds for any $t_1,t_2,\cdots, t_r\in [0,T]$:
	\[\text{vec}(f(t_1),f(t_2),\cdots,f(t_r)) \sim \mathcal{N}(\text{vec}(\mu(t_1),\mu(t_2),\cdots,\mu(t_r)),\Sigma),\]
	where $\Sigma$ is defined as follows. For any $s,t\in[mr]$, there exist unique $j_1,j_2\in[m]$ and $k_1,k_2\in[r]$ such that $s = (j_1-1)r+k_1, t = (j_2-1)r+k_2$. Define
	\[\Sigma[s,t] \coloneqq Cov(f_{j_1}(t_{k_1}),f_{j_2}(t_{k_2})) = R_{j_1j_2}(t_{k_1}, t_{k_2}).\]
	Here $R_{j_1j_2}$ is the auto-covariance function when $j_1 = j_2$ and is the cross-covariance function if $j_1 \neq j_2$. 
\end{definition}
We introduce the following assumption on the distribution of observations $W_{ij}$.
\begin{assumption}[Distributional Assumption of Observations]\label{a2}
	Suppose $W_{ij}$'s satisfy 
	\begin{equation}\label{3}
		\begin{aligned}
			W_{ij}(t_{ijk}) = Z_{ij}(t_{ijk}) + \epsilon_{ijk}\quad\text{for all}\quad t_{ijk} \in \Omega_{ij},\\
			\text{and}\quad Z_{i}\sim \mathcal{MGP}(\mu_i,R),  \epsilon_{ijk} \stackrel{iid}{\sim} \mathcal{N}(0, \sigma_j^2), \quad Z_{i}, \epsilon_{ijk} \text{ are independent}.
		\end{aligned}
	\end{equation}
	Here, $\mu_i\coloneqq(\mu_{ij})_{j\in[m]}: [0,1]\rightarrow\mathbb{R}^m$  is the mean vector function for subject $i$, $Z_i$ follows the multivariate Gaussian process, $\epsilon_{ijk}$ is the Gaussian noise, and $R$ is the covariance function. Suppose that $R$ and $\mu_i$ are Lipschitz continuous for all $i\in[n]$. $Z_{1},\ldots, Z_{n}$ are all independent.  In addition, for all $i\in[n], j\in[m]$, suppose $t_{ijk}\stackrel{iid}{\sim}\mathcal{U}[0,1]$ for all $k\in[r_{ij}]$.
\end{assumption}
Assumption \ref{a2} concerns all features within the period $[0, T]$ and supposes they follow the multivariate Gaussian process with additive noise. Under Assumption \ref{a2}, for different subjects, the auto- and cross-covariance kernels of $W_i$ are identical, while their mean functions, denoted as $\mu_{i}$, may differ across different $i$.

In addition to the actual observations of the data $W_{ij}$, for the convenience of presenting our methods and theory, we also assume $\mathcal{W} = (w_{ij})_{i\in[n], j\in[p]}$ is a collection of $n$ random samples, each generated according to Model \eqref{3}.

%%%%%%%%%%%%%
\subsection{Formulation of the Post-clustering Inference Problem}\label{sec:formulation}
%%%%%%%%%%%%%

In two-component clustering analysis, we apply a functional clustering algorithm, such as two-stage clustering methods with functional basis expansion \citep{abraham2003unsupervised, serban2005cats, kayano2010functional, coffey2014clustering, giacofci2013wavelet} on $W$ to obtain two clusters of subjects, denoted by $\mathcal{C}_1$ and $\mathcal{C}_2$. Here, $\mathcal{C}_1$ and $\mathcal{C}_2$ record the indices of subjects in Clusters 1 and 2, respectively, and form a partition of $[n]$. We aim to test if there is a significant difference in the means of clusters $\mathcal{C}_1, \mathcal{C}_2$.

In previous work on selective inference for matrix data, \cite{gao2024selective} considers the hypothesis test
\[\tilde{H_0}^{\{\mathcal{C}_1,\mathcal{C}_2\}}:\overline{\mu}_{\mathcal{C}_1} =\overline{\mu}_{\mathcal{C}_2}
\quad \text{versus}\quad \tilde{H_1}^{\{\mathcal{C}_1,\mathcal{C}_2\}}:\overline{\mu}_{\mathcal{C}_1} \neq\overline{\mu}_{\mathcal{C}_2},\]
where $\overline{\mu}_{\mathcal{C}_1}\coloneqq \sum_{i\in\mathcal{C}_1}\mu_i/|\mathcal{C}_1|$ and $\overline{\mu}_{\mathcal{C}_2}\coloneqq \sum_{i\in\mathcal{C}_2}\mu_i/|\mathcal{C}_2|$ denote the group means of clusters $\mathcal{C}_1$ and $\mathcal{C}_2$.  {This group-average hypothesis is not directly accommodated by our current functional selective-inference construction when the covariance function $R$ is unknown.} To elaborate, \cite{gao2024selective} assumes $X_i\sim \mathcal{N}(\mu_i,\sigma^2 I_d)$, where $X_i$ is the data in vector format (e.g., sequencing reads), $\mu_i$ is the mean function, and $\sigma^2$ is the unknown variance parameter. In contrast, in the functional setting, the covariance matrix $\sigma^2 I_d$ is replaced by a covariance function $R$, and estimating $R$ without additional assumptions is challenging. This difficulty arises because the estimation of $R$ requires knowledge of the mean function $\mu_i$, and any non-zero difference in $\mu_i$ within a cluster $\mathcal{G}\in{\mathcal{C}_1,\mathcal{C}_2}$ would introduce a nuisance parameter, complicating the estimation process. Similar phenomena and discussions were presented in \cite{yun2023selective}, where they extended the selective inference framework for matrix data proposed in \cite{gao2024selective,chen2023selective} to settings with unknown variance.

To address the aforementioned issue, we propose a modeling approach for the null hypothesis, wherein all subjects in a cluster have the same mean function. The variability of observations across different samples is encapsulated through their covariance function $R$.  We define $\mu_{\mathcal{C}_1} = \mu_i, \forall i\in \mathcal{C}_1$ and $\mu_{\mathcal{C}_2} = \mu_i, \forall i\in \mathcal{C}_2$ as the mean functions of samples in clusters $\mathcal{C}_1$ and $\mathcal{C}_2$, respectively. Consequently, the task of post-clustering selective inference can be formulated as the following hypothesis-testing problem:
\begin{equation}\label{1}
	H_0^{\{\mathcal{C}_1,\mathcal{C}_2\}}:\mu_{\mathcal{C}_1} =\mu_{\mathcal{C}_2}
	\quad \text{versus}\quad H_1^{\{\mathcal{C}_1,\mathcal{C}_2\}}:\mu_{\mathcal{C}_1} \neq\mu_{\mathcal{C}_2}.
\end{equation}

 {The common within-cluster mean is a working assumption of the inferential model rather than a property guaranteed by the clustering algorithm. It still allows subjects in the same cluster to have different realized trajectories through the centered stochastic-process variation governed by $R$ and through measurement error. Systematic subject-specific departures from the common mean fall outside the exact theory; their effect on covariance estimation and whitening is discussed in Section~\ref{sec:covariance-estimation} and examined numerically in Section~\ref{s5}.}

\begin{figure}[t]
	\centering
	\subfigure[]{
		\begin{minipage}[t]{0.47\textwidth}
			\centering
			\includegraphics[width=0.9\textwidth]{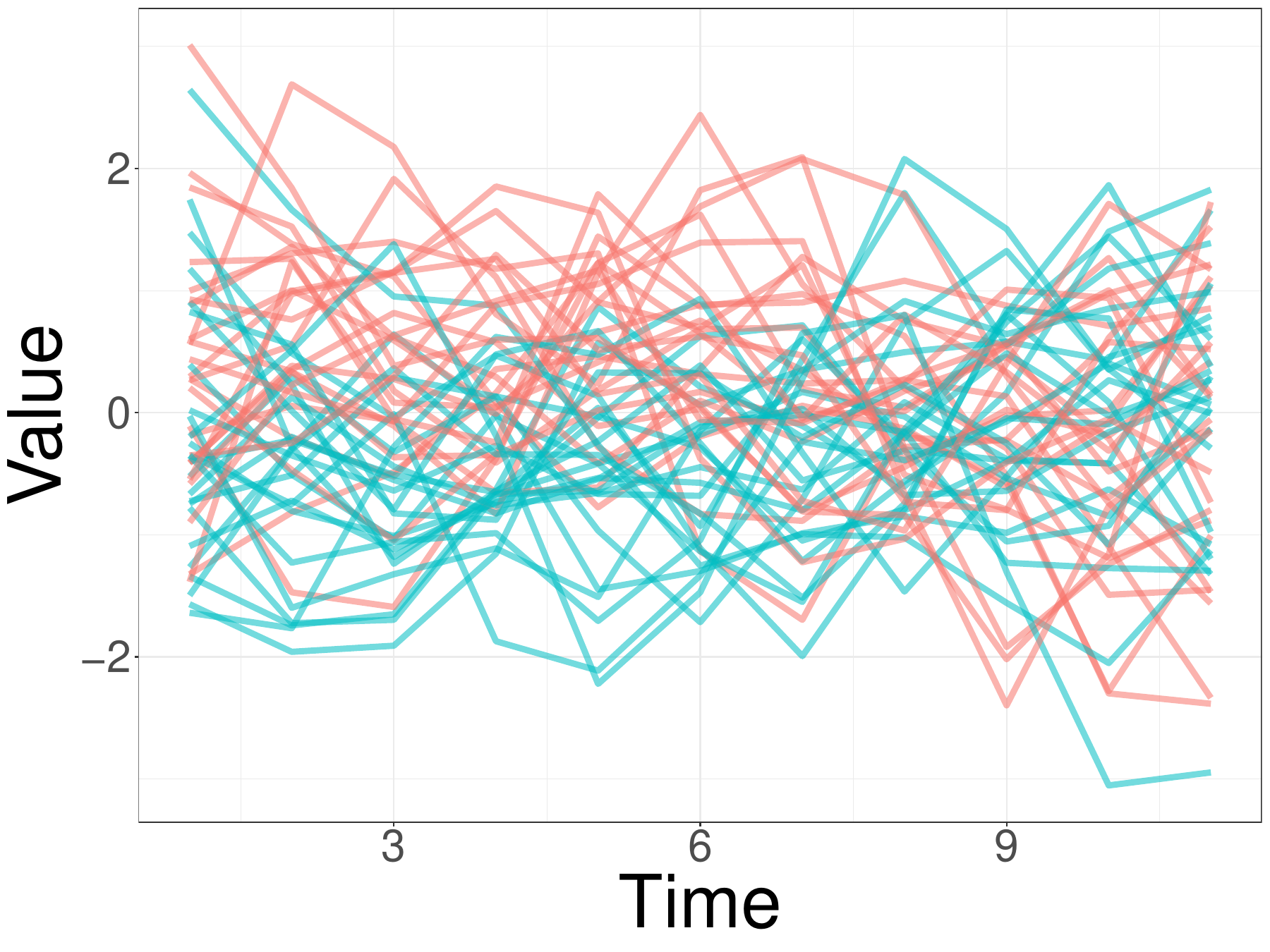}
		\end{minipage}
		\label{fig1a}
	}
	\subfigure[]{
		\begin{minipage}[t]{0.47\textwidth}
			\centering
			\includegraphics[width=0.9\textwidth]{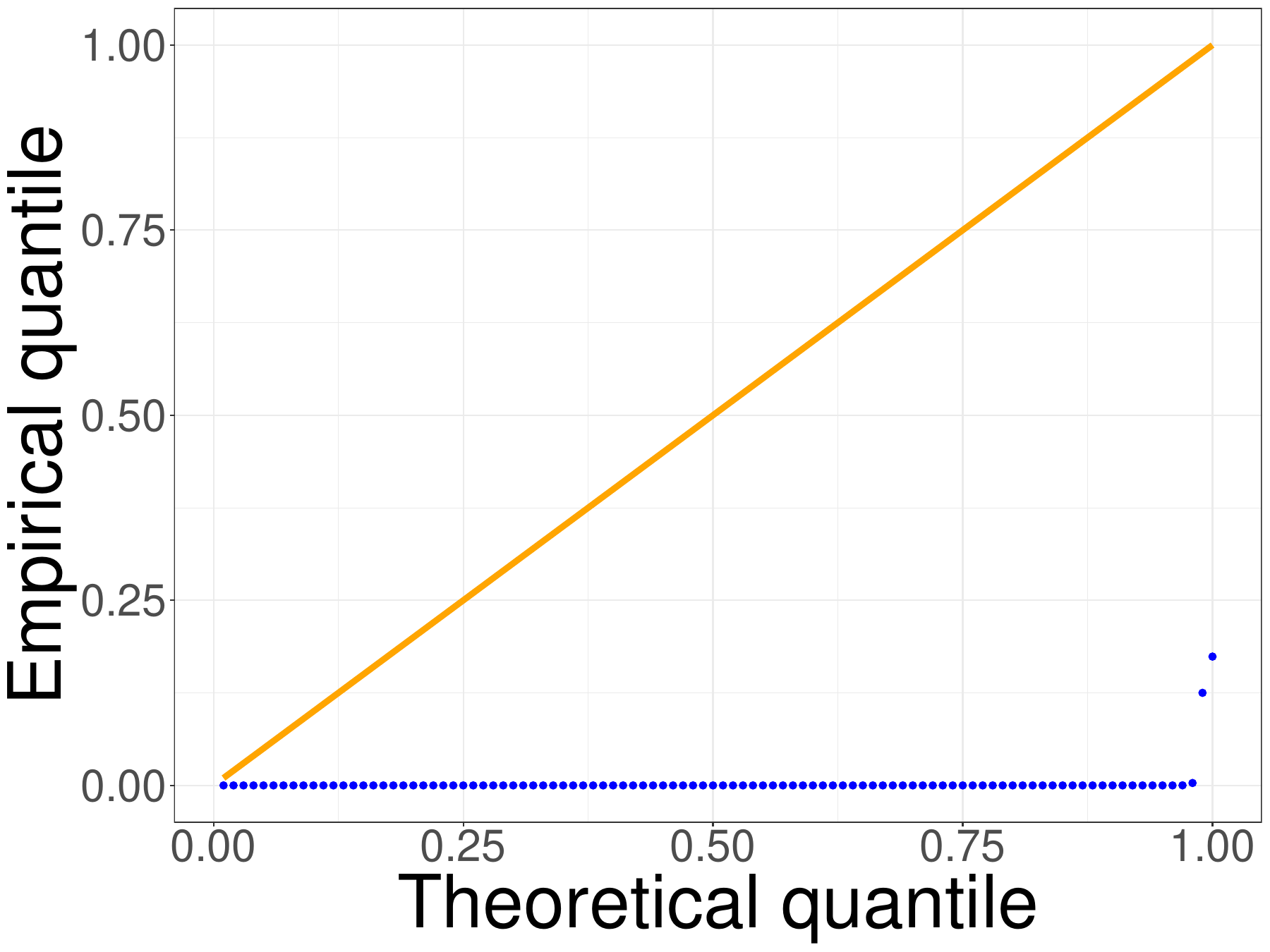} 
		\end{minipage}
		\label{fig1b}
	}
	\caption{\textbf{Failure of the Wald test.} We sample 100 datasets following Model \eqref{3} with zero mean functions and zero noise terms ($\mu_i=0, \sigma_j^2 = 0$ for all $i, j\in[m]$). Each dataset contains 500 subjects, where the record for each subject contains 1 feature with 11 time points scattered uniformly in $[0, 1]$ (i.e., $n = 500$, $m = 1$, $r_{ij} = 11$ for all $i\in[500], j\in[1]$). For each dataset, we apply $k$-means to obtain two clusters. \textbf{(a)} shows the first 100 records of the first dataset labeled by the clustering outputs. \textbf{(b)} is the quantile plot of the $p$-values for the 100 datasets obtained by the Wald test.}
	\label{fig1}
\end{figure}

A natural approach for solving \eqref{1} is to apply Wald test \citep{wald1943tests}. Considering a special case: suppose the time points of measurements $\Omega_i$ are consistent across all subjects. In this case, $w_i = (w_{ij})_{j\in[m]}$ forms a $\left(\sum_{j = 1}^m r_{1j}\right)$-dimensional vector. Denote $\overline{w}_{\mathcal{C}_1}$ and $\overline{w}_{\mathcal{C}_2}$ as the sample means within clusters $\mathcal{C}_1$ and $\mathcal{C}_2$, i.e., $\overline{w}_{\mathcal{G}} = \sum_{i\in\mathcal{G}}w_i/|\mathcal{G}|$ for all $\mathcal{G}\in\{\mathcal{C}_1,\mathcal{C}_2\}$. A straightforward way to evaluate the $p$-value is
\begin{equation}\label{2}
	\mathbb{P}_{H_0^{\{\mathcal{C}_1,\mathcal{C}_2\}}}\left(\lVert \overline{W}_{\mathcal{C}_1}-\overline{W}_{\mathcal{C}_2}\rVert\geq\lVert\overline{w}_{\mathcal{C}_1}-\overline{w}_{\mathcal{C}_2}\rVert\right),
\end{equation}
where $\overline{W}_{\mathcal{G}} = \sum_{i\in\mathcal{G}}W_i/|\mathcal{G}|$ for all $\mathcal{G}\in\{\mathcal{C}_1,\mathcal{C}_2\}$.  {However, this method can be invalid because it does not control the post-clustering selective type-I error: under the null, the resulting naive $p$-values are not approximately uniform and can accumulate near $0$ or $1$, which is problematic in practice. See Figure \ref{fig1} for an illustrative example.}

%%%%%%%%%%%%%%
\section{Selective Inference for Functional Data Clustering}\label{s3}
%%%%%%%%%%%%%%
\begin{figure}[!ht]
	\centering
	\subfigure[]{
		\begin{minipage}[t]{0.3\textwidth}
			\centering
			\includegraphics[width=0.9\textwidth]{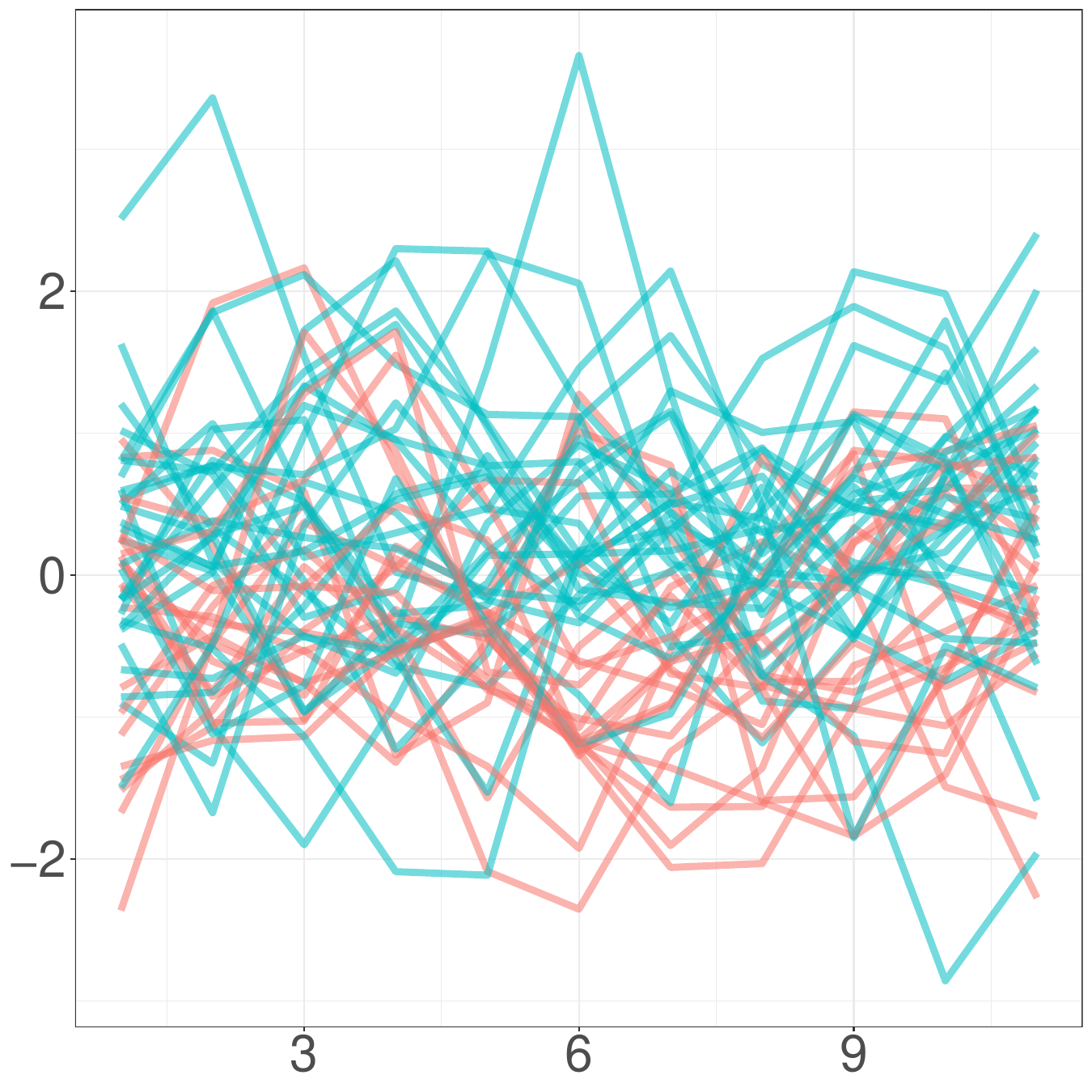}
		\end{minipage}
		\label{fig2a}
	}
	\subfigure[]{
		\begin{minipage}[t]{0.3\textwidth}
			\centering
			\includegraphics[width=0.9\textwidth]{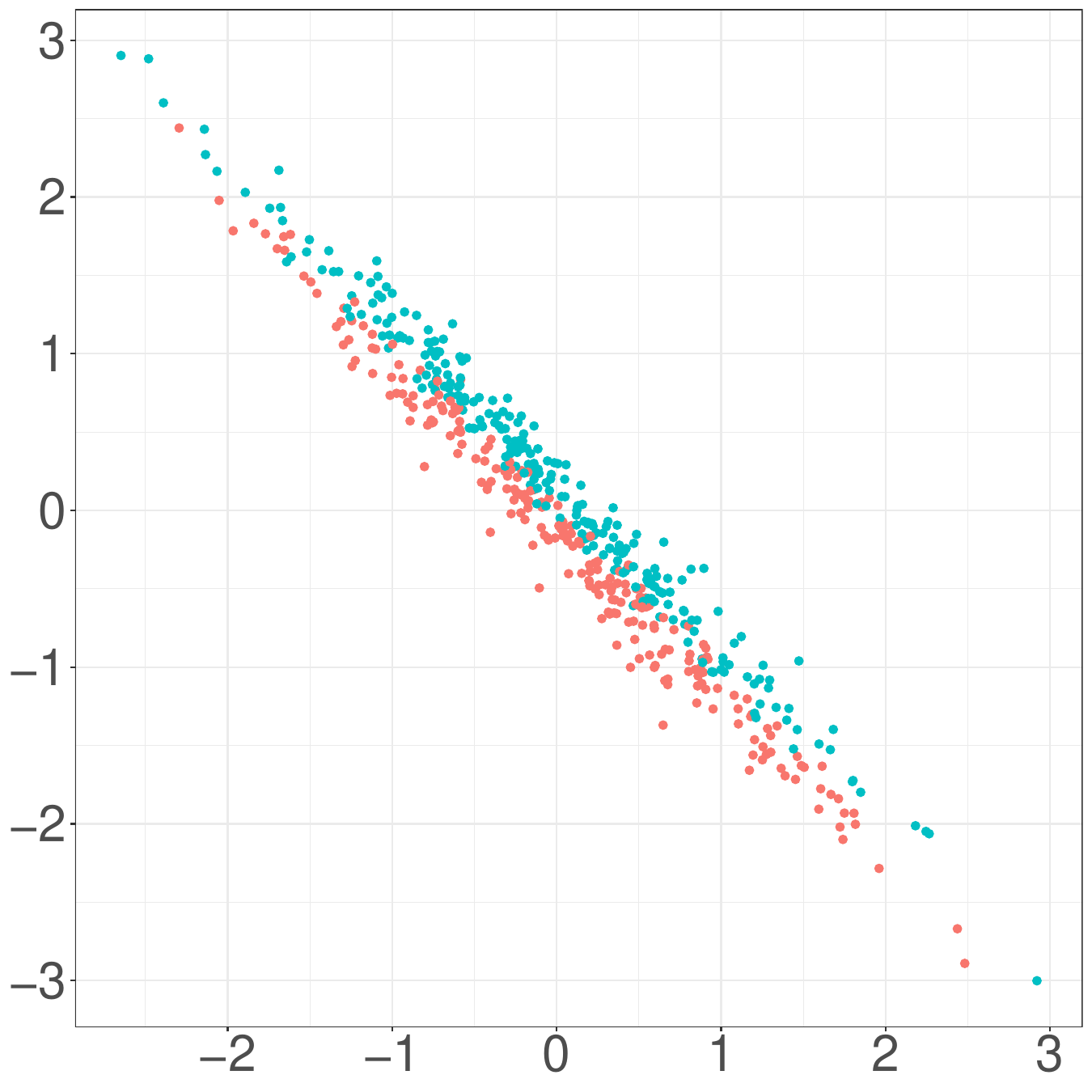} 
		\end{minipage}
		\label{fig2b}
	}
	\subfigure[]{
		\begin{minipage}[t]{0.3\textwidth}
			\centering
			\includegraphics[width=0.9\textwidth]{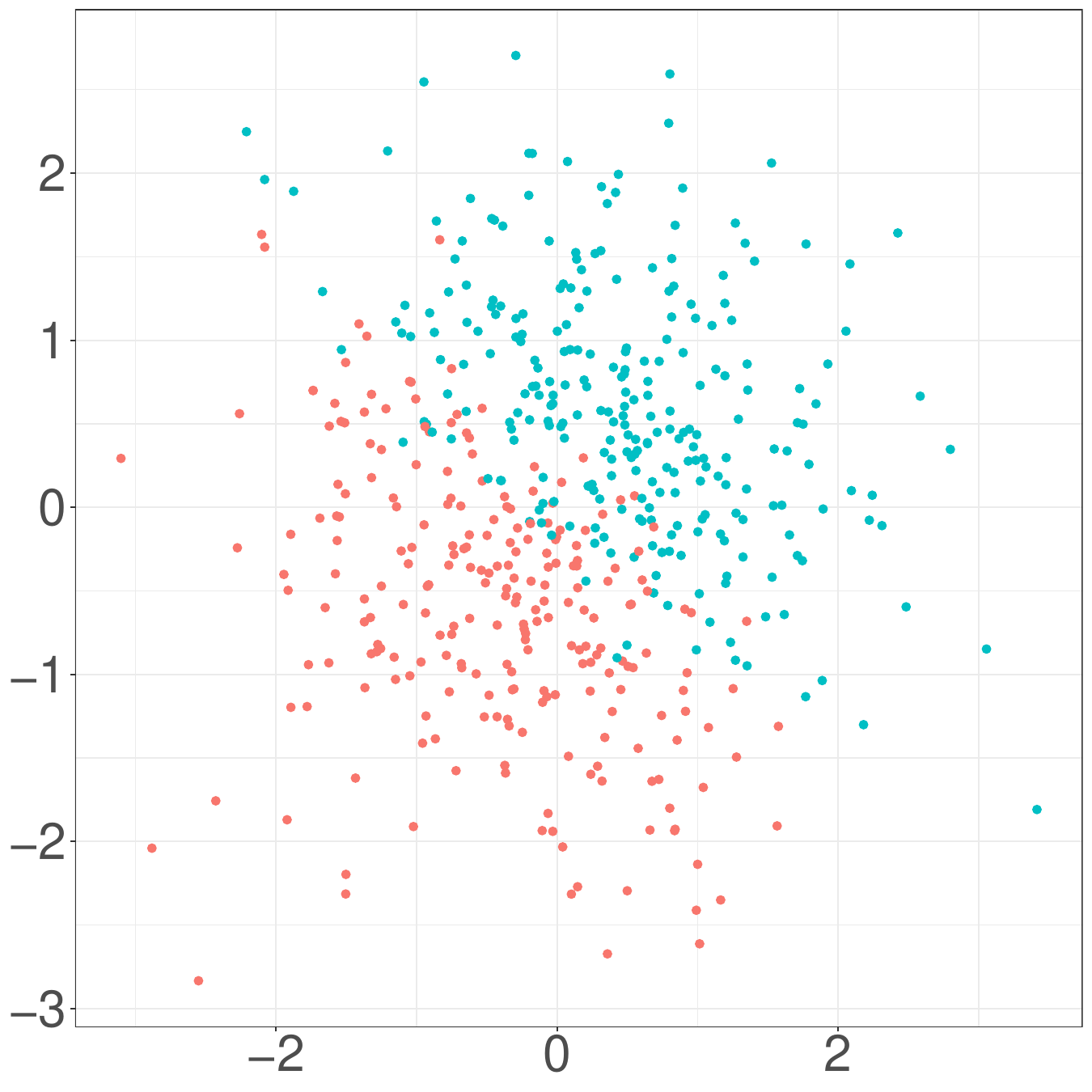} 
		\end{minipage}
		\label{fig2c}
	}
	\subfigure[]{
		\begin{minipage}[t]{0.3\textwidth}
			\centering
			\includegraphics[width=0.9\textwidth]{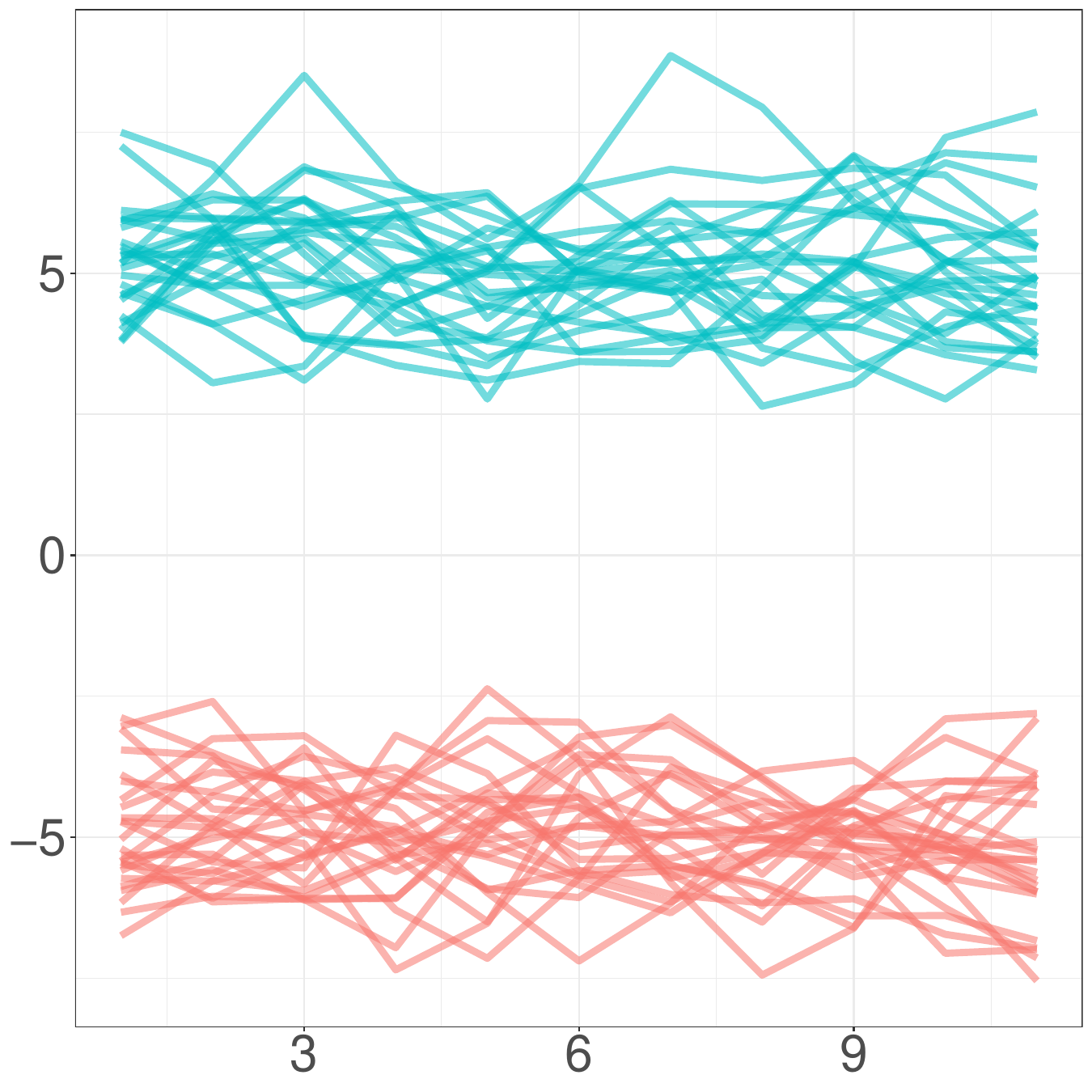}
		\end{minipage}
		\label{fig2d}
	}
	\subfigure[]{
		\begin{minipage}[t]{0.3\textwidth}
			\centering
			\includegraphics[width=0.9\textwidth]{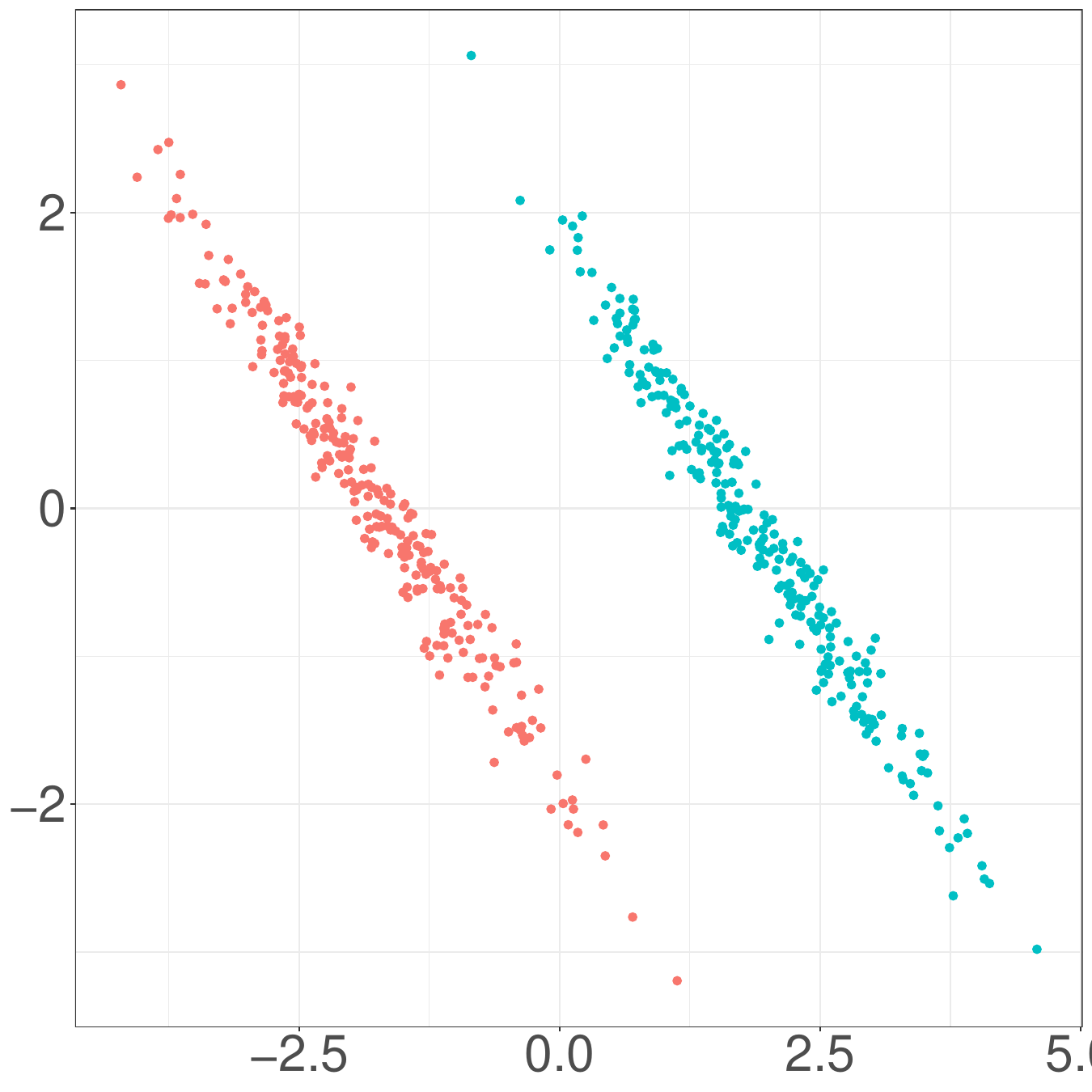} 
		\end{minipage}
		\label{fig2e}
	}
	\subfigure[]{
		\begin{minipage}[t]{0.3\textwidth}
			\centering
			\includegraphics[width=0.9\textwidth]{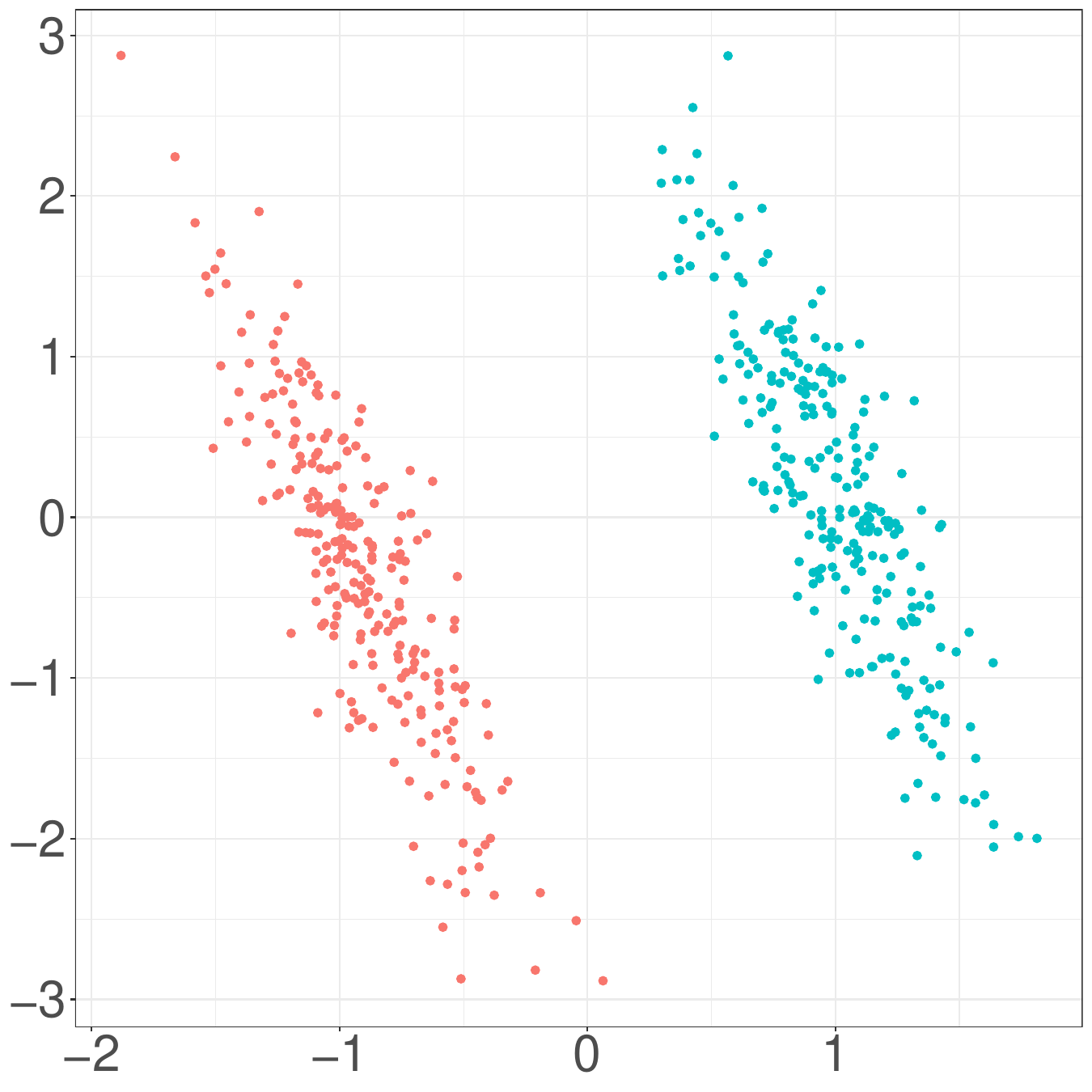} 
		\end{minipage}
		\label{fig2f}
	}
	\caption{\textbf{Overall procedure.} \textbf{(a)}. the first 100 curves of the dataset with the common mean $\mu_i = 0$ and the zero noise $\epsilon_i = 0$ labeled by the clustering results. \textbf{(b)}. scatter plot (first two coordinates) of the low-dimensional embedding for this dataset. \textbf{(c)}. scatter plot of the whitened data; \textbf{(d)}. the first 100 curves of the dataset with the mean $\mu_i = -5$ or  $\mu_i = 5$ and the zero noise term labeled by the clustering results; \textbf{(e)}. scatter plots (first two coordinates) for the low-dimensional embedding and whitened data; \textbf{(f)}. scatter plots (first two coordinates) for the low-dimensional embedding and whitened data.} 
	\label{fig2}
\end{figure}

In this section, we introduce the procedure of \algo\ for testing the difference between post-clustering matrix data following Model \eqref{3} that addresses the challenges described in Section \ref{sec:formulation}. 

Our procedure is based on the selective inference framework, which adjusts the inferential process to account for the selection that has occurred, thereby providing statistically valid conclusions. In the context of our hypothesis test \eqref{1}, we aim to test $\mu_{\mathcal{C}_1} =\mu_{\mathcal{C}_2}$ given $\mathcal{C}_1, \mathcal{C}_2\in\mathcal{C}(w)$, where $w\coloneqq (w_i)_{i\in[n]}$ is a random sample generated by Model \eqref{3}. A natural approach is deriving the selection procedure by constructing a $p$-value conditioning on the clustering outputs $\mathcal{C}_1$ and $\mathcal{C}_2$. Specifically, in the special case where the time points of measurements $\Omega_i$ are consistent across all subjects, the selective $p$-value could be defined as follows:
\begin{equation}\label{8}
	\mathbb{P}_{H_0^{\{\mathcal{C}_1, \mathcal{C}_2\}}}\left(\lVert \overline{W}_{\mathcal{C}_1}-\overline{W}_{\mathcal{C}_2}\rVert\geq\lVert\overline{w}_{\mathcal{C}_1}-\overline{w}_{\mathcal{C}_2}\rVert\Big|~\mathcal{C}_1, \mathcal{C}_2\in\mathcal{C}(\mathcal{W})\right).
\end{equation}
However, in practice, the time records $\Omega_i$ often differ across subjects, rendering the sample means $\overline{w}_{\mathcal{C}_1}$ and $\overline{w}_{\mathcal{C}_2}$ ill-defined, and the direct application of our initial approach \eqref{8} is infeasible. To overcome this, we transform $w_i$ into vectors with the same dimension by considering their low-dimensional representations through basis expansion regression. More details on this can be found in Section \ref{s2.1}. Additionally, the $p$-value as initially defined is numerically infeasible due to the complexity of the condition $\{\mathcal{C}_1, \mathcal{C}_2\in\mathcal{C}(\mathcal{W})\}$. In Section \ref{sec:pval}, we simplify this condition through an orthogonal decomposition and define a formal $p$-value.

Our method includes three main steps: first, we conduct the basis expansion regression to embed the functional data $W_i$ into low-dimensional vectors and form a tensor structure; second, we conduct the whitening transformation to normalize the tensor; third, we calculate the selective $p$-value on the tensor data leveraging the clustering information. We describe these three steps in the following Sections \ref{s2.1}, \ref{sec:covariance-estimation}, \ref{sec:pval}, respectively.

%%%%%%%%%%%%%
\subsection{Low-dimensional Embedding}\label{s2.1}
%%%%%%%%%%%%%

In this step, we aim to identify a low-dimensional representation of the functional data by embedding the multivariate Gaussian process data into a dimension-reduced tensor representation.

Recall Model \eqref{3}, where $W_{ij}$ denotes the record of the $i$th subject and the $j$th feature. Given the time points $\Omega_{ij}$ and $(W_{ij}(t_{ijk}))_{k\in[r_{ij}]}\in\mathbb{R}^{r_{ij}}$, where $r_{ij}$ is the number of time points for this record. We select $q$ basis functions $\{\phi_s\}_{s\in[q]}$. Here, $q$ is a user-specified positive integer and each $\phi_s$ is Lipschitz continuous. We then perform the ridge regression as follows:
\begin{equation}\label{4}
	\alpha_{ij}\coloneqq\mathop{\arg\min}_{\alpha\in\mathbb{R}^{q}}\left\{\sum_{k = 1}^{r_{ij}}\left(W_{ij}(t_{ijk})-\sum_{s = 1}^{q}\alpha_s\cdot\phi_s(t_{ijk})\right)^2 +\lambda\lVert \alpha\rVert^2\right\},
\end{equation}
where $i\in[n]$, $j\in[m]$, $\lambda$ is a regularization parameter, and $(\phi_s(t_{ijk}))_{k\in [r_{ij}]}$ is a $r_{ij}$-dimensional vector of the basis function $\phi_s$. Define the matrix $\Phi_{ij} \coloneqq (\phi_s(t_{ijk}))_{s\in[q], k\in[r_{ij}]}\in\mathbb{R}^{q\times r_{ij}}$, where the $(s,k)$th entry of $\Phi_{ij}$ is $\phi_s(t_{ijk})$ for $s\in[q], k\in[r_{ij}]$. Define $K_{ij}\coloneqq\Phi_{ij}\Phi_{ij}^{\top}\in\mathbb{R}^{q\times q}$. Then, the coefficient vector $\alpha_{ij}$ has the following closed-form expression:
\begin{equation}\label{6}
	\alpha_{ij} = (K_{ij}+\lambda I_q)^{-1}\Phi_{ij}(W_{ij}(t_{ijk}))_{k\in[r_{ij}]},
\end{equation}
 {and the corresponding subject-level basis reconstruction is $\widehat{W}_{ij}^{(q)}(t)=\sum_{s=1}^q\alpha_{ijs}\phi_s(t)$. This reconstruction is used as a linear feature map for the individual record; it is not a pooled estimator of a population or cluster mean function.}  {The basis family and the truncation level $q$ are practical tuning parameters. In finite samples, choosing $q$ too small can omit directions that contain cluster-separation signal, while choosing $q$ too large can introduce noisier components and make covariance estimation and whitening less stable. The regularization parameter $\lambda$ is used to stabilize the ridge embedding and should be small but positive; in practice it can be fixed at a mild regularization level or chosen by a standard rule such as cross-validation. For the clustering step, the number of clusters may be chosen using standard diagnostics such as the silhouette index. Since PSIMF performs selective inference for a selected two-cluster contrast, a multi-cluster structure may be explored in practice through recursive binary splitting, where one first identifies a data-driven binary split and then further subdivides a selected branch if warranted. In the numerical studies below we use the first few eigenfunctions of the Gaussian RBF kernel and set $q=3$, which provides a stable low-dimensional representation while keeping the subsequent covariance estimation tractable.}

The linear transformation \eqref{6} embeds the functional data $W_{ij}$ into a $q$-dimensional vector $\alpha_{ij}$. By applying this low-dimensional embedding to all $i\in[n]$ and $j\in[m]$, we transform each $W_{ij}$ into an $m\times q$ matrix. Consequently, the entire dataset $\mathcal{W}$ is transformed into an $n\times m\times q$ tensor.
\begin{definition}[Low-dimensional embedding]
	Given the basis functions $\{\phi_s\}_{s\in[q]}$ and the time record $\Omega$, we define the linear map $H:\mathcal{W}\rightarrow\mathbb{R}^{n\times m\times q}$, where 
	\begin{equation}\label{lowemb}
		H(\mathcal{W})[i,j,:] \coloneqq (K_{ij}+\lambda I_q)^{-1}\Phi_{ij}(W_{ij}(t_{ijk}))_{k\in[r_{ij}]}.
	\end{equation}
	%Namely, the $(i,j,k)\in[n]\times[m]\times[q]$ entry of $H(\mathcal{W})$ is $\alpha_{ijk}$.
\end{definition}
The following lemma shows that under the distributional assumption, the vectorization of each slice $H(\mathcal{W})[i,:,:]\in\mathbb{R}^{m\times q}$ of the resulting tensor from Step 1 follows a multivariate normal distribution. 
\begin{Lemma}[Distribution of $H(\mathcal{W})$]\label{l1} Under Assumption \ref{a2}, the following statements hold:
	\begin{enumerate}[(I)]
		\item Define the block diagonal matrix $D_i \coloneqq \diag((K_{ij}+\lambda I_q)^{-1}\Phi_{ij})_{j\in[m]}\in\mathbb{R}^{mq\times(\sum_{j = 1}^mr_{ij})}$ and define $\mu_{ij}^{\Omega_i} \coloneqq (\mu_{ij}(t_{ijk}))_{k\in[r_{ij}]}$ as the vector of $\mu_{ij}$ characterized by the time record $\Omega_{i}$. For any $a,b\in[m]$, define $\Sigma_{ab}^{\Omega_i} = (R_{ab}(t_{ijk_1},t_{ijk_2}))_{k_1,k_2\in[r_{ij}]}\in\mathbb{R}^{r_{ij}\times r_{ij}}$ as the covariance matrix, then %the vectorization of $H(\mathcal{W})[i,:,:]$ satisfies multivariate normal distribution:
		\begin{equation}\label{7}
			\begin{split}
				& \text{vec}\left(H(\mathcal{W})[i,:,:]\right)\\
			\sim & \mathcal{N}\left\{\text{vec}\left[\left((K_{ij}+\lambda I_q)^{-1}\Phi_{ij}\mu_{ij}^{\Omega_i}\right)_{j\in[m]}\right],D_i\left[\Sigma_1^{(i)}+\Sigma_2^{(i)}\right]D_i^\top\right\},
			\end{split}
		\end{equation}
		where 
		\[\Sigma_1^{(i)} \coloneqq \left(\Sigma_{ab}^{\Omega_i}\right)_{a,b\in[m]}\quad\text{and}\quad \Sigma_{2}^{(i)}\coloneqq \diag(\sigma_j^2\cdot I_{r_{ij}})_{j\in[m]}.\]
		\item For some $i\in[n]$, if the number of time points $r_{ij}$ goes to infinite for all $j\in[m]$, then the distribution of $\text{vec}(H(\mathcal{W})[i,:,:])$ converges as follow:
		\begin{equation}\label{embasy}
			\text{vec}\left(H(\mathcal{W})[i,:,:]\right)\stackrel{d}{\longrightarrow}\mathcal{N}\left\{\text{vec}\left(\left(K^{-1}\mu_{ij}^{(0)}\right)_{j\in[m]}\right),\Lambda \right\} \text{ as } \min_{j\in[m]}r_{ij}\rightarrow\infty.
		\end{equation}
		Here,
		\begin{equation*}
			\begin{split}
				& K\coloneqq\left(\int_0^1 \phi_{s_1}(t)\phi_{s_2}(t)dt\right)_{s_1,s_2\in[q]},\mu_{ij}^{(0)}\coloneqq\left(\int_0^1\phi_s(t)\mu_{ij}(t)dt\right)_{s\in[q]},\\
				& \Lambda\coloneqq(\Lambda_{ab})_{a,b\in[m]}.
			\end{split}
		\end{equation*}
		We assume $K$ is invertible and define
		\begin{equation*}
			\begin{split}
				& \Lambda_{ab}\coloneqq \\
				& K^{-1} \left(\int_0^1\int_0^1\phi_{s_1}(t_1)\phi_{s_2}(t_2)(R_{ab}(t_1,t_2)+\mathbbm{1}_{a=b}\cdot \sigma_a^2) dt_1dt_2\right)_{s_1,s_2\in[q]}K^{-1}.
			\end{split}
		\end{equation*}
	\end{enumerate}
\end{Lemma}
\begin{proof}
	See Appendix \ref{pl1}.
\end{proof}
 {Lemma~\ref{l1}(I) is the exact conditional Gaussian statement for fixed subject-specific observation times $\Omega_{ij}$, so the framework naturally allows irregular sampling across individuals. By contrast, Lemma~\ref{l1}(II) is an asymptotic approximation whose accuracy improves as $\min_{j\in[m]} r_{ij}$ increases. In sparse longitudinal settings the embedding remains well defined, but fewer observations per feature can reduce the quality of the estimated coefficients, make covariance estimation and whitening noisier, and weaken finite-sample Type~I error calibration and power.}

 {This finite-sample observation concerns the subject-wise ridge embedding used in PSIMF. It does not imply that population mean or covariance estimation generally requires densely observed trajectories. In sparse designs, pooled smoothing methods can borrow information across subjects to estimate these population-level quantities \citep{yao2005functional,zhang2016sparse}. Incorporating such pooled estimators into the present selective pipeline is a promising extension, but the dependence between nuisance estimation and the data-driven clustering event would need to be accounted for.}

%%%%%%%%%%%%%%
\subsection{Whitening Transformation}\label{sec:covariance-estimation}
%%%%%%%%%%%%%%

Next, we perform the whitening transformation to normalize $H(\mathcal{W})$ obtained in Step 1. The goal is to transform the covariance matrix of distribution \eqref{7} into an identity matrix.  {We remark that whitening rescales directions in the embedded space, it can change the geometry seen by a distance-based clustering algorithm and therefore affect the estimated partition and downstream power. In PSIMF, the inferential target is the pair of clusters selected after whitening rather than the partition obtained before the transformation.} 

Define $\beta_{(i)}^{\Omega_i}\coloneqq\left((K_{ij}+\lambda I_q\right)^{-1}\Phi_{ij}\mu_{ij}^{\Omega_i})_{j\in[m]}$ and $\Lambda_{(i)} \coloneqq D_i\left[\Sigma_1^{(i)}+\Sigma_2^{(i)}\right]D_i^\top$. Then the distribution of \eqref{7} reduces to $\text{vec}\left(H(\mathcal{W})[i,:,:]\right)\sim\mathcal{N}\left(\text{vec}(\beta_{(i)}^{\Omega_i}),\Lambda_{(i)}\right)$. We introduce a slice-wise transformation $L$ on the tensor $H(\mathcal{W})$:
\begin{equation}\label{eq16}
	\begin{aligned}
		\text{vec}(L(\mathcal{W})[i,:,:])\coloneqq \Lambda_{(i)}^{-\frac{1}{2}}\text{vec}(H(\mathcal{W})[i,:,:]), \quad i=1,\ldots, n.
	\end{aligned}
\end{equation}
Under the distributional assumption, the transformed tensor follows another normal distribution:
\begin{equation}\nonumber
	\text{vec}(L(\mathcal{W})[i,:,:])\sim \mathcal{N}\left\{\Lambda_{(i)}^{-\frac{1}{2}}\text{vec}(\beta_{(i)}^{\Omega_i}),I_{mq}\right\}, \quad i=1,\ldots, n.
\end{equation}
Analogous to Lemma \ref{l1}, we can characterize the asymptotic distribution of $\text{vec}(L(\mathcal{W})[i,:,:])$: 
\begin{equation}\label{eq19}
	\text{vec}(L(\mathcal{W})[i,:,:])\stackrel{d}{\longrightarrow}\mathcal{N}(\Lambda^{-\frac{1}{2}}\text{vec}(\beta_{(i)}),I_{mq})\quad\text{as}\quad\min_{j\in[m]}r_{ij}\rightarrow\infty,
\end{equation}
where $\beta_{(i)}\coloneqq (K^{-1}\mu_{ij}^{(0)})_{j\in[m]}$ and $K,\Lambda,\mu_{ij}^{(0)}$ are defined in Lemma \ref{l1}. 

%%%%%%%%%%%%%
\paragraph{Covariance Estimation.}
%%%%%%%%%%%%%
In practice, both the covariance function $R$ and the variance of noise $\sigma_j^2$ are typically unknown, rendering the covariance matrix $\Lambda$ unknown. Therefore, we apply the following sample covariance estimator to estimate $\Lambda$:
\begin{equation*}
	\begin{split}
		\hat{\Lambda} \coloneqq & \frac{1}{n-1}\Big[\sum_{i = 1}^n \left(\text{vec}(H(\mathcal{W})[i,:,:])-\text{vec}(\overline{H(\mathcal{W})})\right)\\
		& \cdot \left(\text{vec}(H(\mathcal{W})[i,:,:]-\text{vec}(\overline{H(\mathcal{W})})\right)^{\top}\Big],
	\end{split}
\end{equation*}
where $\overline{H(\mathcal{W})}\coloneqq\sum_{i = 1}^nH(\mathcal{W})[i,:,:]/n$ is the sample mean. 

If the null hypothesis $H_0^{\{\mathcal{C}_1,\mathcal{C}_2\}}$ holds (i.e., $\mu_{\mathcal{C}_1} = \mu_{\mathcal{C}_2}$), we have $\mu_{i_1} = \mu_{i_2}$ and $\beta_{(i_1)} = \beta_{(i_2)}$ for all $i_1,i_2\in[n]$. Lemma \ref{l1} implies that $\text{vec}(H(\mathcal{W})[i,:,:])\sim\mathcal{N}(\text{vec}(\beta_{(i)}),\Lambda)$ as $r_{ij}\rightarrow\infty$ for all $j\in[m]$. Therefore, if $H_0^{\{\mathcal{C}_1,\mathcal{C}_2\}}$ holds and the sample size $n\rightarrow\infty$, we have $\hat{\Lambda}\stackrel{p}{\longrightarrow}\Lambda$.

If the alternative hypothesis $H_1^{\{\mathcal{C}_1,\mathcal{C}_2\}}$ holds (i.e., $\mu_{\mathcal{C}_1}\neq\mu_{\mathcal{C}_2}$), define 
\[\beta_{\mathcal{C}_j}\coloneqq\beta_{(i)}\quad \text{for some}\quad i\in\mathcal{C}_j\quad \text{and}\quad j\in\{1,2\},\]
then we have $\beta_{\mathcal{C}_1}\neq \beta_{\mathcal{C}_2}$. In this situation, we rewrite the sample covariance estimator as follows:
\begin{equation}\label{eq18}
	\begin{aligned}
		\hat{\Lambda} &= \bigg(\sum_{j = 1}^2\bigg[\sum_{i\in\mathcal{C}_j}(\text{vec}(H(\mathcal{W})[i,:,:])-\text{vec}(\overline{H(\mathcal{W})}_{\mathcal{C}_j}))\\
		& \quad \cdot (\text{vec}(H(\mathcal{W})[i,:,:])-\text{vec}(\overline{H(\mathcal{W})}_{\mathcal{C}_j}))^{\top}\\
		&+|\mathcal{C}_j|(\text{vec}(\overline{H(\mathcal{W})}_{\mathcal{C}_j})-\text{vec}(\overline{H(\mathcal{W})}))(\text{vec}(\overline{H(\mathcal{W})}_{\mathcal{C}_j})-\text{vec}(\overline{H(\mathcal{W})}))^{\top}\bigg]\bigg)/(n-1).
	\end{aligned}
\end{equation}
Intuitively, \eqref{eq18} implies that 
\begin{equation}\label{covasy}
	\hat{\Lambda}\stackrel{p}{\longrightarrow} \Lambda+\frac{c}{(c+1)^2}\text{vec}(\beta_{\mathcal{C}_1}-\beta_{\mathcal{C}_2})\cdot\text{vec}(\beta_{\mathcal{C}_1}-\beta_{\mathcal{C}_2})^{\top}
\end{equation}
as $|\mathcal{C}_1|, |\mathcal{C}_2|\rightarrow\infty$, $|\mathcal{C}_1|/|\mathcal{C}_2|\rightarrow c$, and $r_{ij}\rightarrow\infty$. Therefore, the sample covariance estimator $\hat{\Lambda}$ has a constant bias $c/(c+1)^2\cdot\text{vec}(\beta_{\mathcal{C}_1}-\beta_{\mathcal{C}_2})\cdot\text{vec}(\beta_{\mathcal{C}_1}-\beta_{\mathcal{C}_2})^{\top}$. In Section \ref{s4}, we leverage this bias to show that the proposed estimator controls the statistical power under mild additional assumptions.

 {If the common within-cluster mean assumption holds only approximately, write $\beta_{(i)}=\beta_{\mathcal{C}_j}+\delta_i$ for $i\in\mathcal{C}_j$. The sample covariance then contains an additional within-cluster dispersion term generated by the $\delta_i$'s. Hence $\hat{\Lambda}$ need not estimate only $\Lambda$ and the between-cluster term in \eqref{covasy}; the whitening transformation and the scaled-$\chi$ selective pivot are no longer exact. We examine the empirical sensitivity to controlled departures from this assumption in Section~\ref{s5}.}

%%%%%%%%%%%%%%
\subsection{Selective $p$-value}\label{sec:pval}
%%%%%%%%%%%%%%

Suppose $w = (w_i)_{i\in[n]}$ is a sample generated from Model \eqref{3} and $L(w)$ is a tensor obtained by low-dimensional embedding and whitening transformation. Next, we apply a clustering algorithm (such as the hierarchical clustering or $k$-means clustering) on $\{L(w)[i,:,:]\}_{i\in[n]}$ to separate $n$ subjects into two clusters, where the indices are denoted by $\mathcal{C}_1, \mathcal{C}_2$, respectively. Analogous to \eqref{8}, we consider the following selective $p$-value that leverages the clustering information:
\begin{equation}\label{eq9}
	\mathbb{P}_{H_0^{\{\mathcal{C}_1,\mathcal{C}_2\}}}\left(\lVert \overline{L(\mathcal{W})}_{\mathcal{C}_1}-\overline{L(\mathcal{W})}_{\mathcal{C}_2}\rVert_F\geq\lVert\overline{L(w)}_{\mathcal{C}_1}-\overline{L(w)}_{\mathcal{C}_2}\rVert_F\Big|~\mathcal{C}_1, \mathcal{C}_2\in\mathcal{C}(L(\mathcal{W}))\right),
\end{equation}
where $\overline{L(\mathcal{W})}_{\mathcal{C}_j} \coloneqq \sum_{i\in\mathcal{C}_j}L(\mathcal{W})[i,:,:]/|\mathcal{C}_j|$ for all $j\in\{1, 2\}$. We remark that the clustering algorithm $\mathcal{C}(\cdot)$ is implemented on a collection of matrix data $\{L(w)[i,:,:]\}_{i\in[n]}$ instead of the functional data $\mathcal{W}$. Intuitively, the proposed $p$-value is a probability of $L(\mathcal{W})$ conditioning on the region $\{L(\mathcal{W}):\mathcal{C}_1,\mathcal{C}_2\in\mathcal{C}(L(\mathcal{W}))\}$. Following the selective inference theory \citep{fithian2014optimal}, this selective $p$-value leverages the model selection information to eliminate the selection bias and further control the selective type-I error.

\begin{definition}\label{d1}(Post-clustering selective type-I error). Suppose that $\mathcal{W} = (W_i)_{i\in[n]}$ follows Model \eqref{3} and $w$ is a realization of $\mathcal{W}$. Suppose that the partition $\mathcal{C}_1,\mathcal{C}_2$ ($\mathcal{C}_1\cup\mathcal{C}_2=[n]$, $\mathcal{C}_1\cap\mathcal{C}_2=\emptyset$) is the output of a clustering algorithm $\mathcal{C}(\cdot)$. Let $H_0^{\{\mathcal{C}_1, \mathcal{C}_2\}}$ be the null hypothesis defined as (\ref{1}), we say a test of $H_0^{\{\mathcal{C}_1, \mathcal{C}_2\}}$ based on $\mathcal{W}$ controls the selective type-I error for clustering at level $\alpha$ if
	\begin{equation}\label{9}
		\mathbb{P}_{H_0^{\{\mathcal{C}_1, \mathcal{C}_2\}}}\left(\text{reject $H_0^{\{\mathcal{C}_1, \mathcal{C}_2\}}$ based on $\mathcal{W}$ at level $\alpha$}\Big|~\mathcal{C}_1, \mathcal{C}_2\in\mathcal{C}(L(\mathcal{W}))\right)\leq \alpha
	\end{equation}
	for any $\alpha\in[0,1]$.
\end{definition}
However, the selective $p$-value \eqref{eq9} cannot be directly calculated because the condition $\mathcal{C}_1, \mathcal{C}_2\in\mathcal{C}(L(\mathcal{W}))$ is numerically infeasible, meaning the selection set $\{L(\mathcal{W}):\mathcal{C}_1,\mathcal{C}_2\in\mathcal{C}(L(\mathcal{W}))\}$ might be complex and difficult to construct. Inspired by the approach in \cite{gao2024selective}, we aim to modify the selection set to simplify the expression of the selective $p$-value. Given the clustering output $\mathcal{C}_1,\mathcal{C}_2$, we define the indicator vector as follows:
\[\nu(\mathcal{C}_1,\mathcal{C}_2) \coloneqq \left(\frac{\mathbbm{1}_{\{i\in\mathcal{C}_1\}}}{|\mathcal{C}_1|}-\frac{\mathbbm{1}_{\{i\in\mathcal{C}_2\}}}{|\mathcal{C}_2|}\right)_{i\in[n]},\]
where the $i$th coordinate of $\nu(\mathcal{C}_1,\mathcal{C}_2)$ is $1/|\mathcal{C}_1|$ if $i\in\mathcal{C}_1$ and $-1/|\mathcal{C}_2|$ if $i\in\mathcal{C}_2$. Based on $\nu(\mathcal{C}_1,\mathcal{C}_2)$, we decouple $L(\mathcal{W})$ by the following orthogonal decomposition.
\begin{Lemma}\label{l2}(Orthogonal Decomposition). For any tensor $\mathcal{A}\in\mathbb{R}^{n\times m\times q}$ and any partition of $[n]$ denoted by $\mathcal{C}_1,\mathcal{C}_2$, we have the following decomposition:
	\begin{equation}\label{eq12}
		\mathcal{A} = \pi^{\perp}_{\nu(\mathcal{C}_1,\mathcal{C}_2)}\times_1 \mathcal{A}+\left(\frac{\lVert \overline{\mathcal{A}}_{\mathcal{C}_1}-\overline{\mathcal{A}}_{\mathcal{C}_2} \rVert_F}{1/|\mathcal{C}_1|+1/|\mathcal{C}_2|}\right)\nu(\mathcal{C}_1,\mathcal{C}_2)\times_1\text{dir}(\overline{\mathcal{A}}_{\mathcal{C}_1}-\overline{\mathcal{A}}_{\mathcal{C}_2})^{\top},
	\end{equation}
	where $\overline{\mathcal{A}}_{\mathcal{C}_i} = \sum_{j\in\mathcal{C}_i}\mathcal{A}[j,:,:]/|\mathcal{C}_i|$ is the mean of mode-1 slices corresponding to the partition $\mathcal{C}_i$, $\times_1$ denotes the tensor mode-1 product (here we view $\nu(\mathcal{C}_1,\mathcal{C}_2)$ as a $n\times1$ matrix and $\text{dir}(\overline{\mathcal{A}}_{\mathcal{C}_1}-\overline{\mathcal{A}}_{\mathcal{C}_2})$ as a $1\times m\times q$ tensor), $\pi^{\perp}_{\nu} = I-\frac{\nu\nu^{\top}}{\lVert\nu\rVert^2}$ is an orthogonal projection matrix, and $\text{dir}(\omega) = \frac{\omega}{\lVert\omega\rVert_F}\mathbbm{1}_{\{\omega\neq 0\}}$ is the direction of $\omega$ (here $\omega$ is a matrix, $\lVert\omega\rVert_F$ is its Frobenius norm, and $\mathbbm{1}_{\{\omega \neq 0\}}$ is the indicator function takes the value $0$ when all the entries in $\omega$ are zero and takes the value $1$ otherwise). 
\end{Lemma}
\begin{proof}
	See Appendix \ref{pl2}.
\end{proof}
 {The matrix formulation in \eqref{eq12} is equivalent to taking the Euclidean norm of the vectorized matrix difference because $\|A\|_F = \|\mathrm{vec}(A)\|_2$. We retain the matrix notation because it preserves the feature-by-basis structure induced by the embedding and makes the tensor decomposition easier to interpret.}
Plugging $L(\mathcal{W})$ into Lemma \ref{l2}, we have
\begin{equation*}
	\begin{split}
		L(\mathcal{W}) = & \pi^{\perp}_{\nu(\mathcal{C}_1,\mathcal{C}_2)}\times_1 L(\mathcal{W})\\
		& + \left[\frac{\lVert\overline{L(\mathcal{W})}_{\mathcal{C}_1}-\overline{L(\mathcal{W})}_{\mathcal{C}_2}\rVert_F}{1/|\mathcal{C}_1|+1/|\mathcal{C}_2|}\right]\nu(\mathcal{C}_1,\mathcal{C}_2)\times_1\text{dir}\left(\overline{L(\mathcal{W})}_{\mathcal{C}_1}-\overline{L(\mathcal{W})}_{\mathcal{C}_2}\right)^{\top}.
	\end{split}
\end{equation*} 
Based on this decomposition of $L(\mathcal{W})$, we define the following selective $p$-value incorporating additional conditions.
\begin{definition}\label{d2}(Selective $p$-value). Suppose that $\mathcal{W}$ follows Model \eqref{3} and $w$ is a realization of $\mathcal{W}$. Suppose that the partition $\mathcal{C}(L(w)) = \{\mathcal{C}_1,\mathcal{C}_2\}$ ($\mathcal{C}_1\cup\mathcal{C}_2=[n], \mathcal{C}_1\cap\mathcal{C}_2=\emptyset$) is the output of a clustering algorithm $\mathcal{C}(\cdot)$. Let $H_0^{\{\mathcal{C}_1, \mathcal{C}_2\}}$ be the null hypothesis defined as (\ref{1}), we propose the following selective $p$-value:
	\begin{equation}\label{12}
		\begin{aligned}
			& p_{selective} = \\
			& \mathbb{P}_{H_0^{\{\mathcal{C}_1, \mathcal{C}_2\}}}\bigg(\lVert \overline{L(\mathcal{W})}_{\mathcal{C}_1}-\overline{L(\mathcal{W})}_{\mathcal{C}_2}\rVert_F\geq\lVert\overline{L(w)}_{\mathcal{C}_1}-\overline{L(w)}_{\mathcal{C}_2}\rVert_F\bigg|\\
			& \mathcal{C}_1,\mathcal{C}_2\in\mathcal{C}(L(\mathcal{W})),\pi^{\perp}_{\nu(\mathcal{C}_1,\mathcal{C}_2)}\times_1L(\mathcal{W}) = \pi^{\perp}_{\nu(\mathcal{C}_1,\mathcal{C}_2)}\times_1 L(w),\\
			& \text{dir}(\overline{L(\mathcal{W})}_{\mathcal{C}_1}-\overline{L(\mathcal{W})}_{\mathcal{C}_2}) = \text{dir}(\overline{L(w)}_{\mathcal{C}_1}-\overline{L(w)}_{\mathcal{C}_2})\bigg).
		\end{aligned}
	\end{equation}
\end{definition}

Under the null hypothesis $H_0^{\{\mathcal{C}_1, \mathcal{C}_2\}}$, we have $\sum_{i\in\mathcal{C}_1}\beta_{(i)}/|\mathcal{C}_1| = \sum_{i\in\mathcal{C}_2}\beta_{(i)}/|\mathcal{C}_2|$, then \eqref{eq19} implies that 
\begin{equation}\label{eq17}
	\text{vec}\left(\overline{L(\mathcal{W})}_{\mathcal{C}_1}-\overline{L(\mathcal{W})}_{\mathcal{C}_2}\right)\stackrel{d}{\longrightarrow} \sqrt{1/|\mathcal{C}_1|+1/|\mathcal{C}_2|}\cdot\mathcal{N}(0,I_{mq})\text{ as } \min_{i\in[n],j\in[m]}r_{ij}\rightarrow\infty.
\end{equation}
Next, we plug in \eqref{eq17} to reform the selective $p$-value \eqref{12}. To elaborate, \eqref{eq17} implies that $\lVert\overline{L(\mathcal{W})}_{\mathcal{C}_1}-\overline{L(\mathcal{W})}_{\mathcal{C}_2}\rVert_F$ follows the distribution $\sqrt{1/|\mathcal{C}_1|+1/|\mathcal{C}_2|}\cdot\chi_{mq}$ if the null hypothesis holds (and the number of time points goes to infinite). Therefore, we can reform the selective $p$-value as a survival function of a truncated chi-squared distribution. 
\begin{Lemma}\label{l3} Suppose that $\mathcal{W}$ follows Model \eqref{3} and $w$ is a realization of $\mathcal{W}$. Suppose that $\text{vec}(\overline{L(\mathcal{W})}_{\mathcal{C}_1}-\overline{L(\mathcal{W})}_{\mathcal{C}_2})\sim \sqrt{1/|\mathcal{C}_1|+1/|\mathcal{C}_2|}\cdot\mathcal{N}(0,I_{mq})$. Then the selective $p$-value \eqref{12} can be rewritten as follows:
	\begin{equation}\label{11}
		p_{selective} = 1-\mathbb{F}\left(\lVert\overline{L(w)}_{\mathcal{C}_1}-\overline{L(w)}_{\mathcal{C}_2}\rVert_F;\sqrt{\frac{1}{|\mathcal{C}_1|}+\frac{1}{|\mathcal{C}_2|}},\mathcal{S}(w;\mathcal{C}_1,\mathcal{C}_2)\right),
	\end{equation}
	where $\mathbb{F}(t;c,\mathcal{S})$ denotes the cumulative distribution function of the random variable $c\cdot \chi_{mq}$ truncated to the set $\mathcal{S}$ defined by
	\begin{equation*}
		\begin{split}
			& \mathcal{S}(w;\mathcal{C}_1,\mathcal{C}_2)
			\coloneqq \Bigg\{\varphi\geq 0: \mathcal{C}_1,\mathcal{C}_2\in\mathcal{C}\Big(\pi^{\perp}_{\nu(\mathcal{C}_1,\mathcal{C}_2)}\times_1 L(w)\\
			& \quad  + \left[\frac{\varphi}{1/|\mathcal{C}_1|+1/|\mathcal{C}_2|}\right]\nu(\mathcal{C}_1,\mathcal{C}_2)\times_1\text{dir}(\overline{L(w)}_{\mathcal{C}_1}-\overline{L(w)}_{\mathcal{C}_2})^{\top}\Big) \Bigg\}.
		\end{split}
	\end{equation*}
\end{Lemma}
\begin{proof}
	See Appendix \ref{pl3}.
\end{proof}

%%%%%%%%%%%%%
\subsection{Numerical Approximation of Selective $p$-value}\label{s323}
%%%%%%%%%%%%%%%%

Now we introduce the Monte Carlo method to compute the truncated survival function \eqref{11}, which serves as an approximation of the selective $p$-value \eqref{12}. 

To begin with, we briefly discuss the geometric intuition of $\mathcal{S}(w; \mathcal{C}_1,\mathcal{C}_2)$. Given a partition $\mathcal{C}_1, \mathcal{C}_2$ obtained by a certain clustering algorithm, we consider the linear map $F:\mathbb{R}\rightarrow\mathbb{R}^{n\times m\times q}$:
\begin{equation}\label{eq28}
	F(\varphi)\coloneqq\pi^{\perp}_{\nu(\mathcal{C}_1,\mathcal{C}_2)}\times_1 L(w)+\left[\frac{\varphi}{1/|\mathcal{C}_1|+1/|\mathcal{C}_2|}\right]\nu(\mathcal{C}_1,\mathcal{C}_2)\times_1\text{dir}\left(\overline{L(w)}_{\mathcal{C}_1}-\overline{L(w)}_{\mathcal{C}_2}\right)^{\top}.
\end{equation}
Intuitively speaking, $F$ operates the orthogonal projection $\pi^{\perp}_{\nu(\mathcal{C}_1,\mathcal{C}_2)}\times_1 L(w)$ along a ``vector" $\nu(\mathcal{C}_1,\mathcal{C}_2) \times_1 \text{dir}\left(\overline{L(w)}_{\mathcal{C}_1}-\overline{L(w)}_{\mathcal{C}_2}\right)^{\top}$ with the length $\varphi$. The set $\mathcal{S}(w;\mathcal{C}_1,\mathcal{C}_2)$ contains all the ``length" $\varphi\in\mathbb{R}$ such that the transformed tensor has the same clustering outputs as $L(w)$ (i.e., the partition is equal to $\mathcal{C}_1,\mathcal{C}_2$). Lemma \ref{l2} shows
\begin{equation*}
	\begin{split}
		L(w) = & \pi^{\perp}_{\nu(\mathcal{C}_1,\mathcal{C}_2)}\times_1 L(w)\\
		& +\left[\frac{\lVert\overline{L(w)}_{\mathcal{C}_1}-\overline{L(w)}_{\mathcal{C}_2}\rVert_F}{1/|\mathcal{C}_1|+1/|\mathcal{C}_2|}\right]\nu(\mathcal{C}_1,\mathcal{C}_2)\times_1\text{dir}\left(\overline{L(w)}_{\mathcal{C}_1}-\overline{L(w)}_{\mathcal{C}_2}\right)^{\top}.
	\end{split}
\end{equation*}
Therefore, we conclude that $\lVert\overline{L(w)}_{\mathcal{C}_1}-\overline{L(w)}_{\mathcal{C}_2}\rVert_F\in\mathcal{S}(w;\mathcal{C}_1,\mathcal{C}_2)$. Furthermore, when $\varphi>\mathcal{S}(w;\mathcal{C}_1,\mathcal{C}_2)$, the transformation $F$ ``pushes away" the sets $\{L(w)[i,:,:]\}_{i\in\mathcal{C}_1}$ and $\{L(w)[i,:,:]\}_{i\in\mathcal{C}_2}$ along the vector $\nu(\mathcal{C}_1,\mathcal{C}_2)\times_1\text{dir}(\overline{L(w)}_{\mathcal{C}_1}-\overline{L(w)}_{\mathcal{C}_2})^{\top}$, and vice versa. If $\varphi$ is too large or small, the clustering output of $F(\varphi)$ will be different from $\mathcal{C}_1,\mathcal{C}_2$. Therefore, $\mathcal{S}(w;\mathcal{C}_1,\mathcal{C}_2)$ concentrates near $\lVert\overline{L(w)}_{\mathcal{C}_1}-\overline{L(w)}_{\mathcal{C}_2}\rVert_F$.

\paragraph{Monte Carlo Approximation.}  We use the Monte Carlo method to approximate \eqref{11}, which is the survival function of the distribution $\sqrt{1/|\mathcal{C}_1|+1/|\mathcal{C}_2|}\cdot\chi_{mq}$ truncated on the set $\mathcal{S}(w;\mathcal{C}_1,\mathcal{C}_2)$. Mathematically, we rewrite \eqref{11} as follows: 
\begin{equation*}
	\begin{split}
		p_{selective} = & \frac{\mathbb{P}\left(\varphi\geq\lVert\overline{L(w)}_{\mathcal{C}_1}-\overline{L(w)}_{\mathcal{C}_2}\rVert_F,\mathcal{C}_1,\mathcal{C}_2\in\mathcal{C}(F(\varphi))\right)}{\mathbb{P}(\mathcal{C}_1,\mathcal{C}_2\in\mathcal{C}(F(\varphi)))} \\
		= & \frac{\mathbb{E}\left[\mathbbm{1}_{\{\varphi\geq\lVert\overline{L(w)}_{\mathcal{C}_1}-\overline{L(w)}_{\mathcal{C}_2}\rVert_F,\mathcal{C}_1,\mathcal{C}_2\in\mathcal{C}(F(\varphi))\}}\right]}{\mathbb{E}[\mathbbm{1}_{\{\mathcal{C}_1,\mathcal{C}_2\in\mathcal{C}(F(\varphi))\}}]},
	\end{split}
\end{equation*}
where $\varphi$ follows the distribution $\sqrt{1/|\mathcal{C}_1|+1/|\mathcal{C}_2|}\cdot\chi_{mq}$, $\mathbb{P}$ is the corresponding probability mass function, and $\mathbb{E}$ is the expectation with respect to $\varphi$. We will sample some $\varphi\in\mathbb{R}$ and check if $\varphi\in\mathcal{S}(w;\mathcal{C}_1,\mathcal{C}_2)$ to approximate $\mathcal{S}(w;\mathcal{C}_1,\mathcal{C}_2)$ and further estimate $p_{selective}$.

We apply importance sampling to approximate this conditional probability. As aforementioned, $\mathcal{S}(w;\mathcal{C}_1,\mathcal{C}_2)$ concentrates near $\lVert\overline{L(w)}_{\mathcal{C}_1}-\overline{L(w)}_{\mathcal{C}_2}\rVert_F$. Therefore, we set $g(x) = f_1(x)/f_2(x)$, where $f_1$ is the density of $\sqrt{1/|\mathcal{C}_1|+1/|\mathcal{C}_2|}\cdot \chi_{mq}$ and $f_2$ is the density function of $\mathcal{N}\left(\lVert\overline{L(w)}_{\mathcal{C}_1}-\overline{L(w)}_{\mathcal{C}_2}\rVert_F,1/|\mathcal{C}_1|+ 1/|\mathcal{C}_2|\right)$. For a positive integer $S$, we sample $S$ values $\gamma_1, \ldots,\gamma_S\sim \mathcal{N}(\lVert\overline{L(w)}_{\mathcal{C}_1}-\overline{L(w)}_{\mathcal{C}_2}\rVert_F,1/|\mathcal{C}_1|+1/|\mathcal{C}_2|)$, then the selective $p$-value can be approximately by
\begin{equation}\label{eq29}
	p_{selective}\approx\frac{\sum\pi_i\mathbbm{1}_{\{\gamma_i\geq\lVert\overline{L(w)}_{\mathcal{C}_1}-\overline{L(w)}_{\mathcal{C}_2}\rVert_F,\mathcal{C}_1,\mathcal{C}_2\in\mathcal{C}(F(\gamma_i))\}}}{\sum\pi_i\mathbbm{1}_{\{\mathcal{C}_1,\mathcal{C}_2\in\mathcal{C}(F(\gamma_i))\}}},~~~~\pi_i = \frac{f_1(\gamma_i)}{f_2(\gamma_i)}.
\end{equation}

%%%%%%%%%%%%%%
\subsection{Overall Procedures}\label{sec:overall-procedure}
%%%%%%%%%%%%%%
We summarize the three steps for computing the selective $p$-value as an overall procedure \algo{} in Algorithm \ref{al1}.

{\footnotesize
	\begin{algorithm}[H]
		\caption{\underbar{P}ost \underbar{S}elective \underbar{I}nference for \underbar{M}ultiple \underbar{F}unctional Data (\algo)}\label{al1}
		\textbf{Step I: Low-dimensional embedding}\;
		
		\KwIn{Data of $n$ subjects $\mathcal{W}$, time record $\Omega$, basis functions $\{\phi_s\}_{s\in[q]}$, regularization term $\lambda$.}
		1. Compute the matrices $K_{ij}$ and $\Phi_{ij}$ by $\Omega$ and $\{\phi_s\}_{s\in[q]}$\;
		2. (Basis expansion regression). $X_i \leftarrow (K_{ij}+\lambda I_q)^{-1}\Phi_{ij}W_{ij}$, for $i\in[n]$\;
		\KwOut{Low-dimensional embedding $\{X_i\}_{i\in[n]}$.}
		\vskip.3cm
		
		\textbf{Step II: Covariance estimation}\;
		
		\KwIn{Low-dimensional embedding $\{X_i\}_{i\in[n]}$.}
		
		3. Compute the sample covariance $\hat{\Lambda} \coloneqq \frac{\sum_{i = 1}^n (\text{vec}(X_i)-\text{vec}(\overline{X}))(\text{vec}(X_i)-\text{vec}(\overline{X}))^{\top}}{n-1}$\;
		
		\KwOut{Estimated covariance matrix $\hat{\Lambda}$.}
		\vskip.3cm

		\textbf{Step III: Whitening and Clustering}\;
		\KwIn{Low-dimensional embedding $\{X_i\}_{i\in[n]}$, covariance matrix $\hat{\Lambda}$.}
		4. (Whitening). Conduct the linear transformation $\text{vec}(Y_i) \leftarrow (\hat{\Lambda})^{-\frac{1}{2}}\text{vec}(X_i)$\;
		5. (Clustering). Apply certain clustering algorithm on $\{Y_i\}_{i\in[n]}$ and obtain a partition $\mathcal{C}_1,\mathcal{C}_2$, where the number of clusters is $2$\;
		\KwOut{Whitened data $\{Y_i\}_{i\in[n]}$, partition $\mathcal{C}_1,\mathcal{C}_2$.}
		\vskip.3cm

		\textbf{Step IV: Numerical approximation of the selective $p$-value}\;
		
		\KwIn{Whitened data $\{Y_i\}_{i\in[n]}$, partition $\mathcal{C}_1,\mathcal{C}_2$, sampling horizon $S$.}
		\For{$s = 1\to S$}{
			6. Generate $\gamma_s\sim\sqrt{1/|\mathcal{C}_1|+1/|\mathcal{C}_2|}\cdot\chi_{mq}$, compute $\pi_s = f_1(\gamma_s)/f_2(\gamma_s)$\;
			7. Apply the same clustering algorithm to obtain the partition $\mathcal{C}(F(\gamma_s))$\;
		}
		\KwOut{Selective $p$-value $\frac{\sum_{s = 1}^S\pi_s\mathbbm{1}_{\{\omega_s\geq\lVert\overline{Y}_{\mathcal{C}_1}-\overline{Y}_{\mathcal{C}_2}\rVert,\mathcal{C}_1,\mathcal{C}_2\in\mathcal{C}(F(\gamma_s))\}}}{\sum_{s = 1}^S\pi_s\mathbbm{1}_{\{\mathcal{C}_1,\mathcal{C}_2\in\mathcal{C}(F(\gamma_s))\}}}$.}
	\end{algorithm}
}
 {In practice, the current paper treats the number of clusters $K$ as fixed in advance and sets $K=2$ throughout the theory, simulations, and real-data analyses. A fully data-adaptive choice of $K$ would introduce an additional layer of selection that is not modeled by the current selective pivot.}

%%%%%%%%%%%%%

%%%%%%%%%%%%%%
\section{Theoretical Guarantees}\label{s4}
%%%%%%%%%%%%%%
In this section, we present theoretical results for \algo. We first focus on the selective type-I error. Recall that the selective $p$-value is the survival function of a truncated chi-squared distribution. Therefore, if the selective $p$-value conditioning on the selection set follows a uniform distribution on $[0, 1]$, then the selective type-I error can be controlled accordingly. The following theorem provides a formal statement:
\begin{theorem}(Selective Type-I error control).\label{t2}
	Suppose $\mathcal{W}$ follows Model \eqref{3} and $w$ is a realization of $\mathcal{W}$. Suppose the partition $\mathcal{C}(L(w)) = \{\mathcal{C}_1,\mathcal{C}_2\}$ ($\mathcal{C}_1\cup\mathcal{C}_2=[n], \mathcal{C}_1\cap\mathcal{C}_2=\emptyset$) is the output of the clustering algorithm $\mathcal{C}(\cdot)$. Suppose $\text{vec}(\overline{L(\mathcal{W})}_{\mathcal{C}_1}-\overline{L(\mathcal{W})}_{\mathcal{C}_2})\sim \sqrt{1/|\mathcal{C}_1|+1/|\mathcal{C}_2|}\cdot\mathcal{N}(0,I_{mq})$ if the null hypothesis $H_0^{\{\mathcal{C}_1,\mathcal{C}_2\}}$ holds, then the selective type-I error is controlled by $\alpha\in[0,1]$:
	\begin{equation}\label{eq20}
		\mathbb{P}_{H_0^{\{\mathcal{C}_1,\mathcal{C}_2\}}}\left(p(\mathcal{W};\mathcal{C}_1,\mathcal{C}_2)\leq\alpha\big|~\mathcal{C}_1,\mathcal{C}_2\in\mathcal{C}(L(\mathcal{W}))\right) = \alpha,
	\end{equation}
	where $p(\mathcal{W};\mathcal{C}_1,\mathcal{C}_2)$ denotes the selective $p$-value given the data $\mathcal{W}$ and the partition $\{\mathcal{C}_1,\mathcal{C}_2\}$.
\end{theorem}

\begin{proof}
	See Appendix \ref{pt2}.
\end{proof}

Next, we study the statistical power of \algo, beginning with an intuitive analysis. Under the alternative hypothesis $H_1^{\{\mathcal{C}_1,\mathcal{C}_2\}}$, if $|\mathcal{C}_1|,|\mathcal{C}_2|\rightarrow\infty$, $|\mathcal{C}_1|/|\mathcal{C}_2|\rightarrow c\in(0,1)$, \eqref{covasy} implies that as $n\rightarrow\infty$,
\[\hat{\Lambda}\stackrel{p}{\longrightarrow} \Lambda+\frac{c}{(c+1)^2}\text{vec}(\beta_{\mathcal{C}_1}-\beta_{\mathcal{C}_2})\cdot\text{vec}(\beta_{\mathcal{C}_1}-\beta_{\mathcal{C}_2})^{\top}.\]
%If there are sufficient time records, %i.e., $\min_{i\in[n],j\in[m]}r_{ij}\rightarrow\infty$, 
Also, suppose there are infinitely many time points, i.e., $\min_{i\in[n],j\in[m]}r_{ij}\rightarrow\infty$, Lemma \ref{l1} implies that
\[\text{vec}(H(\mathcal{W})[i,:,:])\sim\mathcal{N}(\text{vec}(\beta_{(i)}),\Lambda).\]
Recall the whitening transformation $\text{vec}(L(\mathcal{W})[i,:,:])= \hat{\Lambda}^{-\frac{1}{2}}\text{vec}(H(\mathcal{W})[i,:,:])$, as $n\rightarrow\infty$, we have
\[\text{vec}(\overline{L(\mathcal{W})}_{\mathcal{C}_1}-\overline{L(\mathcal{W})}_{\mathcal{C}_2})\stackrel{p}{\longrightarrow}\hat{\Lambda}^{-\frac{1}{2}}\cdot\text{vec}(\beta_{\mathcal{C}_1}-\beta_{\mathcal{C}_2}).\]
Intuitively, since $\mu_i$ are Lipschitz continuous for $i\in[n]$ according to Assumption \ref{a2}, as the difference between clusters increases, i.e., $\lVert\mu_{\mathcal{C}_1}-\mu_{\mathcal{C}_2}\rVert_\infty\rightarrow\infty$, we have $\lVert\beta_{\mathcal{C}_1}-\beta_{\mathcal{C}_2}\rVert_F\rightarrow\infty$. Following the Sherman–Morrison formula $(A+uv^\top)^{-1} = A^{-1}-A^{-1}uv^\top A^{-1}/(1+v^\top A^{-1}u)$, we obtain that
$\lVert\overline{L(\mathcal{W})}_{\mathcal{C}_1}-\overline{L(\mathcal{W})}_{\mathcal{C}_2}\rVert_F\stackrel{p}{\longrightarrow}(c+1)/\sqrt{c}.$ Therefore, suppose $w$ is a realization of $\mathcal{W}$, plugging the above asymptotic property into Lemma \ref{l3}, the selective $p$-value is
\[p_{selective} \approx 1-\mathbb{F}\left((c+1)/\sqrt{c};\sqrt{1/|\mathcal{C}_1|+1/|\mathcal{C}_2|},\mathcal{S}(w;\mathcal{C}_1,\mathcal{C}_2)\right),\]
 {which converges to $0$ as the sample size $n$ increases.} The following theorem provides a formal statement.

\begin{figure}[t]
	\centering
	\subfigure[]{
		\begin{minipage}[t]{0.3\textwidth}
			\centering
			\includegraphics[width=0.9\textwidth]{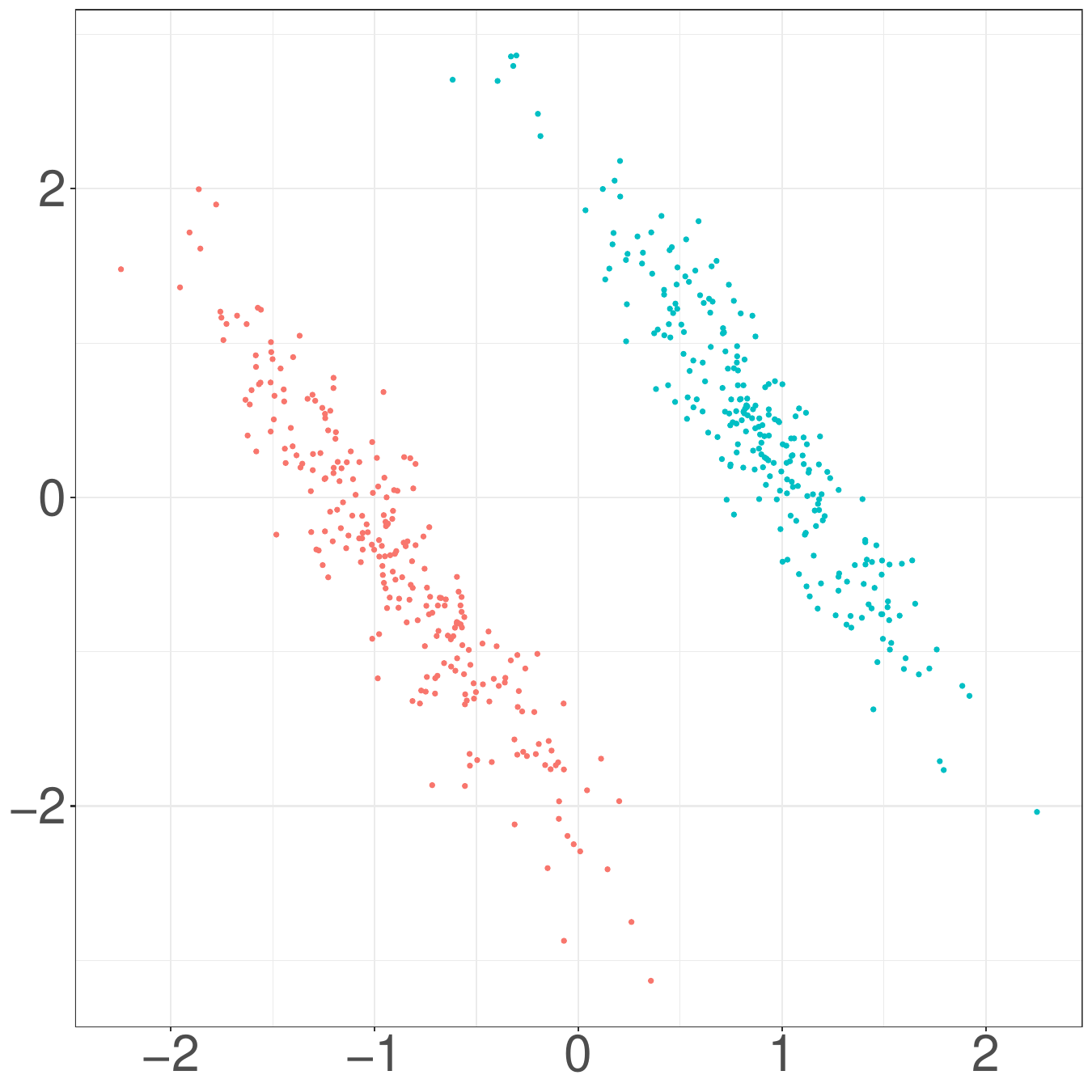}
		\end{minipage}
		\label{fig3a}
	}
	\subfigure[]{
		\begin{minipage}[t]{0.3\textwidth}
			\centering
			\includegraphics[width=0.9\textwidth]{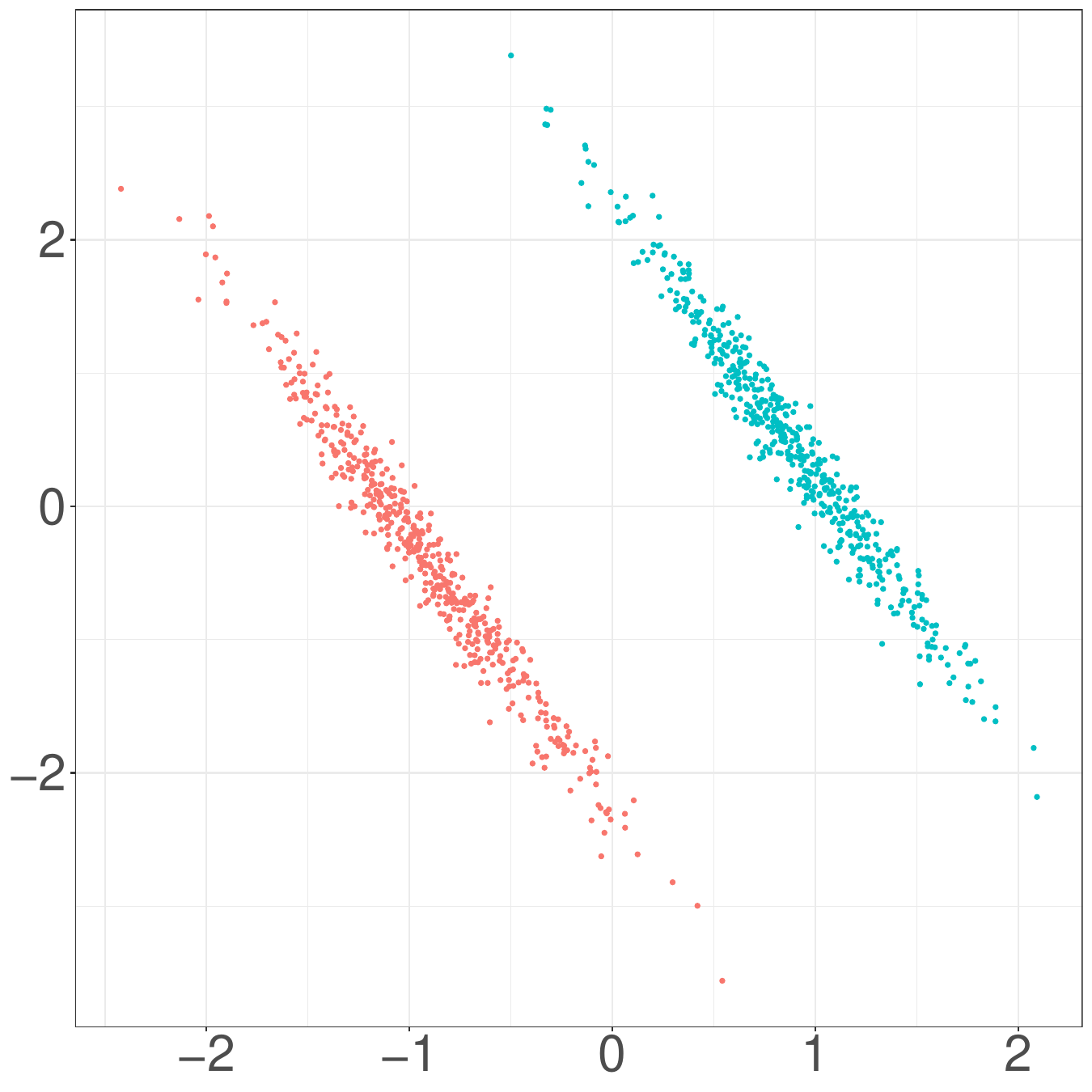} 
		\end{minipage}
		\label{fig3b}
	}
	\subfigure[]{
		\begin{minipage}[t]{0.3\textwidth}
			\centering
			\includegraphics[width=0.9\textwidth]{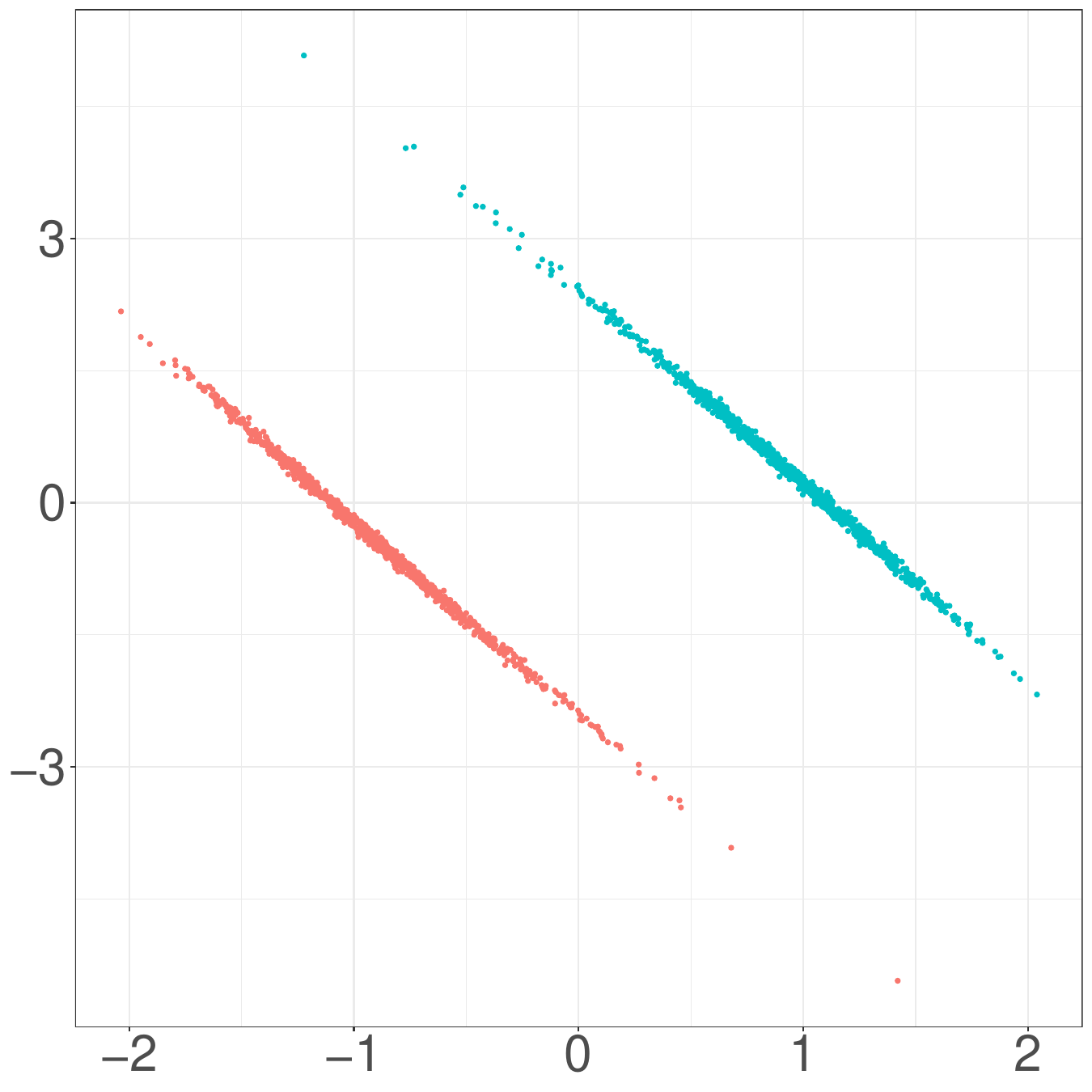} 
		\end{minipage}
		\label{fig3c}
	}
	\caption{\textbf{Scatter plot of $L(\mathcal{W})$.} Set $m = 1, q = 2,r_{ij} = 20$ and $n\in\{500,1000,2000\}$ for all subjects and features, set $|\mathcal{C}_1| = |\mathcal{C}_2| = n/2, \mu_{\mathcal{C}_1} \in\{5,10,50\}$ and $\mu_{\mathcal{C}_2} = -\mu_{\mathcal{C}_1}$. \textbf{(a)}. scatter plot of $L(\mathcal{W})$ with $n = 500$ and $\lVert\mu_{\mathcal{C}_1}-\mu_{\mathcal{C}_2}\rVert_\infty = 10$. \textbf{(b)}. scatter plot of $L(\mathcal{W})$ with $n = 1000$ and $\lVert\mu_{\mathcal{C}_1}-\mu_{\mathcal{C}_2}\rVert_\infty = 20$.  \textbf{(c)}. scatter plot of $L(\mathcal{W})$ with $n = 2000$ and $\lVert\mu_{\mathcal{C}_1}-\mu_{\mathcal{C}_2}\rVert_\infty = 100$.}
	\label{fig3}
\end{figure}

\begin{theorem}\label{t3}(Statistical power). Suppose that $\mathcal{W}$ follows Model \eqref{3} and $w$ is a realization of $\mathcal{W}$. Suppose that the partition $\mathcal{C}(L(w)) = \{\mathcal{C}_1,\mathcal{C}_2\}$ ($\mathcal{C}_1\cup\mathcal{C}_2=[n]$, $\mathcal{C}_1\cap\mathcal{C}_2=\emptyset$) is the output of a clustering algorithm $\mathcal{C}(\cdot)$. Suppose that $\text{vec}(H(\mathcal{W})[i,:,:])\sim\mathcal{N}(\text{vec}(\beta_{(i)}),\Lambda)$ if the alternative hypothesis $H_1^{\{\mathcal{C}_1,\mathcal{C}_2\}}$ holds. Then for all $\alpha\in(0,1]$, we have
	\[\lim_{\lVert\mu_{\mathcal{C}_1}-\mu_{\mathcal{C}_2}\rVert_\infty\rightarrow\infty,n\rightarrow\infty}\mathbb{P}_{H_1^{\{\mathcal{C}_1,\mathcal{C}_2\}}}\left(p(\mathcal{W};\mathcal{C}_1,\mathcal{C}_2)\leq\alpha\big|~\mathcal{C}_1,\mathcal{C}_2\in\mathcal{C}(L(\mathcal{W}))\right) = 1\]
	if (i): $|\mathcal{C}_1|, |\mathcal{C}_2|\rightarrow\infty$ and $|\mathcal{C}_1|/|\mathcal{C}_2|\rightarrow c\in(0,1)$ as $n\rightarrow\infty$, (ii): for any $\epsilon,\delta>0$, there exists $M>0$ such that $\mathbb{P}_{H_1^{\{\mathcal{C}_1,\mathcal{C}_2\}}}\left(\sqrt{1/|\mathcal{C}_1|+1/|\mathcal{C}_2|}\cdot s\in\mathcal{S}(\mathcal{W};\mathcal{C}_1,\mathcal{C}_2)\right)\geq 1-\epsilon$ for any $s>\delta$, $\lVert\mu_{\mathcal{C}_1}-\mu_{\mathcal{C}_2}\rVert_\infty> M$ and $n>M$. Here, the covariance function $R$ is fixed. 
\end{theorem}
\begin{proof}
	See Appendix \ref{pt3}.
\end{proof}

We briefly explain Assumptions (i) and (ii) in the above theorem. First, Assumption (i) implies that two clusters are asymptotically balanced as the sample size increases. Secondly, Assumption (ii) suggests that the lower bound of $\mathcal{S}(\mathcal{W};\mathcal{C}_1,\mathcal{C}_2)$ converges to 0 as $n$ and $\lVert\mu_{\mathcal{C}_1}-\mu_{\mathcal{C}_2}\rVert_\infty$ increase. Recall the geometric intuition of $\mathcal{S}(\mathcal{W};\mathcal{C}_1,\mathcal{C}_2)$ in Section \ref{s323}, the lower bound of $\mathcal{S}(\mathcal{W};\mathcal{C}_1,\mathcal{C}_2)$ is the maximum range of ``push back" under the linear operator $F(\cdot)$ that remains the same clustering outputs. Figure \ref{fig3} shows the shape of two clusters becomes flatter as $n$ and $\lVert\mu_{\mathcal{C}_1}-\mu_{\mathcal{C}_2}\rVert_\infty$ increase. Thus, the clustering algorithm may recover $\mathcal{C}_1,\mathcal{C}_2$ even if the distance of two clusters and the lower bound of $\mathcal{S}(\mathcal{W};\mathcal{C}_1,\mathcal{C}_2)$ tend to zero when $n$ and $\lVert\mu_{\mathcal{C}_1}-\mu_{\mathcal{C}_2}\rVert_\infty$ grow, namely Assumption (ii) is feasible. In addition, we intuitively explain why clusters become flatter as $n$ and $\lVert\mu_{\mathcal{C}_1}-\mu_{\mathcal{C}_2}\rVert_\infty$ grow: recall that $\text{vec}(L(\mathcal{W})[i,:,:])= \hat{\Lambda}^{-\frac{1}{2}}\text{vec}(H(\mathcal{W})[i,:,:])\sim\mathcal{N}(\hat{\Lambda}^{-\frac{1}{2}}\text{vec}(\beta_{(i)}),\hat{\Lambda}^{-\frac{1}{2}}\Lambda\hat{\Lambda}^{-\frac{1}{2}})$. Therefore, as $n$ and $\lVert\mu_{\mathcal{C}_1}-\mu_{\mathcal{C}_2}\rVert_\infty$ increase, $\hat{\Lambda}^{-\frac{1}{2}}\Lambda\hat{\Lambda}^{-\frac{1}{2}}$ is approximately singular and the shape of clusters becomes flatter.  {Theorem~\ref{t3} does not prove that the truncation set $\mathcal{S}(\mathcal{W};\mathcal{C}_1,\mathcal{C}_2)$ is automatically unbounded above for an arbitrary clustering rule. Rather, this behavior is encoded in Assumption~(ii): asymptotically, sufficiently large perturbations along the selected contrast remain in the selection event with high probability. If one could keep increasing the perturbation until it left the set, then Assumption~(ii) would fail and the theorem would no longer apply. This is one reason the result should be interpreted as a two-cluster statement rather than a general guarantee for clustering algorithms with $K>2$.}

\section{Simulation Studies}\label{s5}
%%%%%%%%%%%%%

In this section, we conduct experiments on synthetic data to evaluate the performance of the proposed procedure \algo. We first assess the selective type-I error to verify the consistency of \algo's performance with Theorem \ref{t2}. Subsequently, we examine the statistical power and explore the robustness of the proposed selective inference framework in Section \ref{misval}. Due to page constraints, we provide the basic setup and discussion of the experiments in this section, while the figures of the experiments are presented in Appendix \ref{ap1}.  {We also add a supplementary sparsity experiment in which subject-specific observation times are removed at random, creating unequal observed time grids across individuals; the corresponding figures are reported in Appendix~\ref{ap1}.}

%%%%%%%%%%%%%%%%
\paragraph{Selective type-I error under a global null.}\label{s511}
%%%%%%%%%%%%%%%%

We generate a dataset containing $100$ instances following Model \eqref{3}, where each instance contains $n = 10000$ subjects, $m = 1$ feature, and $r_{ij} = 15$ time points. Specifically, for all $s\in[100]$, $i\in[n]$, and $j \in [m]$, we generate $\Omega_i^{(s)} = \{t_{ijk}\}_{k\in [r_{ij}]}$, where $t_{ijk}\stackrel{iid}{\sim}\mathcal{U}[0,1]$ and $W_{ij}^{(s)}(t_{ijk}) = Z_{ij}^{(s)}(t_{ijk}) + \epsilon_{ijk}^{(s)}$, where $Z_i^{(s)}\sim\mathcal{MGP}(\mu_i,R)$, $\epsilon_{ijk}^{(s)} \stackrel{iid}{\sim} \mathcal{N}(0, \sigma_j^2)$, and $Z_i^{(s)}, \epsilon_{ijk}^{(s)}$ are independent. Here we set $\sigma_j^2 = 0.1$, $\mu_i = 0$ and conduct the simulation for three different covariance functions:
\begin{enumerate}[(i)]
	\item Rational quadratic kernel
	\[R(x,y) = \left(1+\frac{(x-y)^2}{\ell^2}\right)^{-1/2};\]
	\item Periodic kernel
	\[R(x,y) = e^{-{8\sin^2(2\pi|x-y|)}};\]
	\item Truncated local periodic kernel
	\begin{equation*}
		\begin{split}
			R(x,y) = & \mathbbm{1}_{\{1/3<|x-y|<2/3\}}\cdot e^{-8\sin^2(2\pi|x-y|)}e^{-2(x-y)^2}\\
			& +\mathbbm{1}_{\{|x-y|\leq 1/3~\text{or}~|x-y|\geq2/3\}}\cdot 0.01.
		\end{split}
	\end{equation*}
\end{enumerate}
Next, we set the basis functions $\{\phi_s\}_{s\in[q]}$ as the eigenfunctions of the Gaussian RBF (Radial Basis Function) kernel $R(x,y) = e^{-\frac{\rho}{1-\rho^2}(x-y)^2}$, where $\rho\in(0,1)$.
By the Mercer expansion \citep{fasshauer2012stable}, the $i$-th eigenfunction of $R(x, y)$ is
\begin{equation}\label{e30}
	\phi_i(x) = \frac{1}{\sqrt{N_i}}H_i(x)e^{-\frac{\rho}{1+\rho}x^2},
\end{equation}
where $N_i = 2^ii!\sqrt{\frac{1-\rho}{1+\rho}}$ and $H_i(x)$ is the $i$-th order physicist's Hermite polynomial. In this experiment, we set $\rho = 0.99$ and set the truncation number $q$ to be $3$, i.e., we use the first three eigenfunctions to conduct the low-dimensional embedding.

Now, we apply the proposed selective inference framework to the generated datasets. Figure \ref{f4} displays quantile plots of the selective $p$-values for datasets corresponding to the three kernels above. The plots demonstrate that the selective $p$-values approximately follow a uniform distribution under the global null hypothesis, thereby validating the statement of Theorem \ref{t2}.

%%%%%%%%%%%%%%%
\paragraph{Statistical power.}\label{s512}
%%%%%%%%%%%%%%%
Next, we present numerical results to verify Theorem \ref{t3}. Specifically, we generate datasets following Model \eqref{a2} under the alternative hypothesis and compute the corresponding statistical power. We compute the selective $p$-value for datasets generated with varying cluster means $\lVert\mu_{\mathcal{C}_1}-\mu_{\mathcal{C}_2}\rVert_\infty$ and sample sizes $n$.

We set the sample size $n = 10 k$ for $k\in\{4,\ldots,10\}$. We fix $m = 1$ and set $r_{ij} = 15$, $\sigma_j^2 = 0.1$ for all $i\in[n], j\in[m]$. For each sample size $n$, we generate a dataset containing $n$ subjects following the alternative hypothesis: $\mu_i(\cdot) = -10$ for $i\leq m/2$ and $\mu_i(\cdot) = 10$ for $i> m/2$. We use the same basis as in \eqref{e30} with the parameter $q = 3$ to conduct the low-dimensional embedding. Figure \ref{f5b} presents the statistical power with a fixed mean difference and increasing sample sizes, showing that the statistical power increases as the sample size increases. 

To investigate the statistical power for different cluster means, we fix the sample size at $n = 80$ and the other parameters remaining the same as in the previous paragraph. For each $k\in\{3.5, 4, 4.5, 5, 5.5, 6, 6.5\}$, we generate a dataset with $n$ records with population means $\mu_i(\cdot) = k$ for $i\leq n/2$ and $\mu_i(\cdot) = -k$ for $i>n/2$. Figure \ref{f5c} presents the statistical power with the same sample size and the increasing difference between cluster means, it shows that the statistical power increases as the difference between cluster means increases.

\paragraph{Empirical robustness analysis.}\label{misval}

We consider three misspecification cases and compute the selective $p$-value under a global null: Wiener process (Figure \ref{f7a}), exponential Brownian motion (Figure \ref{f7c}), and Ornstein–Uhlenbeck (OU) process (Figure \ref{f7e}). We also present the QQ-plot of the selective $p$-value in Appendix \ref{ap1}.  {This experiment serves as a robustness analysis under model misspecification. In particular, the exponential Brownian motion case provides a genuinely non-Gaussian departure from the main working model, while the Wiener and OU cases probe robustness to alternative stochastic process structures.}

\paragraph{Unequal observation times across individuals.} {To study the setting that motivates the kernel embedding directly, we generated trajectories on a common 100-point grid and then removed observation times independently for each subject, yielding subject-specific incomplete time records. We considered loss rates of 10\%, 30\%, 50\%, and 70\%, and recomputed the embedding using only the observed time locations. Figures~\ref{fsparse1} and~\ref{fsparse2} display the first five observed sample curves together with the QQ plots of the selective $p$-values for Gaussian/RBF, RQ, and LPE embeddings. Across all four sparsity levels, the QQ plots remain reasonably close to the diagonal, indicating that the selective $p$-values stay well calibrated even when the observed time points differ substantially across individuals.}

\paragraph{ {Within-cluster mean-function heterogeneity.}} {We next assess sensitivity to the common within-cluster mean assumption while keeping the intended coarse two-cluster contrast fixed. We generated $n=80$ subjects from three latent subtypes $A,B,C$ with sizes $40,20,20$ and mean functions
\[
\mu_A(t)=d,\qquad \mu_B(t)=-d+\tau h(t),\qquad \mu_C(t)=-d-\tau h(t),
\]
where $d=4.5$ and $h(t)=\sqrt{3}(2t-1)$. The intended coarse comparison is $A$ versus $B\cup C$. Because $B$ and $C$ have equal sizes, the average mean function of $B\cup C$ remains $-d$, and the coarse contrast is unchanged as the within-cluster heterogeneity varies. We define
\[
\eta=\frac{\lVert\mu_B-\mu_C\rVert_{L^2}}{\lVert\mu_A-\overline{\mu}_{B\cup C}\rVert_{L^2}}=\frac{\tau}{d}
\]
and consider $\eta\in\{0,0.1,0.2,0.3,0.4,0.5\}$. Thus, $\eta=0$ satisfies the common-mean assumption within $B\cup C$, while $\eta>0$ gives a controlled violation. For each level, we generated 100 datasets with 15 subject-specific uniform observation times, a periodic covariance kernel, $q=3$, $\lambda=10^{-7}$, and Ward hierarchical clustering, and applied the full PSIMF pipeline.

Table~\ref{tab:heterogeneity} reports the results. Across all heterogeneity levels, the mean coarse-label accuracy remained between $0.966$ and $0.984$, and the proportion of replicates with accuracy at least $0.90$ remained between $0.94$ and $1.00$. Conditional on this high-accuracy recovery, the selective rejection rate ranged from $0.713$ to $0.863$, with no systematic decrease as $\eta$ increased. These results suggest that the intended coarse two-cluster comparison can remain empirically stable under moderate violations in this design. For $\eta>0$, however, the strict working model is misspecified, so this experiment is a sensitivity analysis rather than a Type~I error study or an extension of the exact guarantee.}

\paragraph{Recursive three-cluster illustration.} {To illustrate how PSIMF can be used when more than two latent groups are present, we also considered a simple recursive three-cluster experiment. The idea is to apply PSIMF sequentially: first, cluster the full sample into two groups and test the Stage~1 split; then, if the first split is significant, re-apply the full pipeline to the selected branch and test the Stage~2 split. We considered three settings corresponding to this recursive structure: a null setting with no true split, a \texttt{coarse\_only} setting in which only the first split is present, and a \texttt{three\_cluster} setting in which both the coarse split and the finer split are present. 

Appendix~\ref{ap1} reports a representative simulated dataset in Figure~\ref{fig:threecluster_illustrative}, the recursive splitting in the whitened embedding in Figure~\ref{fig:threecluster_scatter}, the corresponding evolution of the longitudinal curves across stages in Figure~\ref{fig:threecluster_curves}, and the stagewise summary in Table~\ref{tab:threecluster}. Figure~\ref{fig:threecluster_illustrative} shows the basic three-cluster geometry in both the longitudinal curves and the whitened low-dimensional embedding. Figure~\ref{fig:threecluster_scatter} then visualizes the recursive process itself, displaying the truth, the Stage~1 split on the full sample, the Stage~2 split after re-applying the full pipeline to the selected branch, and the final recursive labels. Figure~\ref{fig:threecluster_curves} presents the same recursive progression in the original longitudinal trajectories. Table~\ref{tab:threecluster} summarizes the stagewise behavior across repeated simulations. 

In the \texttt{coarse\_only} setting, the Stage~1 correct-split rate and selective rejection rate are both high, while the conditional Stage~2 selective rejection rate remains low, which is the expected pattern when only the first split is truly present. In the \texttt{three\_cluster} setting, the Stage~1 split is recovered in a substantial fraction of replicates, and conditional on proceeding to Stage~2, the finer split is recovered almost perfectly with a high conditional selective rejection rate. These results suggest that PSIMF can serve as a recursive building block for multi-cluster structure by reducing the problem to a sequence of selected two-cluster comparisons.}

%%%%%%%%%%%%%%%
\section{Phenotyping of AKI based on EHR}\label{sec:AKI}
%%%%%%%%%%%%%%%
Now we present a real-data application of our selective inference framework. Acute Kidney Injury (AKI) is a common clinical syndrome characterized by a complex treatment process and high mortality rates. The pathology of AKI exhibits a high degree of heterogeneity, posing significant challenges to the formulation of treatment plans. Consequently, identifying new AKI subtypes is crucial for improving patient outcomes. The severity of the disease in AKI patients tends to vary over time, making the problem of hypothesis testing for functional disease subtypes of significant practical importance.

We specifically focus on the MIMIC-IV EHR dataset from PhysioNet \citep{johnson2020mimic,johnson2023mimic,goldberger2000physiobank}, which contains de-identified medical data from patients admitted to the Intensive Care Units (ICU) at Beth Israel Deaconess Medical Center from 2008 to 2019. The database provides a variety of medical data, including vital signs, medications, laboratory measurements, diagnostic codes, and hospital length of stay. This dataset is rich in individual patient-level information and is freely accessible, making it feasible for clinical research worldwide.

\begin{figure*}[!hbt]
	\centering
	\captionsetup{justification=centering}
	
	\includegraphics[width=.95\textwidth]{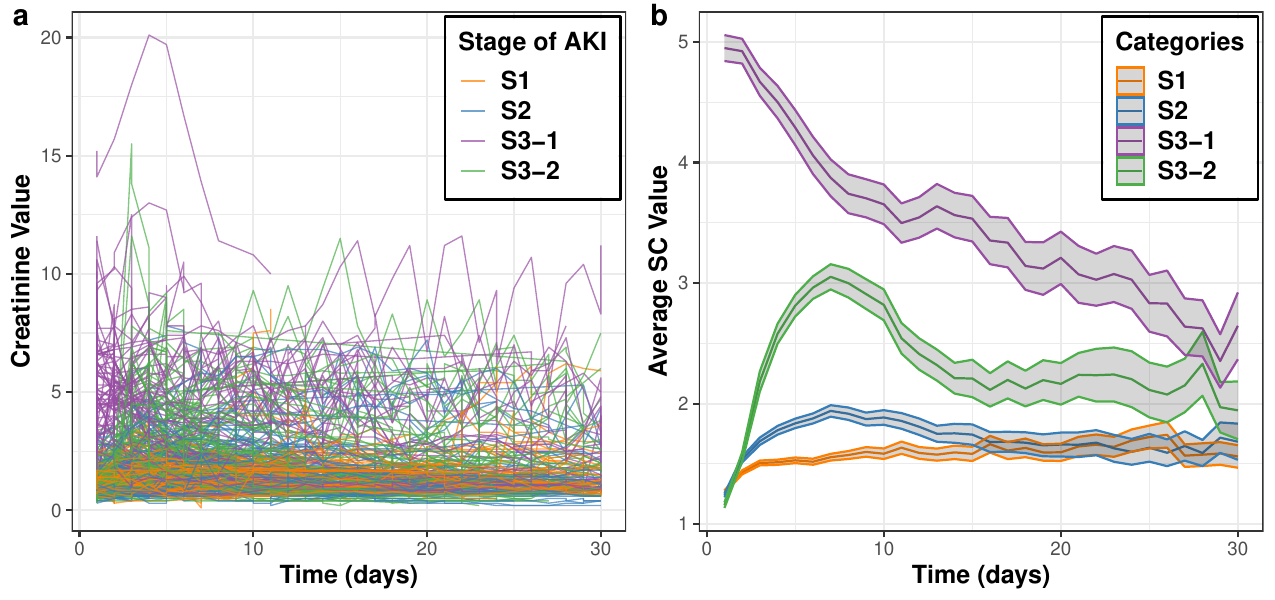}
	
	\caption[Trajectories of AKI subtypes.]{
		a. Real trajectories for 100 randomly selected patients from each category. 
		b. Trajectories (Mean $\pm$ $1.96\times$ standard deviation/$\sqrt{\text{sample size}}$) of the four AKI subtypes.
	}
	\label{fig:3}
\end{figure*}

We focus on data from adult patients with AKI admitted to the ICU. We identify patients with ICD codes with explanations including ``acute kidney failure." Then we preprocess the data similarly to the framework provided by \citep{song2020cross}, excluding patients at or before admission with 1) End Stage Renal Disease, 2) Burns, and 3) Renal Dialysis. Subsequently, according to the clinical practice guidelines for Acute Kidney Injury designated by Kidney Disease Improving Global Outcomes (KDIGO)\footnote{The original Kdigo's definition of AKI staging includes two key quantities: serum Creatinine and urine output. We focus on the criterion of serum creatinine since the urine output information may be unavailable in other studies  \citep{song2020cross} and using SCr is inconsistent with the later analysis. Our baseline creatinine value is chosen as the earliest creatinine measurement recorded within the first 48 hours following the patient's admission to the ICU. The original definition of Kdigo stages \citep{khwaja2012kdigo} does not specify a period of observing increases over baseline. Due to the heterogeneity of ICU stays of these patients, we set the observation period to 48 hours (maximum SCr value) or 7 days (multiplication from the baseline), following the AKI definition used in \citep{song2020cross}. Additionally, considering that a relatively small increase in SCr might be due to random variation, the second condition in the definition of Stage 1 could lead to false positives \citep{makris2016acute,lin2015false}. Therefore, this condition is not considered in this study.
}:
\begin{itemize}
	\item Stage 1: Serum creatinine (SCr) value rises to 1.5-1.9 times the baseline value within 7 days or SCr value increases $\geq$ 0.3 mg/dl within 48 hours.
	\item Stage 2: SCr value rises to 2.0-2.9 times the baseline value within 7 days.
	\item Stage 3: meets at least one of the following two conditions: 
	\begin{itemize}
		\item SCr value rises to 3 times the baseline value or more within 7 days;
		\item increase in SCr value to $\geq$ 4.0 mg/dl within 48 hours.%the maximum SCr value over 48 hours is greater than 4.0mg/dl. 
	\end{itemize}
\end{itemize}
We further divide Stage 3 into two subclasses: S3-1, where there is an increase in SCr value to $\geq$ 4.0 mg/dl within 48 hours; and S3-2, where the SCr value increases to three times the baseline or more within 7 days without meeting the conditions of S3-1. The specific shape of this longitudinal data is shown in Figure \ref{fig:3}. For consistency in definition, we selected data from the first seven days as our study subjects. We then used hierarchical clustering based on squared Euclidean distance to cluster each category combination, specifying the number of clusters as 2. In this clustering scenario, we compared the $p$-values under two distinct test methods: \algo, performing the post-clustering hypothesis test \eqref{1}, and the Wald test.  {Although the clinical data contain more than two predefined AKI categories, each application of PSIMF in this section is still a separate two-cluster analysis with $K=2$, which matches the scope of the current theory.}

\begin{comment}
	\begin{table}[h!]
	\centering
	\begin{tabular}{l|cccc|ccc} 
		\toprule
		Included data & S1 & S2 & S3-1 & S3-2 & (S1, S3-1) & (S1, S3-2)  & (S3-1, S3-2) \\
		\midrule
		\algo   ~$p$-value & 0.2070 & 0.4504 & 0.5324 & 0.9033 & 0.0095 & 0.0254  & 0.5455 \\
		Wald $p$-value & $<10^{-307}$ & $<10^{-307}$ & $<10^{-307}$ & $<10^{-307}$ & $<10^{-307}$ & $<10^{-307}$ & $<10^{-307}$  \\
		\bottomrule
	\end{tabular}
	\caption{Comparison of $p$-values under different clustering scenarios.}
	\label{Table_1}
\end{table}
\end{comment}

\begin{table}[h!]
	\centering
	\begin{tabular}{l|cccc} 
		\toprule
		Included data & S1 & S2 & S3-1 & S3-2  \\
		\midrule
		\algo   ~$p$-value & 0.2070 & 0.4504 & 0.5324 & 0.9033 \\
		Wald $p$-value & $<10^{-307}$ & $<10^{-307}$ & $<10^{-307}$ & $<10^{-307}$  \\
		\midrule

	Included data & (S1, S3-1) & (S1, S3-2)  & (S3-1, S3-2) \\
	\midrule
	\algo   ~$p$-value & 0.0095 & 0.0254  & 0.5455 \\
	Wald $p$-value & $<10^{-307}$ & $<10^{-307}$ & $<10^{-307}$  \\
	\bottomrule
	\end{tabular}
	\caption{Comparison of $p$-values under different clustering scenarios.}
	\label{Table_1}
\end{table}	

As indicated in Table \ref{Table_1}, when the input data comprises only one cohort of patients (S1, S2, S3-1, or S3-2), the $p$-values produced by \algo\;are relatively high, whereas those from the Wald test are significantly lower. Our approach appropriately refrains from rejecting the null hypothesis, unlike the Wald test.  {Under the working model, these results indicate that \algo\;does not detect a significant selected mean-function difference within any of the four predefined cohorts.} 

When the input data includes patients from different AKI stages, the $p$-value for the combination of S1 and S3-1 is low, correctly indicating the heterogeneity of this combined class. Clinically, given the definitions of AKI, S1 and S3-1 exhibit significant differences in distribution and mean, justifying the rejection of the null hypothesis. Similarly, the $p$-value for the combination of S1 and S3-2 is also low. We recognize that these two categories are not adjacent in the clinical staging of AKI---with S1 involving an SCr rise to 1.5-1.9 times the baseline within 7 days, and S3-2 defined by an SCr increase to three times the baseline or more. Therefore, the significant results under \algo\; are justified due to the non-adjacent staging definitions.

Lastly, the $p$-value produced by \algo ~for the combination of S3-1 and S3-2 is relatively high.  {Thus, under the working model, the analysis does not provide evidence of a selected mean-function difference between S3-1 and S3-2.} This is clinically plausible, as both S3-1 and S3-2 are categorized under Stage-3 AKI and are not sufficiently distinct to warrant classification into separate subtypes.

%%%%%%%%%%%%%
\section{Discussions}\label{sec:dis}
%%%%%%%%%%%%%
This paper focuses on the post-clustering inference problem for functional data. We establish a selective inference framework and propose a selective $p$-value for functional data that reduces the selective bias induced by the clustering algorithm. Our theoretical results show that the proposed method controls the selective type-I error and statistical power when there are sufficient time records and sufficient subjects. We use numerical simulation to verify our theory and further apply the method to the phenotyping of Acute Kidney Injury (AKI), in which the selective $p$-value between different stages matches the medical consensus.

Our work opens up several avenues for future research. In this paper, we primarily focus on scenarios where $t_{ijk} \sim \mathcal{U}[0,1]$. However, in real-world applications, time records may be more concentrated during certain periods, potentially invalidating this assumption. A valuable direction for future research would be to extend our theoretical results to accommodate a broader class of distributions for $t_{ijk}$. Additionally, our analysis of selective type-I error and statistical power currently relies on asymptotic approximations, using the limiting distribution of $H(\mathcal{W})$ with an infinite number of time records. Addressing the challenge of deriving finite-sample results for selective type-I error and statistical power remains an open problem and an important next step.  {Another limitation is that the low-dimensional embedding is not lossless: projecting trajectories onto a finite basis inevitably discards part of the functional information. Extending the post-clustering procedure either to direct trajectory-level selective inference or to a regime in which $q$ diverges with $n$ would therefore be highly valuable, but would require new theory for selective inference in infinite-dimensional or increasing-dimensional settings, including control of truncation error, covariance estimation, whitening, and the geometry of the clustering event.}

 {The Gaussian-process assumption and the common within-cluster mean assumption play distinct roles. Gaussianity is not necessary for pooled estimation of population mean and covariance functions; under suitable regularity conditions, pooled estimators can be asymptotically normal \citep{yao2005functional,zhang2016sparse}. It is, however, central to the selective pivot underlying our theoretical construction: after the linear embedding and whitening, joint Gaussianity yields the independence of orthogonal projections and of the norm and direction of the selected contrast, as well as the scaled-$\chi$ law used in Lemma~\ref{l3} and Theorem~\ref{t2}. Replacing the current nuisance estimation by pooled asymptotically normal estimators may lead to an asymptotically valid extension, but would require additional theory accounting for nuisance estimation and the clustering selection event. Likewise, the sensitivity experiment suggests that the coarse two-cluster comparison can remain empirically stable under moderate within-cluster mean heterogeneity, while the exact selective Type~I error guarantee continues to rely on the stated common-mean model.}

In addition, we study the setting where clustering algorithms output two clusters, and it would be interesting to extend the selective inference framework to the setting of multiple clusters. For this problem, a key challenge would be estimating the unknown kernel function with multiple clusters. To elaborate, in Section \ref{sec:covariance-estimation}, we leverage all the data within two clusters and consider the sample covariance estimator. When there are multiple clusters, the transformed data $\text{vec}(L(\mathcal{W})[i,:,:])$ within each cluster follows a truncated multivariate normal distribution. Therefore, combining two clusters and leveraging the sample covariance estimator would be biased. 

Furthermore, the proposed method necessitates that the clustering algorithm $\mathcal{C}(\cdot)$ be applicable to a collection of matrix data ${L(w)[i,:,:]}_{i\in[n]}$. Extending our framework to accommodate a broader class of clustering algorithms is another interesting area for future research.

% Acknowledgements and Disclosure of Funding should go at the end, before appendices and references

%\acks{All acknowledgements go at the end of the paper before appendices and references.
%	Moreover, you are required to declare funding (financial activities supporting the
%	submitted work) and competing interests (related financial activities outside the submitted work).
%	More information about this disclosure can be found on the JMLR website.}


\newpage

\appendix

%%%%%%%%%%
\section{Proof of Main Theorems}\label{sec:additional-proofs}
%%%%%%%%%%

\subsection{Proof of Theorem \ref{t2}}\label{pt2}
Suppose $w$ is a realization of \eqref{3} and $\mathcal{C}_1,\mathcal{C}_2$ is the partition obtain by a clustering algorithm based on $L(w)$. For any $\alpha\in[0,1]$, we consider the following conditional probability:
\begin{equation}\label{a2.1}
	\begin{aligned}
		 \mathbb{P}_{H_0^{\{\mathcal{C}_1,\mathcal{C}_2\}}}\bigg( & p(\mathcal{W};\mathcal{C}_1,\mathcal{C}_2)\leq\alpha\bigg| \mathcal{C}_1,\mathcal{C}_2\in\mathcal{C}(L(\mathcal{W})),\\
		& \pi^{\perp}_{\nu(\mathcal{C}_1,\mathcal{C}_2)}\times_1L(\mathcal{W}) = \pi^{\perp}_{\nu(\mathcal{C}_1,\mathcal{C}_2)}\times_1 L(w),\\
		& \text{dir}(\overline{L(\mathcal{W})}_{\mathcal{C}_1}-\overline{L(\mathcal{W})}_{\mathcal{C}_2}) = \text{dir}(\overline{L(w)}_{\mathcal{C}_1}-\overline{L(w)}_{\mathcal{C}_2})\bigg). 
	\end{aligned}
\end{equation}
Given the partition $\mathcal{C}_1,\mathcal{C}_2$, for any $\mathcal{W}$ follows Model \eqref{3} and satisfies $\mathcal{C}_1,\mathcal{C}_2\in\mathcal{C}(L(\mathcal{W}))$, \eqref{11} implies that
\[p(\mathcal{W};\mathcal{C}_1,\mathcal{C}_2) = 1-\mathbb{F}\left(\lVert\overline{L(\mathcal{W})}_{\mathcal{C}_1}-\overline{L(\mathcal{W})}_{\mathcal{C}_2}\rVert_F;\sqrt{\frac{1}{|\mathcal{C}_1|}+\frac{1}{|\mathcal{C}_2|}},\mathcal{S}(\mathcal{W};\mathcal{C}_1,\mathcal{C}_2)\right).\]
Therefore, we rewrite \eqref{a2.1} as follows:
\begin{equation}\label{a2.2}
	\begin{aligned}
		\mathbb{P}_{H_0^{\{\mathcal{C}_1,\mathcal{C}_2\}}}\bigg(&1-\mathbb{F}\left(\lVert\overline{L(\mathcal{W})}_{\mathcal{C}_1}-\overline{L(\mathcal{W})}_{\mathcal{C}_2}\rVert_F;\sqrt{\frac{1}{|\mathcal{C}_1|}+\frac{1}{|\mathcal{C}_2|}},\mathcal{S}(\mathcal{W};\mathcal{C}_1,\mathcal{C}_2)\right)\leq\alpha\bigg|\\
		&\mathcal{C}_1,\mathcal{C}_2\in\mathcal{C}(L(\mathcal{W})), \pi^{\perp}_{\nu(\mathcal{C}_1,\mathcal{C}_2)}\times_1L(\mathcal{W}) = \pi^{\perp}_{\nu(\mathcal{C}_1,\mathcal{C}_2)}\times_1 L(w),\\
		& \text{dir}(\overline{L(\mathcal{W})}_{\mathcal{C}_1}-\overline{L(\mathcal{W})}_{\mathcal{C}_2}) = \text{dir}(\overline{L(w)}_{\mathcal{C}_1}-\overline{L(w)}_{\mathcal{C}_2})\bigg). 
	\end{aligned}
\end{equation}
Under the conditions $\pi^{\perp}_{\nu(\mathcal{C}_1,\mathcal{C}_2)}\times_1L(\mathcal{W}) = \pi^{\perp}_{\nu(\mathcal{C}_1,\mathcal{C}_2)}\times_1 L(w),
\text{dir}(\overline{L(\mathcal{W})}_{\mathcal{C}_1}-\overline{L(\mathcal{W})}_{\mathcal{C}_2}) = \text{dir}(\overline{L(w)}_{\mathcal{C}_1}-\overline{L(w)}_{\mathcal{C}_2})$,  the truncation sets $\mathcal{S}(\mathcal{W};\mathcal{C}_1,\mathcal{C}_2)$ and $\mathcal{S}(w;\mathcal{C}_1,\mathcal{C}_2)$ are equivalent:
\begin{equation}\nonumber
	\begin{aligned}
		\mathcal{S}&(\mathcal{W};\mathcal{C}_1,\mathcal{C}_2)\\
		&= \Bigg\{\varphi\geq 0:\mathcal{C}_1,\mathcal{C}_2\in\mathcal{C}\Bigg(\pi^{\perp}_{\nu(\mathcal{C}_1,\mathcal{C}_2)}\times_1 L(\mathcal{W})\\
		& \quad +\left[\frac{\varphi}{\frac{1}{|\mathcal{C}_1|}+\frac{1}{|\mathcal{C}_2|}}\right]\nu(\mathcal{C}_1,\mathcal{C}_2)\times_1\text{dir}(\overline{L(\mathcal{W})}_{\mathcal{C}_1}-\overline{L(\mathcal{W})}_{\mathcal{C}_2})^{\top}\Bigg) \Bigg\}\\
		&= \Bigg\{\varphi\geq 0:\mathcal{C}_1,\mathcal{C}_2\in\mathcal{C}\Big(\pi^{\perp}_{\nu(\mathcal{C}_1,\mathcal{C}_2)}\times_1 L(w)\\
		& \quad +\left[\frac{\varphi}{\frac{1}{|\mathcal{C}_1|} +\frac{1}{|\mathcal{C}_2|}}\right]\nu(\mathcal{C}_1,\mathcal{C}_2)\times_1\text{dir}(\overline{L(w)}_{\mathcal{C}_1}-\overline{L(w)}_{\mathcal{C}_2})^{\top}\Big) \Bigg\}\\
		&= \mathcal{S}(w;\mathcal{C}_1,\mathcal{C}_2).
	\end{aligned}
\end{equation}
Note that the random variable $\lVert\overline{L(\mathcal{W})}_{\mathcal{C}_1}-\overline{L(\mathcal{W})}_{\mathcal{C}_2}\rVert_F$ is independent of $\pi^{\perp}_{\nu(\mathcal{C}_1,\mathcal{C}_2)}\times_1L(\mathcal{W})$ and $\text{dir}(\overline{L(w)}_{\mathcal{C}_1}-\overline{L(w)}_{\mathcal{C}_2})$ (we refer readers to Appendix \ref{pl3} for the proof). Therefore, the conditional probability \eqref{a2.2} can be rewritten as follows
\begin{equation}\label{a2.3}
	\begin{aligned}
		&\mathbb{P}_{H_0^{\{\mathcal{C}_1,\mathcal{C}_2\}}}\bigg\{1-\mathbb{F}\left(\lVert\overline{L(\mathcal{W})}_{\mathcal{C}_1}-\overline{L(\mathcal{W})}_{\mathcal{C}_2}\rVert_F;\sqrt{\frac{1}{|\mathcal{C}_1|}+\frac{1}{|\mathcal{C}_2|}},\mathcal{S}(w;\mathcal{C}_1,\mathcal{C}_2)\right)\leq\alpha\bigg|~\\
		& \quad \mathcal{C}_1,\mathcal{C}_2\in\mathcal{C}\bigg(\pi^{\perp}_{\nu(\mathcal{C}_1,\mathcal{C}_2)}\times_1 L(w)\\
		& \quad +\left[\frac{\lVert \overline{L(\mathcal{W})}_{\mathcal{C}_1}-\overline{L(\mathcal{W})}_{\mathcal{C}_2} \rVert_F}{1/|\mathcal{C}_1|+1/|\mathcal{C}_2|}\right]
		\nu(\mathcal{C}_1,\mathcal{C}_2)\times_1\text{dir}(\overline{L(w)}_{\mathcal{C}_1}-\overline{L(w)}_{\mathcal{C}_2})^{\top}\bigg),\\&\quad\pi^{\perp}_{\nu(\mathcal{C}_1,\mathcal{C}_2)}\times_1L(\mathcal{W}) = \pi^{\perp}_{\nu(\mathcal{C}_1,\mathcal{C}_2)}\times_1 L(w),\\
		& \quad \text{dir}(\overline{L(\mathcal{W})}_{\mathcal{C}_1}-\overline{L(\mathcal{W})}_{\mathcal{C}_2}) = \text{dir}(\overline{L(w)}_{\mathcal{C}_1}-\overline{L(w)}_{\mathcal{C}_2})\bigg\}\\
		=& \mathbb{P}_{H_0^{\{\mathcal{C}_1,\mathcal{C}_2\}}}\bigg\{1-\mathbb{F}\left(\lVert\overline{L(\mathcal{W})}_{\mathcal{C}_1}-\overline{L(\mathcal{W})}_{\mathcal{C}_2}\rVert_F;\sqrt{\frac{1}{|\mathcal{C}_1|}+\frac{1}{|\mathcal{C}_2|}},\mathcal{S}(w;\mathcal{C}_1,\mathcal{C}_2)\right)\leq\alpha\bigg|\\
		& \quad \mathcal{C}_1,\mathcal{C}_2\in\mathcal{C}\bigg(\pi^{\perp}_{\nu(\mathcal{C}_1,\mathcal{C}_2)}\times_1 L(w)\\
		& \quad +\left[\frac{\lVert \overline{L(\mathcal{W})}_{\mathcal{C}_1}-\overline{L(\mathcal{W})}_{\mathcal{C}_2} \rVert_F}{1/|\mathcal{C}_1|+1/|\mathcal{C}_2|}\right]
		\nu(\mathcal{C}_1,\mathcal{C}_2)\times_1\text{dir}(\overline{L(w)}_{\mathcal{C}_1}-\overline{L(w)}_{\mathcal{C}_2})^{\top}\bigg)\bigg\}\\
		=&\mathbb{P}_{H_0^{\{\mathcal{C}_1,\mathcal{C}_2\}}}\bigg(1-\mathbb{F}\left(\lVert\overline{L(\mathcal{W})}_{\mathcal{C}_1}-\overline{L(\mathcal{W})}_{\mathcal{C}_2}\rVert_F;\sqrt{\frac{1}{|\mathcal{C}_1|}+\frac{1}{|\mathcal{C}_2|}},\mathcal{S}(w;\mathcal{C}_1,\mathcal{C}_2)\right)\leq\alpha\bigg|\\&\quad\lVert \overline{L(\mathcal{W})}_{\mathcal{C}_1}-\overline{L(\mathcal{W})}_{\mathcal{C}_2} \rVert_F\in\mathcal{S}(w;\mathcal{C}_1,\mathcal{C}_2)\bigg).
	\end{aligned}
\end{equation}
Plugging \eqref{a2.3} into \eqref{a2.1}, we have
\begin{equation}\label{a2.4}
	\begin{aligned}
		&\mathbb{P}_{H_0^{\{\mathcal{C}_1,\mathcal{C}_2\}}}\bigg(p(\mathcal{W};\mathcal{C}_1,\mathcal{C}_2)\leq\alpha\bigg|\mathcal{C}_1,\mathcal{C}_2\in\mathcal{C}(L(\mathcal{W})),\\
		& \quad \pi^{\perp}_{\nu(\mathcal{C}_1,\mathcal{C}_2)}\times_1L(\mathcal{W}) = \pi^{\perp}_{\nu(\mathcal{C}_1,\mathcal{C}_2)}\times_1 L(w),\\
		&\quad\text{dir}(\overline{L(\mathcal{W})}_{\mathcal{C}_1}-\overline{L(\mathcal{W})}_{\mathcal{C}_2}) = \text{dir}(\overline{L(w)}_{\mathcal{C}_1}-\overline{L(w)}_{\mathcal{C}_2})\bigg)\\
		=&\mathbb{P}_{H_0^{\{\mathcal{C}_1,\mathcal{C}_2\}}}\bigg(1-\mathbb{F}\left(\lVert\overline{L(\mathcal{W})}_{\mathcal{C}_1}-\overline{L(\mathcal{W})}_{\mathcal{C}_2}\rVert_F;\sqrt{\frac{1}{|\mathcal{C}_1|}+\frac{1}{|\mathcal{C}_2|}},\mathcal{S}(w;\mathcal{C}_1,\mathcal{C}_2)\right)\leq\alpha\bigg|\\&\quad\lVert \overline{L(\mathcal{W})}_{\mathcal{C}_1}-\overline{L(\mathcal{W})}_{\mathcal{C}_2} \rVert_F\in\mathcal{S}(w;\mathcal{C}_1,\mathcal{C}_2)\bigg)= \alpha.
	\end{aligned}
\end{equation}
Now we use \eqref{a2.4} to compute the selective $p$-value. By the law of iterated expectation, we rewrite the selective type-I error as follows:
\begin{equation}\nonumber
	\begin{aligned}
		&\mathbb{P}_{H_0^{\{\mathcal{C}_1,\mathcal{C}_2\}}}\left(p(\mathcal{W};\mathcal{C}_1,\mathcal{C}_2)\leq\alpha\big|~\mathcal{C}_1,\mathcal{C}_2\in\mathcal{C}(L(\mathcal{W}))\right)\\
		=& \mathbb{E}_{H_0^{\{\mathcal{C}_1,\mathcal{C}_2\}}}\left(\mathbbm{1}_{p(\mathcal{W};\mathcal{C}_1,\mathcal{C}_2)\leq\alpha}\big|~\mathcal{C}_1,\mathcal{C}_2\in\mathcal{C}(L(\mathcal{W}))\right)\\
		= & \mathbb{E}_{H_0^{\{\mathcal{C}_1,\mathcal{C}_2\}}}\Bigg\{\mathbb{E}\bigg[\mathbbm{1}_{p(\mathcal{W};\mathcal{C}_1,\mathcal{C}_2)\leq\alpha}\bigg|\mathcal{C}_1,\mathcal{C}_2\in\mathcal{C}(L(\mathcal{W})),\\
		& \quad\quad\quad \pi^{\perp}_{\nu(\mathcal{C}_1,\mathcal{C}_2)}\times_1L(\mathcal{W}) = \pi^{\perp}_{\nu(\mathcal{C}_1,\mathcal{C}_2)}\times_1 L(w),\\
		&
		\quad\quad\quad\text{dir}(\overline{L(\mathcal{W})}_{\mathcal{C}_1}-\overline{L(\mathcal{W})}_{\mathcal{C}_2}) = \text{dir}(\overline{L(w)}_{\mathcal{C}_1}-\overline{L(w)}_{\mathcal{C}_2})\bigg]\Bigg|~\mathcal{C}_1,\mathcal{C}_2\in\mathcal{C}(L(\mathcal{W}))\Bigg\}.
	\end{aligned}
\end{equation}
Plugging in \eqref{a2.4}, we obtain
\begin{equation}\nonumber
	\begin{aligned}
		&\mathbb{P}_{H_0^{\{\mathcal{C}_1,\mathcal{C}_2\}}}\left(p(\mathcal{W};\mathcal{C}_1,\mathcal{C}_2)\leq\alpha\big|~\mathcal{C}_1,\mathcal{C}_2\in\mathcal{C}(L(\mathcal{W}))\right)\\
		= & \mathbb{E}_{H_0^{\{\mathcal{C}_1,\mathcal{C}_2\}}}\bigg(\alpha\big|~\mathcal{C}_1,\mathcal{C}_2\in\mathcal{C}(L(\mathcal{W}))\bigg) = \alpha.
	\end{aligned}
\end{equation}

\subsection{Proof of Theorem \ref{t3}}\label{pt3}
Plug in \eqref{11}, we have
\begin{equation}\label{Eq31}
	\begin{aligned}
		&\mathbb{P}_{H_1^{\{\mathcal{C}_1,\mathcal{C}_2\}}}\left(p(\mathcal{W};\mathcal{C}_1,\mathcal{C}_2)\leq\alpha\big|~\mathcal{C}_1,\mathcal{C}_2\in\mathcal{C}(L(\mathcal{W}))\right)\\
		=& \mathbb{P}_{H_1^{\{\mathcal{C}_1,\mathcal{C}_2\}}}\Bigg(1- \mathbb{F}\left(\lVert\overline{L(\mathcal{W})}_{\mathcal{C}_1}-\overline{L(\mathcal{W})}_{\mathcal{C}_2}\rVert_F;\sqrt{\frac{1}{|\mathcal{C}_1|}+\frac{1}{|\mathcal{C}_2|}},\mathcal{S}(\mathcal{W};\mathcal{C}_1,\mathcal{C}_2)\right)\leq\alpha\\
		&\quad\quad\big|~\mathcal{C}_1,\mathcal{C}_2\in\mathcal{C}(L(\mathcal{W}))\Bigg).
	\end{aligned}
\end{equation}
We rewrite the survival function $\mathbb{F}(\cdot)$ as follows:
\begin{equation}\nonumber
	\begin{split} 
		& \mathbb{F}\left(\lVert\overline{L(\mathcal{W})}_{\mathcal{C}_1}-\overline{L(\mathcal{W})}_{\mathcal{C}_2}\rVert_F;\sqrt{\frac{1}{|\mathcal{C}_1|}+\frac{1}{|\mathcal{C}_2|}},\mathcal{S}(\mathcal{W};\mathcal{C}_1,\mathcal{C}_2)\right) \\
		= & \mathbb{P}\left(\varphi\leq\lVert\overline{L(\mathcal{W})}_{\mathcal{C}_1}-\overline{L(\mathcal{W})}_{\mathcal{C}_2}\rVert_F\big|\varphi\in\mathcal{S}(\mathcal{W};\mathcal{C}_1,\mathcal{C}_2)\right),
	\end{split}
\end{equation}
where $\varphi$ follows the distribution $\sqrt{1/|\mathcal{C}_1|+1/|\mathcal{C}_2|}\cdot\chi_{mq}$. Using the fact that $\mathbb{P}(A|B)\geq \mathbb{P}(A)-\mathbb{P}(B^c)$, we have
\begin{equation}\label{eq35}
	\begin{aligned}
		&\mathbb{P}_{H_1^{\{\mathcal{C}_1,\mathcal{C}_2\}}}\left(p(\mathcal{W};\mathcal{C}_1,\mathcal{C}_2)\leq\alpha\big|~\mathcal{C}_1,\mathcal{C}_2\in\mathcal{C}(L(\mathcal{W}))\right)\\
		\leq& \mathbb{P}_{H_1^{\{\mathcal{C}_1,\mathcal{C}_2\}}}\left(1- \mathbb{F}\left(\lVert\overline{L(\mathcal{W})}_{\mathcal{C}_1}-\overline{L(\mathcal{W})}_{\mathcal{C}_2}\rVert_F;\sqrt{\frac{1}{|\mathcal{C}_1|}+\frac{1}{|\mathcal{C}_2|}},\mathcal{S}(\mathcal{W};\mathcal{C}_1,\mathcal{C}_2)\right)\leq\alpha\right)\\
		&-\mathbb{P}_{H_1^{\{\mathcal{C}_1,\mathcal{C}_2\}}}\left(\mathcal{C}_1,\mathcal{C}_2\notin\mathcal{C}(L(\mathcal{W}))\right)\\
		=& \mathbb{P}_{H_1^{\{\mathcal{C}_1,\mathcal{C}_2\}}}\left(\mathbb{P}\left(\varphi\leq\lVert\overline{L(\mathcal{W})}_{\mathcal{C}_1}-\overline{L(\mathcal{W})}_{\mathcal{C}_2}\rVert_F\big|\varphi\in\mathcal{S}(\mathcal{W};\mathcal{C}_1,\mathcal{C}_2)\right)\geq 1-\alpha\right)\\
		&-\mathbb{P}_{H_1^{\{\mathcal{C}_1,\mathcal{C}_2\}}}\left(\mathcal{C}_1,\mathcal{C}_2\notin\mathcal{C}(L(\mathcal{W}))\right).\\
	\end{aligned}
\end{equation}
Following assumption (ii), we obtain that
\begin{equation}\label{eq34}
	\lim_{\lVert\mu_{\mathcal{C}_1}-\mu_{\mathcal{C}_2}\rVert_\infty\rightarrow\infty,n\rightarrow\infty}\mathbb{P}_{H_1^{\{\mathcal{C}_1,\mathcal{C}_2\}}}\left(\mathcal{C}_1,\mathcal{C}_2\notin\mathcal{C}(L(\mathcal{W}))\right) = 0.
\end{equation}
To derive the bound for 
\[\mathbb{P}_{H_1^{\{\mathcal{C}_1,\mathcal{C}_2\}}}\left(\mathbb{P}\left(\varphi\leq\lVert\overline{L(\mathcal{W})}_{\mathcal{C}_1}-\overline{L(\mathcal{W})}_{\mathcal{C}_2}\rVert_F\big|\varphi\in\mathcal{S}(\mathcal{W};\mathcal{C}_1,\mathcal{C}_2)\right)\geq 1-\alpha\right),\] we leverage the inequality $\mathbb{P}(A|B)\geq \mathbb{P}(A)-\mathbb{P}(B^c)$ again and obtain
\begin{equation}\label{Eq32}
	\begin{aligned}
		&\mathbb{P}_{H_1^{\{\mathcal{C}_1,\mathcal{C}_2\}}}\left(\mathbb{P}\left(\varphi\leq\lVert\overline{L(\mathcal{W})}_{\mathcal{C}_1}-\overline{L(\mathcal{W})}_{\mathcal{C}_2}\rVert_F\big|\varphi\in\mathcal{S}(\mathcal{W};\mathcal{C}_1,\mathcal{C}_2)\right)\geq 1-\alpha\right)\\
		\geq & \mathbb{P}_{H_1^{\{\mathcal{C}_1,\mathcal{C}_2\}}}\left(\mathbb{P}\left(\varphi\leq\lVert\overline{L(\mathcal{W})}_{\mathcal{C}_1}-\overline{L(\mathcal{W})}_{\mathcal{C}_2}\rVert_F\right)-\mathbb{P}(\varphi\notin\mathcal{S}(\mathcal{W};\mathcal{C}_1,\mathcal{C}_2))\geq 1-\alpha\right).
	\end{aligned}
\end{equation}
\paragraph{Asymptotic behavior of $\lVert\overline{L(\mathcal{W})}_{\mathcal{C}_1}-\overline{L(\mathcal{W})}_{\mathcal{C}_2}\rVert_F$.}
Recall that 
\[\text{vec}(H(\mathcal{W})[i,:,:])\sim\mathcal{N}(\text{vec}(\beta_{(i)}),\Lambda),\] we rewrite $\text{vec}(H(\mathcal{W})[i,:,:])$ as follows:
\[\text{vec}(H(\mathcal{W})[i,:,:])\coloneqq\text{vec}(\beta_{(i)})+e_i,\]
where $e_i\sim\mathcal{N}(0,\Lambda)$. Under the alternative hypothesis $H_1^{\{\mathcal{C}_1,\mathcal{C}_2\}}$ and assumption (i): $|\mathcal{C}_1|,|\mathcal{C}_2|\rightarrow\infty$, $|\mathcal{C}_1|/|\mathcal{C}_2|\rightarrow c\in(0,1)$, \eqref{covasy} implies that
\begin{equation}\label{Eq29}
	\hat{\Lambda}\stackrel{p}{\longrightarrow} \Lambda+\frac{c}{(c+1)^2}\text{vec}(\beta_{\mathcal{C}_1}-\beta_{\mathcal{C}_2})\cdot\text{vec}(\beta_{\mathcal{C}_1}-\beta_{\mathcal{C}_2})^{\top}.
\end{equation} 
Recall that the whitening transformation outputs $\text{vec}(L(\mathcal{W})[i,:,:])= \hat{\Lambda}^{-\frac{1}{2}}\text{vec}(H(\mathcal{W})[i,:,:])$. Plugging in \eqref{Eq29}, Slutsky's theorem implies that
\begin{equation*}
	\begin{split}
		& \text{vec}(L(\mathcal{W})[i,:,:])\stackrel{d}{\longrightarrow}\\
		& \left(\Lambda+\frac{c}{(c+1)^2}\text{vec}(\beta_{\mathcal{C}_1}-\beta_{\mathcal{C}_2})\cdot\text{vec}(\beta_{\mathcal{C}_1}-\beta_{\mathcal{C}_2})^{\top}\right)^{-1/2}(\text{vec}(\beta_{(i)})+e_i)
	\end{split}
\end{equation*}
as $n\rightarrow\infty$, which further implies that
\begin{equation*}
	\begin{split}
		& \text{vec}(\overline{L(\mathcal{W})}_{\mathcal{C}_1}-\overline{L(\mathcal{W})}_{\mathcal{C}_2}) \stackrel{d}{\longrightarrow}\\ 
	& \left(\Lambda+\frac{c}{(c+1)^2}\text{vec}(\beta_{\mathcal{C}_1}-\beta_{\mathcal{C}_2})\cdot\text{vec}(\beta_{\mathcal{C}_1}-\beta_{\mathcal{C}_2})^{\top}\right)^{-1/2}\text{vec}(\beta_{\mathcal{C}_1}-\beta_{\mathcal{C}_2})
	\end{split}
\end{equation*}
as $n\rightarrow\infty$. Therefore, as $\lVert\mu_{\mathcal{C}_1}-\mu_{\mathcal{C}_2}\rVert_\infty\rightarrow\infty$, we have $\lVert\beta_{\mathcal{C}_1}-\beta_{\mathcal{C}_2}\rVert_F\rightarrow\infty$ and the Sherman–Morrison formula $(A+uv^\top)^{-1} = A^{-1}-A^{-1}uv^\top A^{-1}/(1+v^\top A^{-1}u)$ implies that
\begin{equation}\nonumber
	\lVert\overline{L(\mathcal{W})}_{\mathcal{C}_1}-\overline{L(\mathcal{W})}_{\mathcal{C}_2}\rVert_F\stackrel{p}{\longrightarrow}(c+1)/\sqrt{c}.
\end{equation}
Since $\varphi\sim\sqrt{1/|\mathcal{C}_1|+1/|\mathcal{C}_2|}\cdot\chi_{mq}$, then for any $\epsilon>0$, there exists $M>0$ such that for any $\lVert{\mu}_{\mathcal{C}_1}-{\mu}_{\mathcal{C}_2}\rVert_\infty>M$ and $n>M$, the following inequality holds
\begin{equation}\label{Eq30}
	\mathbb{P}_{H_1^{\{\mathcal{C}_1,\mathcal{C}_2\}}}\left(\mathbb{P}\left(\varphi\leq\lVert\overline{L(\mathcal{W})}_{\mathcal{C}_1}-\overline{L(\mathcal{W})}_{\mathcal{C}_2}\rVert_F\right)\geq 1-\alpha/2\right)\geq 1-\epsilon.
\end{equation}
\paragraph{Asymptotic behaviour of $\mathbb{P}(\varphi\notin\mathcal{S}(\mathcal{W};\mathcal{C}_1,\mathcal{C}_2))$.} We rewrite $\mathbb{P}(\varphi\notin\mathcal{S}(\mathcal{W};\mathcal{C}_1,\mathcal{C}_2))$ as follows:
\[\mathbb{P}(\varphi\notin\mathcal{S}(\mathcal{W};\mathcal{C}_1,\mathcal{C}_2)) = \mathbb{P}(\sqrt{1/|\mathcal{C}_1|+1/|\mathcal{C}_2|}\cdot s\notin\mathcal{S}(\mathcal{W};\mathcal{C}_1,\mathcal{C}_2)),\]
where $s\sim\chi_{mq}$. Following assumption (ii), for any $\epsilon>0$ set $\delta = z_{\alpha/2}$ as the $\alpha/2$ quantile of distribution $\chi_{mq}$,  there exists $M>0$ such that 
\[\mathbb{P}_{H_1^{\{\mathcal{C}_1,\mathcal{C}_2\}}}\left(\sqrt{1/|\mathcal{C}_1|+1/|\mathcal{C}_2|}\cdot s\in\mathcal{S}(\mathcal{W};\mathcal{C}_1,\mathcal{C}_2)\right)\geq 1-\epsilon\]
for any $s>\delta$, $\lVert\mu_{\mathcal{C}_1}-\mu_{\mathcal{C}_2}\rVert_\infty> M$ and $n>M$. This further implies that
\begin{equation}\label{eq31}
	\mathbb{P}_{H_1^{\{\mathcal{C}_1,\mathcal{C}_2\}}}\left(\mathbb{P}(\varphi\notin\mathcal{S}(\mathcal{W};\mathcal{C}_1,\mathcal{C}_2))\leq \alpha/2\right)\geq 1-\epsilon.
\end{equation}
Plugging \eqref{Eq30} and \eqref{eq31} into \eqref{Eq32}, we obtain that 
\begin{equation}\label{eq32}
	\begin{split}
	& \lim_{\lVert\mu_{\mathcal{C}_1}-\mu_{\mathcal{C}_2}\rVert_\infty\rightarrow\infty,n\rightarrow\infty}\\
	& \mathbb{P}_{H_1^{\{\mathcal{C}_1,\mathcal{C}_2\}}}\left(\mathbb{P}\left(\varphi\leq\lVert\overline{L(\mathcal{W})}_{\mathcal{C}_1}-\overline{L(\mathcal{W})}_{\mathcal{C}_2}\rVert_F\right)-\mathbb{P}(\varphi\notin\mathcal{S}(\mathcal{W};\mathcal{C}_1,\mathcal{C}_2))\geq 1-\alpha\right) = 1.
	\end{split}
\end{equation}
Combine \eqref{eq34} and \eqref{eq32} with \eqref{eq35}, we come to the result that
\[\lim_{\lVert\mu_{\mathcal{C}_1}-\mu_{\mathcal{C}_2}\rVert_\infty\rightarrow\infty,n\rightarrow\infty}\mathbb{P}_{H_1^{\{\mathcal{C}_1,\mathcal{C}_2\}}}\left(p(\mathcal{W};\mathcal{C}_1,\mathcal{C}_2)\leq\alpha\big|~\mathcal{C}_1,\mathcal{C}_2\in\mathcal{C}(L(\mathcal{W}))\right) = 1.\]

%%%%%%%%%%
\section{Proof of Auxiliary Lemmas}\label{sec:auxiliary-lemmas}
%%%%%%%%%%

\subsection{Proof of Lemma \ref{l1}}\label{pl1}
\paragraph{Proof of (I).} Under Assumption \ref{a2}, we have
\[(W_{ij}(t_{ijk}))_{k\in[r_{ij}]}\sim\mathcal{N}(\mu_{ij}^{\Omega_i},\Sigma_{jj}^{\Omega_i}+\sigma_j^2\cdot I_{r_{ij}})\]
and
\[\text{vec}((W_{ij}(t_{ijk}))_{j\in[m],k\in[r_{ij}]})\sim \mathcal{N}(\text{vec}((\mu_{ij}^{\Omega_i})_{j\in[m]}),\Sigma_1^{(i)}+\Sigma_2^{(i)}).\]
Recall the low-dimensional embedding \eqref{6}, we have
\[\text{vec}(H(\mathcal{W})[i,:,:]) = D_i\cdot\text{vec}((W_{ij}(t_{ijk}))_{j\in[m],k\in[r_{ij}]}).\]
Therefore, the transformed data $H(\mathcal{W})[i,:,:]$ follows the distribution
\[\text{vec}(H(\mathcal{W})[i,:,:])\sim\mathcal{N}(\text{vec}(((K_{ij}+\lambda I_q)^{-1}\Phi_{ij}\mu_{ij}^{\Omega_i})_{j\in[m]}),D_i\left[\Sigma_1^{(i)}+\Sigma_2^{(i)}\right]D_i^\top).\]

\paragraph{Proof of (II).}
Notice that $D_i = \text{diag}((K_{ij}+\lambda I_q)^{-1})_{j\in[m]}\cdot\text{diag}(\Phi_{ij})_{j\in[m]}$, we have
\begin{equation}\label{e28}
	\begin{aligned}
		& ~~ \text{vec}(H(\mathcal{W})[i,:,:]) \\
		&= \text{diag}((K_{ij}+\lambda I_q)^{-1})_{j\in[m]}\cdot\text{diag}(\Phi_{ij})_{j\in[m]}\cdot\text{vec}((W_{ij}(t_{ijk}))_{j\in[m],k\in[r_{ij}]})\\
		&= \underbrace{\left[\text{diag}((K_{ij}/r_{ij}+\lambda/r_{ij}\cdot I_q)^{-1})_{j\in[m]}\right]}_{X^{\Omega_i}}\\
		& \quad  \cdot\underbrace{\left[\text{diag}(\Phi_{ij}/r_{ij})_{j\in[m]}\cdot\text{vec}((W_{ij}(t_{ijk}))_{j\in[m],k\in[r_{ij}]})\right]}_{Y^{\Omega_i}}.
	\end{aligned}
\end{equation}
\begin{itemize}
	\item Convergence of $X^{\Omega_i}$
\end{itemize}
Recall that $K_{ij} = \Phi_{ij}\Phi_{ij}^\top$, namely
\[K_{ij} = \left(\sum_{k = 1}^{r_{ij}}\phi_a(t_{ijk})\phi_b(t_{ijk})\right)_{a,b\in[q]}.\]
By the law of large numbers, we obtain that
\[K_{ij}/r_{ij} = \left(\sum_{k = 1}^{r_{ij}}\phi_a(t_{ijk})\phi_b(t_{ijk})/r_{ij}\right)_{a,b\in[q]}\stackrel{p}{\longrightarrow}\left(\int_{0}^1\phi_a(t)\phi_b(t)dt\right)_{a,b\in[q]} = K.\]
Combine the above equation with the continuous mapping theorem, we have
\begin{equation}
	(K_{ij}/r_{ij}+\lambda/r_{ij}\cdot I_q)^{-1}\stackrel{p}{\longrightarrow}K^{-1}\quad\text{as}\quad r_{ij}\rightarrow\infty.
\end{equation}
Then $X^{\Omega_i}$ converge to $\diag(K^{-1})_{j\in[m]}$ in probability if $r_{ij}\rightarrow\infty$ for all $j\in[m]$. 
\begin{itemize}
	\item Convergence of $Y^{\Omega_i}$
\end{itemize}
We leverage the following lemma:
\begin{Lemma}\label{asy}
	For any $n\in\mathbb{N}$, suppose $\Omega_n\coloneqq\{t_i\}_{i\in[n]}$ is a set of random variables, where $t_i\stackrel{\text{iid}}{\sim}\mathcal{U}[0,1]$. Suppose $\mu_n:\mathbb{R}^n\rightarrow\mathbb{R}^m$ and $\Sigma_n:\mathbb{R}^n\rightarrow\mathcal{S}^{m\times m}_+$, where $m\in\mathbb{N}$ is a fixed integer. Suppose $\{Y_n\}_{n\in\mathbb{N}^+}$ is a sequence of random vectors satisfying $Y_n\in\mathbb{R}^m$ and
	\[Y_n|\Omega_n\sim\mathcal{N}(\mu_n(\Omega_n),\Sigma_n(\Omega_n)).\]
	If there exists a fixed vector $\mu$ and a fixed matrix $\Sigma$ such that
	\[\mu_n(\Omega_n)\stackrel{p}{\longrightarrow}\mu \quad\text{and}\quad \Sigma_n(\Omega_n)\stackrel{p}{\longrightarrow}\Sigma\quad\text{as}\quad n\rightarrow\infty,\]
	then the following statement holds:
	\[Y_n\stackrel{d}{\longrightarrow}\mathcal{N}(\mu,\Sigma).\]
	\begin{proof}
		See Appendix \ref{pasy} for the complete proof.
	\end{proof}
\end{Lemma}
For a time record $\Omega_i = (\Omega_{ij})_{j\in[m]}$ (where $|\Omega_{ij}| = r_{ij}$), define $\mu_{\Omega_i}\coloneqq\text{vec}((\Phi_{ij}\mu_{ij}^{\Omega_i}/r_{ij})_{j\in[m]})$ and $\Sigma_{\Omega_i} \coloneqq \text{diag}(\Phi_{ij}/r_{ij})_{j\in[m]}\left[\Sigma_1^{(i)}+\Sigma_2^{(i)}\right]\text{diag}(\Phi_{ij}/r_{ij})_{j\in[m]}^\top$, then we have
\[Y^{\Omega_i}|\Omega_i\sim\mathcal{N}(\mu_{\Omega_i},\Sigma_{\Omega_i}).\]
Next, we are going to prove that
\begin{equation}
	\begin{aligned}
		& \displaystyle\mathrm{1)}\quad \mu_{\Omega_i}\stackrel{p}{\longrightarrow}\text{vec}((\mu_{ij}^{(0)})_{j\in[m]}),\\
		& \displaystyle\mathrm{2)}\quad\Sigma_{\Omega_i}\stackrel{p}{\longrightarrow} (K\Lambda_{ab}K)_{a,b\in[m]},
	\end{aligned}
\end{equation}
as $\min_{j\in[m]}r_{ij}\rightarrow\infty$. 
\paragraph{Proof of 1).} Since $\mu_{ij}^{\Omega_i} = (\mu_{ij}(t_{ijk}))_{k\in[r_{ij}]}$, we have $\Phi_{ij}[s,:]\mu_{ij}^{\Omega_i}/r_{ij} = \sum_{k = 1}^{r_{ij}}\mu_{ij}(t_{ijk})\phi_s(t_{ijk})/r_{ij}$ for any $s\in[q]$, then law of large number implies that
\[\Phi_{ij}\mu_{ij}^{\Omega_i}/r_{ij}\rightarrow\left(\int_0^1 \phi_s(t)\mu_{ij}(t)dt\right)_{s\in[q]}=\mu_{ij}^{(0)}\quad\text{and}\quad\mu_{\Omega_i}\rightarrow\text{vec}((\mu_{ij}^{(0)})_{j\in[m]}).\]
\paragraph{Proof of 2).} By the definition of $\Sigma_{\Omega_i}$, we have $\Sigma_{\Omega_i} = (D_{ab}^{\Omega_i})_{a,b\in[m]}$, where
\[D_{ab}^{\Omega_i} = \frac{1}{r_{ia}r_{ib}}\Phi_{ia}\left[\Sigma_{ab}^{\Omega_i}+\mathbbm{1}_{a = b}\cdot \sigma_a^2 I_{ia}\right]\Phi_{ib}^\top.\] 
For any $s_1,s_2\in[q]$, the $(s_1,s_2)$ entry of $D_{ab}^{\Omega_i}$ is
\[D_{ab}^{\Omega_i}[s_1,s_2]=\frac{1}{r_{ia}r_{ib}}\sum_{k_1 = 1}^{r_{ia}}\sum_{k_2 = 1}^{r_{ib}}\phi_{s_1}(t_{iak_1})\phi_{s_2}(t_{ibk_2})(R_{ab}(t_{iak_1},t_{ibk_2})+\mathbbm{1}_{a = b}\cdot\sigma_a^2).\]
\begin{Lemma}\label{lcon}
	Suppose $a_1,a_2,\cdots,a_m,b_1,\cdots,b_n\stackrel{\text{iid}}{\sim}\mathcal{U}[0,1]$. Define
	\begin{equation*}
		\begin{split}
			& C_{m,n} \coloneqq \frac{1}{mn}\sum_{k_1 = 1}^m\sum_{k_2 = 1}^nf(a_{k_1})g(b_{k_2})\psi(a_{k_1},b_{k_2}),\\
			& C\coloneqq\int_0^1\int_0^1f(t_1)g(t_2)\psi(t_1,t_2)dt_1dt_2,
		\end{split}
	\end{equation*}
	where $f,g:[0,1]\rightarrow\mathbb{R}$ and $\psi:[0,1]\times[0,1]\rightarrow\mathbb{R}$ are Lipschitz continuous. Then we have
	\[C_{m,n}\stackrel{p}{\longrightarrow}C\quad\text{as}\quad \min\{m,n\}\rightarrow\infty.\]
\end{Lemma}
\begin{proof}
	See Appendix \ref{pcon} for the complete proof.
\end{proof}
Lemma \ref{lcon} implies that
\[D_{ab}^{\Omega_i}[s_1,s_2]\stackrel{p}{\longrightarrow}\int_0^1\int_0^1\phi_{s_1}(t_1)\phi_{s_2}(t_2)(R_{ab}(t_1,t_2)+\mathbbm{1}_{a=b}\cdot \sigma_a^2) dt_1dt_2\]
as $\min\{r_{ia},r_{ib}\}\rightarrow\infty$. Therefore, we have 
\[\Sigma_{\Omega_i}\stackrel{p}{\longrightarrow}(K\Lambda_{ab}K)_{a,b\in[m]}\quad\text{as}\quad\min_{j\in[m]}r_{ij}\rightarrow\infty.\]

Combine 1), 2) and leverage Lemma \ref{asy}, we obtain that
\[Y^{\Omega_i}\stackrel{d}{\longrightarrow}\mathcal{N}(\mu_{\Omega_i},\Sigma_{\Omega_i}).\]
Plugging the convergence of $X^{\Omega_i}$ and $Y^{\Omega_i}$ into \eqref{e28} and leveraging Slutsky's theorem, we have
\[\text{vec}(H(\mathcal{W})[i,:,:])\stackrel{d}{\longrightarrow}\mathcal{N}(\text{vec}((K^{-1}\mu_{ij}^{(0)})_{j\in[m]}),\Lambda)\quad\text{as}\quad\min_{j\in[m]}r_{ij}\rightarrow\infty.\]

\subsection{Proof of Lemma \ref{l2}}\label{pl2}
To begin with, we have
\[\mathcal{A} = \pi^{\perp}_{\nu(\mathcal{C}_1,\mathcal{C}_2)}\times_1 \mathcal{A} + (I-\pi^{\perp}_{\nu(\mathcal{C}_1,\mathcal{C}_2)})\times_1 \mathcal{A}.\]
By the definition of $\pi^{\perp}_{\nu(\mathcal{C}_1,\mathcal{C}_2)}$, we have $I-\pi^{\perp}_{\nu(\mathcal{C}_1,\mathcal{C}_2)} = \nu(\mathcal{C}_1,\mathcal{C}_2)\nu(\mathcal{C}_1,\mathcal{C}_2)^{\top}/\lVert\nu(\mathcal{C}_1,\mathcal{C}_2)\rVert^2$ and $\lVert\nu(\mathcal{C}_1,\mathcal{C}_2)\rVert^2 = 1/|\mathcal{C}_1|+ 1/|\mathcal{C}_2|$. As a result, we can rewrite the second term in the above equation as follows:
\begin{equation}\label{eq13}
	\begin{aligned}
		(I-\pi^{\perp}_{\nu(\mathcal{C}_1,\mathcal{C}_2)})\times_1 \mathcal{A} &= \frac{\nu(\mathcal{C}_1,\mathcal{C}_2)\nu(\mathcal{C}_1,\mathcal{C}_2)^{\top}}{1/|\mathcal{C}_1|+ 1/|\mathcal{C}_2|}\times_1\mathcal{A}\\
		&= \frac{\nu(\mathcal{C}_1,\mathcal{C}_2)}{1/|\mathcal{C}_1|+ 1/|\mathcal{C}_2|}\times_1(\overline{\mathcal{A}}_{\mathcal{C}_1}-\overline{\mathcal{A}}_{\mathcal{C}_2})^{\top},
	\end{aligned}
\end{equation}
where the last equation holds by the property of tensor mode product. The equation \eqref{eq13} further leads to \eqref{eq12} and finishes the proof.

\subsection{Proof of Lemma \ref{l3}}\label{pl3}
Combine \eqref{12} with the orthogonal decomposition \eqref{eq12}, we have
\begin{equation}\nonumber
	\begin{aligned}
		&p_{selective}= \mathbb{P}_{H_0^{\{\mathcal{C}_1, \mathcal{C}_2\}}}\bigg\{\lVert \overline{L(\mathcal{W})}_{\mathcal{C}_1}-\overline{L(\mathcal{W})}_{\mathcal{C}_2} \rVert_F\geq \lVert \overline{L(w)}_{\mathcal{C}_1}-\overline{L(w)}_{\mathcal{C}_2} \rVert_F\bigg|\mathcal{C}_1,\mathcal{C}_2\in\\
		&\quad \mathcal{C}\bigg(\pi^{\perp}_{\nu(\mathcal{C}_1,\mathcal{C}_2)}\times_1 L(\mathcal{W})\\
		& \quad +\left[\frac{\lVert \overline{L(\mathcal{W})}_{\mathcal{C}_1}-\overline{L(\mathcal{W})}_{\mathcal{C}_2} \rVert_F}{1/|\mathcal{C}_1|+1/|\mathcal{C}_2|}\right]
		\nu(\mathcal{C}_1,\mathcal{C}_2)\times_1\text{dir}(\overline{L(\mathcal{W})}_{\mathcal{C}_1}-\overline{L(\mathcal{W})}_{\mathcal{C}_2})^{\top}\bigg),\\
		&\quad \pi^{\perp}_{\nu(\mathcal{C}_1,\mathcal{C}_2)}\times_1 L(\mathcal{W}) = \pi^{\perp}_{\nu(\mathcal{C}_1,\mathcal{C}_2)}\times_1 L(w),\\
		&\quad \text{dir}(\overline{L(\mathcal{W})}_{\mathcal{C}_1}-\overline{L(\mathcal{W})}_{\mathcal{C}_2}) = \text{dir}(\overline{L(w)}_{\mathcal{C}_1}-\overline{L(w)}_{\mathcal{C}_2})\bigg\}\\
		&=\mathbb{P}_{H_0^{\{\mathcal{C}_1, \mathcal{C}_2\}}}\Bigg\{\lVert \overline{L(\mathcal{W})}_{\mathcal{C}_1}-\overline{L(\mathcal{W})}_{\mathcal{C}_2} \rVert_F\geq \lVert \overline{L(w)}_{\mathcal{C}_1}-\overline{L(w)}_{\mathcal{C}_2} \rVert_F\bigg|\mathcal{C}_1,\mathcal{C}_2\in\\
		&\quad \mathcal{C}\bigg(\pi^{\perp}_{\nu(\mathcal{C}_1,\mathcal{C}_2)}\times_1 L(w)\\
		& \quad +\left[\frac{\lVert \overline{L(\mathcal{W})}_{\mathcal{C}_1}-\overline{L(\mathcal{W})}_{\mathcal{C}_2} \rVert_F}{1/|\mathcal{C}_1|+1/|\mathcal{C}_2|}\right]
		\nu(\mathcal{C}_1,\mathcal{C}_2)\times_1\text{dir}(\overline{L(w)}_{\mathcal{C}_1}-\overline{L(w)}_{\mathcal{C}_2})^{\top}\bigg),\\
		&\quad \pi^{\perp}_{\nu(\mathcal{C}_1,\mathcal{C}_2)}\times_1L(\mathcal{W}) = \pi^{\perp}_{\nu(\mathcal{C}_1,\mathcal{C}_2)}\times_1 L(w),\\
		& \quad \text{dir}(\overline{L(\mathcal{W})}_{\mathcal{C}_1}-\overline{L(\mathcal{W})}_{\mathcal{C}_2}) = \text{dir}(\overline{L(w)}_{\mathcal{C}_1}-\overline{L(w)}_{\mathcal{C}_2})\Bigg\}.
	\end{aligned}
\end{equation}
Next, we are going to show that
\begin{equation}\nonumber
	\begin{aligned}
		&\text{1)}\quad\lVert \overline{L(\mathcal{W})}_{\mathcal{C}_1}-\overline{L(\mathcal{W})}_{\mathcal{C}_2} \rVert_F\perp\pi^{\perp}_{\nu(\mathcal{C}_1,\mathcal{C}_2)}\times_1 L(\mathcal{W}),\\
		&\text{2)}\quad\lVert \overline{L(\mathcal{W})}_{\mathcal{C}_1}-\overline{L(\mathcal{W})}_{\mathcal{C}_2} \rVert_F\perp\text{dir}(\overline{L(\mathcal{W})}_{\mathcal{C}_1}-\overline{L(\mathcal{W})}_{\mathcal{C}_2}).
	\end{aligned}
\end{equation}
\paragraph{Proof of 1).}
By the property of vectorization operator, we have $\text{vec}(\pi^{\perp}_{\nu(\mathcal{C}_1,\mathcal{C}_2)}\times_1L(\mathcal{W})) = (I_{mq}\otimes\pi^{\perp}_{\nu(\mathcal{C}_1,\mathcal{C}_2)})\text{vec}(\mathcal{M}_1(L(\mathcal{W})))$, where $(I_{mq}\otimes\pi^{\perp}_{\nu(\mathcal{C}_1,\mathcal{C}_2)})$ is the orthogonal projection matrix that projects $\text{vec}(\mathcal{M}_1(L(\mathcal{W})))$ onto a subspace orthogonal to $I_{mq}\otimes\nu(\mathcal{C}_1,\mathcal{C}_2)$. By the property of multivariate normal distribution, the projections of a multivariate normal vector onto two orthogonal subspaces are independent. Therefore, $(I_{mq}\otimes\pi^{\perp}_{\nu(\mathcal{C}_1,\mathcal{C}_2)})\text{vec}(\mathcal{M}_1(L(\mathcal{W})))$ is independent to $(I_{mq}\otimes\nu(\mathcal{C}_1,\mathcal{C}_2))\text{vec}(\mathcal{M}_1(L(\mathcal{W})))$, i.e., $\text{vec}(\pi^{\perp}_{\nu(\mathcal{C}_1,\mathcal{C}_2)}\times_1L(\mathcal{W}))$ is independent to $\text{vec}(\nu(\mathcal{C}_1,\mathcal{C}_2)^{\top}\times_1L(\mathcal{W})) = \text{vec}(\overline{L(\mathcal{W})}_{\mathcal{C}_1}-\overline{L(\mathcal{W})}_{\mathcal{C}_2})$.

\paragraph{Proof of 2).} 
\paragraph{Proof of 2).}
Since $
\mathrm{vec}\left(\overline{L(\mathcal{W})}_{\mathcal{C}_1}
-\overline{L(\mathcal{W})}_{\mathcal{C}_2}\right)
\sim
\mathcal{N}\left(0,
\left(\frac{1}{|\mathcal{C}_1|}+\frac{1}{|\mathcal{C}_2|}\right)I_{mq}\right)$,
its Euclidean norm and direction are independent by spherical symmetry of the isotropic multivariate normal distribution. Because the Frobenius norm of a matrix equals the Euclidean norm of its vectorization, it follows that
$
\left\|
\overline{L(\mathcal{W})}{\mathcal{C}_1}
-\overline{L(\mathcal{W})}{\mathcal{C}_2}
\right\|_F
$
is independent of
$
\operatorname{dir}\left(
\overline{L(\mathcal{W})}{\mathcal{C}_1}
-\overline{L(\mathcal{W})}{\mathcal{C}_2}
\right).$

%Since $\text{vec}(\overline{L(\mathcal{W})}_{\mathcal{C}_1}-\overline{L(\mathcal{W})}_{\mathcal{C}_2})$ follows the scaled standard normal distribution $\mathcal{N}(0,(1/|\mathcal{C}_1|+1/|\mathcal{C}_2|)I_{mq})$, then $\lVert \overline{L(\mathcal{W})}_{\mathcal{C}_1}-\overline{L(\mathcal{W})}_{\mathcal{C}_2} \rVert_F$ is independent to $\text{vec}(\overline{L(\mathcal{W})}_{\mathcal{C}_1}-\overline{L(\mathcal{W})}_{\mathcal{C}_2})$. Because the length and direction of a multivariate normal distribution are independent (if the covariance matrix is the scaled identity matrix).

We have shown that $\lVert \overline{L(\mathcal{W})}_{\mathcal{C}_1}-\overline{L(\mathcal{W})}_{\mathcal{C}_2} \rVert_F$ is independent of $\pi^{\perp}_{\nu(\mathcal{C}_1,\mathcal{C}_2)}\times_1L(\mathcal{W})$ and $\text{dir}(\overline{L(\mathcal{W})}_{\mathcal{C}_1}-\overline{L(\mathcal{W})}_{\mathcal{C}_2})$. Therefore, we can drop these conditions and rewrite the selective $p$-value as follows
\begin{equation}\nonumber
	\begin{aligned}
		&p_{selective} = \mathbb{P}_{H_0^{\{\mathcal{C}_1, \mathcal{C}_2\}}}\bigg(\lVert \overline{L(\mathcal{W})}_{\mathcal{C}_1}-\overline{L(\mathcal{W})}_{\mathcal{C}_2} \rVert_F\geq \lVert \overline{L(w)}_{\mathcal{C}_1}-\overline{L(w)}_{\mathcal{C}_2} \rVert_F\bigg|\\ &\quad \mathcal{C}_1,\mathcal{C}_2\in\mathcal{C}\bigg(\pi^{\perp}_{\nu(\mathcal{C}_1,\mathcal{C}_2)}\times_1 L(w)\\
		& \qquad +\left[\frac{\lVert \overline{L(\mathcal{W})}_{\mathcal{C}_1}-\overline{L(\mathcal{W})}_{\mathcal{C}_2} \rVert_F}{1/|\mathcal{C}_1|+1/|\mathcal{C}_2|}\right]
		\nu(\mathcal{C}_1,\mathcal{C}_2)\times_1\text{dir}(\overline{L(w)}_{\mathcal{C}_1}-\overline{L(w)}_{\mathcal{C}_2})^{\top}\bigg)\bigg).
	\end{aligned}
\end{equation}
Define $\varphi = \lVert \overline{L(\mathcal{W})}_{\mathcal{C}_1}-\overline{L(\mathcal{W})}_{\mathcal{C}_2} \rVert_F$ and $\mathcal{S}(w,\mathcal{C}_1,\mathcal{C}_2) = \{\varphi\geq0:\mathcal{C}_1,\mathcal{C}_2\in\mathcal{C}(\pi^{\perp}_{\nu(\mathcal{C}_1,\mathcal{C}_2)}\times_1 L(w)+\left(\varphi/(1/|\mathcal{C}_1|+1/|\mathcal{C}_2|)\right)
\nu(\mathcal{C}_1,\mathcal{C}_2)\times_1\text{dir}(\overline{L(w)}_{\mathcal{C}_1}-\overline{L(w)}_{\mathcal{C}_2})^{\top})\}$, then the selective $p$-value has the form \[p_{selective} = \mathbb{P}_{H_0^{\{\mathcal{C}_1, \mathcal{C}_2\}}}(\varphi\geq \lVert \overline{L(w)}_{\mathcal{C}_1}-\overline{L(w)}_{\mathcal{C}_2} \rVert_F|\mathcal{C}_1,\mathcal{C}_2\in\mathcal{S}(w,\mathcal{C}_1,\mathcal{C}_2)).\]
Since $\text{vec}(\overline{L(\mathcal{W})}_{\mathcal{C}_1}-\overline{L(\mathcal{W})}_{\mathcal{C}_2})\sim \sqrt{1/|\mathcal{C}_1|+1/|\mathcal{C}_2|}\cdot\mathcal{N}(0,I_{mq})$, the random variable $\varphi$ follows the $\sqrt{1/|\mathcal{C}_1|+1/|\mathcal{C}_2|}\cdot\chi_{mq}$ distribution and further finishes the proof.

\subsection{Proof for Lemma \ref{asy}}\label{pasy}
Define a sequence of random vectors $\{\tilde{Y}_n\}_{n\in\mathbb{N}^+}$ as follows
\[\tilde{Y}_n\coloneqq \mu_n(\Omega_n)+\Sigma_n(\Omega_n)^{1/2}\cdot Z\quad\text{where}\quad Z\sim\mathcal{N}(0,I_m)\perp\Omega_n,\]
then $\tilde{Y}_n$ and $Y_n$ follow the same distribution. By Slutsky's theorem, we have
\[\tilde{Y}_n = \mu_n(\Omega_n)+\Sigma_n(\Omega_n)^{1/2}\cdot Z\stackrel{p}{\longrightarrow} \mu+\Sigma^{1/2}\cdot Z,\]
which further implies that
\[Y_n\stackrel{d}{\longrightarrow}\mathcal{N}(\mu,\Sigma).\]

\subsection{Proof for Lemma \ref{lcon}}\label{pcon}
We decompose $C_{m,n}-C$ as follows:
\begin{equation}\nonumber
	\begin{aligned}
		& ~~ C_{m,n}-C \\
		&= \frac{1}{mn}\sum_{k_1 = 1}^m\sum_{k_2 = 1}^nf(a_{k_1})g(b_{k_2})\psi(a_{k_1},b_{k_2})-\int_0^1\int_0^1f(t_1)g(t_2)\psi(t_1,t_2)dt_1dt_2\\
		&= \underbrace{\frac{1}{m}\sum_{k_1 = 1}^mf(a_{k_1})\left[\frac{1}{n}\sum_{k_2 = 1}^ng(b_{k_2})\psi(a_{k_1},b_{k_2})-\int_0^1g(t)\psi(a_{k_1},t)dt\right]}_{\displaystyle\mathrm{A}}\\
		&+ \underbrace{\int_0^1 g(t_2)\left[\frac{1}{m}\sum_{k_1 = 1}^mf(a_{k_1})\psi(a_{k_1},t_2)-\int_0^1f(t_1)\psi(t_1,t_2)dt_1\right]dt_2}_{\displaystyle\mathrm{B}}.
	\end{aligned}
\end{equation}
Therefore, for any $\epsilon>0$, we have
\begin{equation}\label{E30}
	\begin{aligned}
		\mathbb{P}(|C_{m,n}-C|\geq\epsilon)\leq\mathbb{P}(\displaystyle\mathrm{|A|}\geq\epsilon/2)+\mathbb{P}(\displaystyle\mathrm{|B|}\geq\epsilon/2).
	\end{aligned}
\end{equation}
Next we set $\epsilon<2\min\{\sup_{(a,b)\in[0,1]^2}|g(b)\psi(a,b)|,\sup_{(a,b)\in[0,1]^2}|f(a)\psi(a,b)|\}/3$ and leverage the following lemmas:
\begin{Lemma}[\cite{yukich1985laws}]\label{lcov}
	If P is a probability measure and f is a function, denote $Pf \coloneqq \int f(z)dP(z)$. Given $Z_1,\cdots,Z_n\stackrel{iid}{\sim}P$, let $P_n$ be the empirical measure and denote $P_nf \coloneqq \sum_{i = 1}^n f(Z_i)/n$. Given a function class $\mathcal{F}$, let $A = \sup_f \int |f|dP$ and $B = \sup_f\lVert f\rVert_\infty$. For any $\epsilon<2A/3$,
	\[\mathbb{P}\left(\sup_{f\in\mathcal{F}}|P_n(f)-P(f)|>\epsilon\right)\leq 4N_{[~]}(\epsilon/8,\mathcal{F},L_1(P))e^{-\frac{96n\epsilon^2}{76AB}}.\]
\end{Lemma}
\begin{Lemma}[\cite{van2000asymptotic}]\label{lbr}
	Let $\mathcal{F} = \{f_\theta:\theta\in\Theta\}$ where $\Theta$ is a bounded subset of $\mathbb{R}^d$. Suppose there exists a function $m$ such that for every $\theta_1,\theta_2$,
	\[|f_{\theta_1}(x)-f_{\theta_2}(x)|\leq m(x)\lVert\theta_1-\theta_2\rVert.\]
	Then,
	\[N_{[~]}(\epsilon,\mathcal{F},L_q(P))\leq\left(\frac{4\sqrt{d}\cdot\text{diam}(\Theta)\int|m(x)|^qdP(x)}{\epsilon}\right)^d.\]
\end{Lemma}
\paragraph{Bounding $\mathbb{P}(\displaystyle\mathrm{|A|}\geq\epsilon/2)$.}
Since $g$ and $\psi$ are Lipschitz continuous, the conditions in Lemma \ref{lcov} and Lemma \ref{lbr} are satisfied. Therefore, there exists a constant $U>0$ such that
\begin{equation}\label{e33}
	\begin{aligned}
		\mathbb{P}\left(\sup_{a\in[0,1]}\left|\frac{1}{n}\sum_{k = 1}^ng(b_k)\psi(a,b_k)-\int_0^1g(t)\psi(a,t)dt\right|\geq\epsilon\right)\lesssim e^{-U\cdot n\epsilon^2}/\epsilon.
	\end{aligned}
\end{equation}

\paragraph{Bounding $\mathbb{P}(\displaystyle\mathrm{|B|}\geq\epsilon/2)$.} Following the same proof steps, there exists a constant $V>0$ such that
\begin{equation}\label{e34}
	\begin{aligned}
		\mathbb{P}\left(\sup_{b\in[0,1]}\left|\frac{1}{m}\sum_{k = 1}^mf(a_k)\psi(a_k,b)-\int_0^1f(t)\psi(t,b)dt\right|\geq\epsilon\right)\lesssim e^{-V\cdot m\epsilon^2}/\epsilon.
	\end{aligned}
\end{equation}
Plugging \eqref{e33} and \eqref{e34} into \eqref{E30}, we obtain that
\[\mathbb{P}(|C_{m,n}-C|\geq\epsilon)\lesssim \frac{e^{-U\cdot n\epsilon^2}+e^{-V\cdot m\epsilon^2}}{\epsilon}.\]
Therefore, for any $\epsilon,\delta>0$ ($\epsilon$ is sufficiently small), there exists $M,N$ such that for any $m>M$ and $n>N$, 
\[\mathbb{P}(|C_{m,n}-C|\geq\epsilon)\leq \delta.\]

\section{Supplementary Figures}\label{ap1}

This section presents the auxiliary figures for numerical simulation and EHR-dataset application in Section \ref{s5} and Section \ref{sec:AKI}.
\begin{figure}[p]
	\centering
	\textbf{Q-Q plot of the selective $p$-value under null hypothesis}\par\medskip\par\medskip
	\subfigure[]{
		\begin{minipage}[h]{0.46\textwidth}
			\centering
			\includegraphics[width=0.9\textwidth]{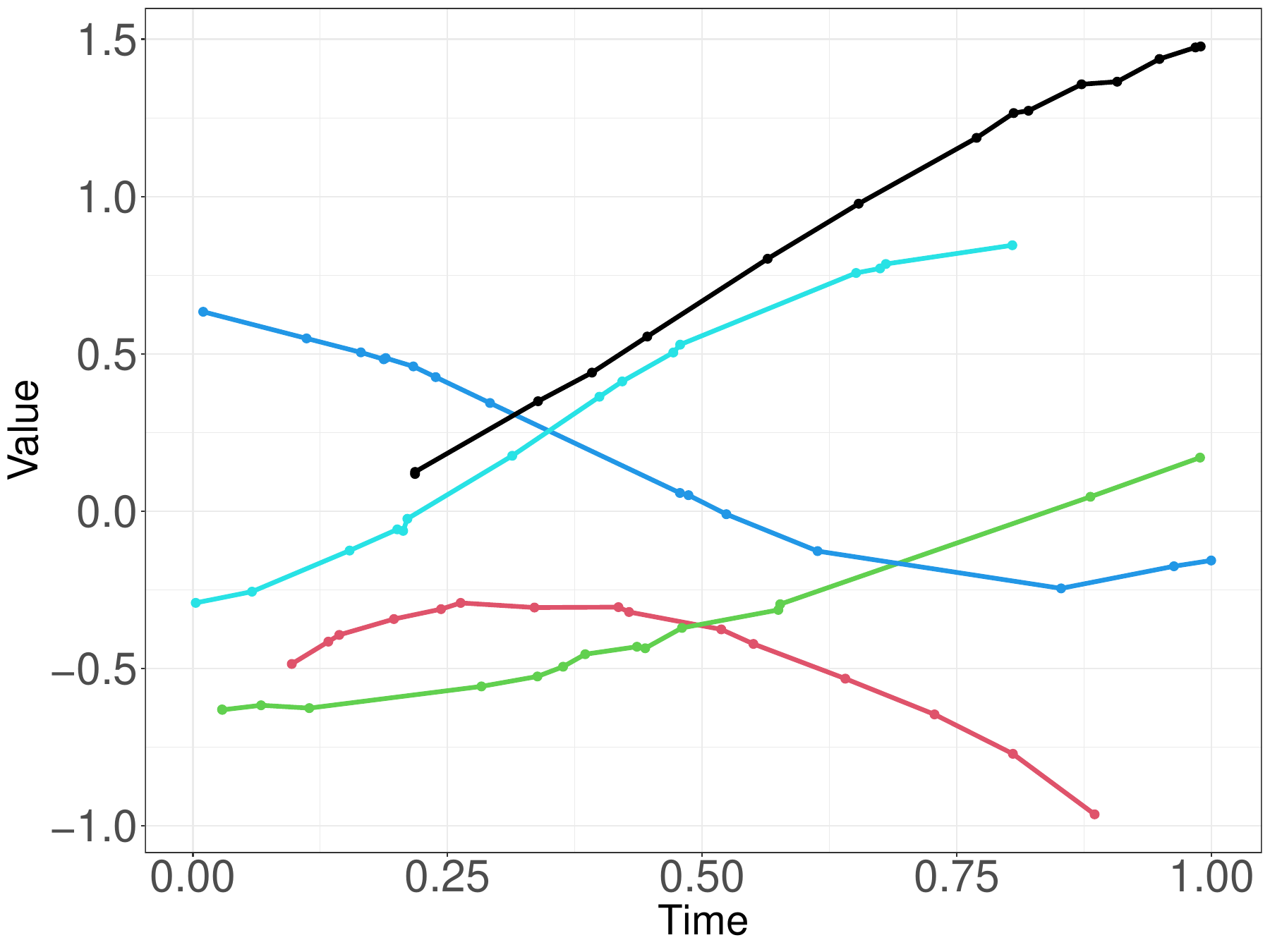}
		\end{minipage}
	}
	\subfigure[]{
		\begin{minipage}[h]{0.46\textwidth}
			\centering
			\includegraphics[width=0.9\textwidth]{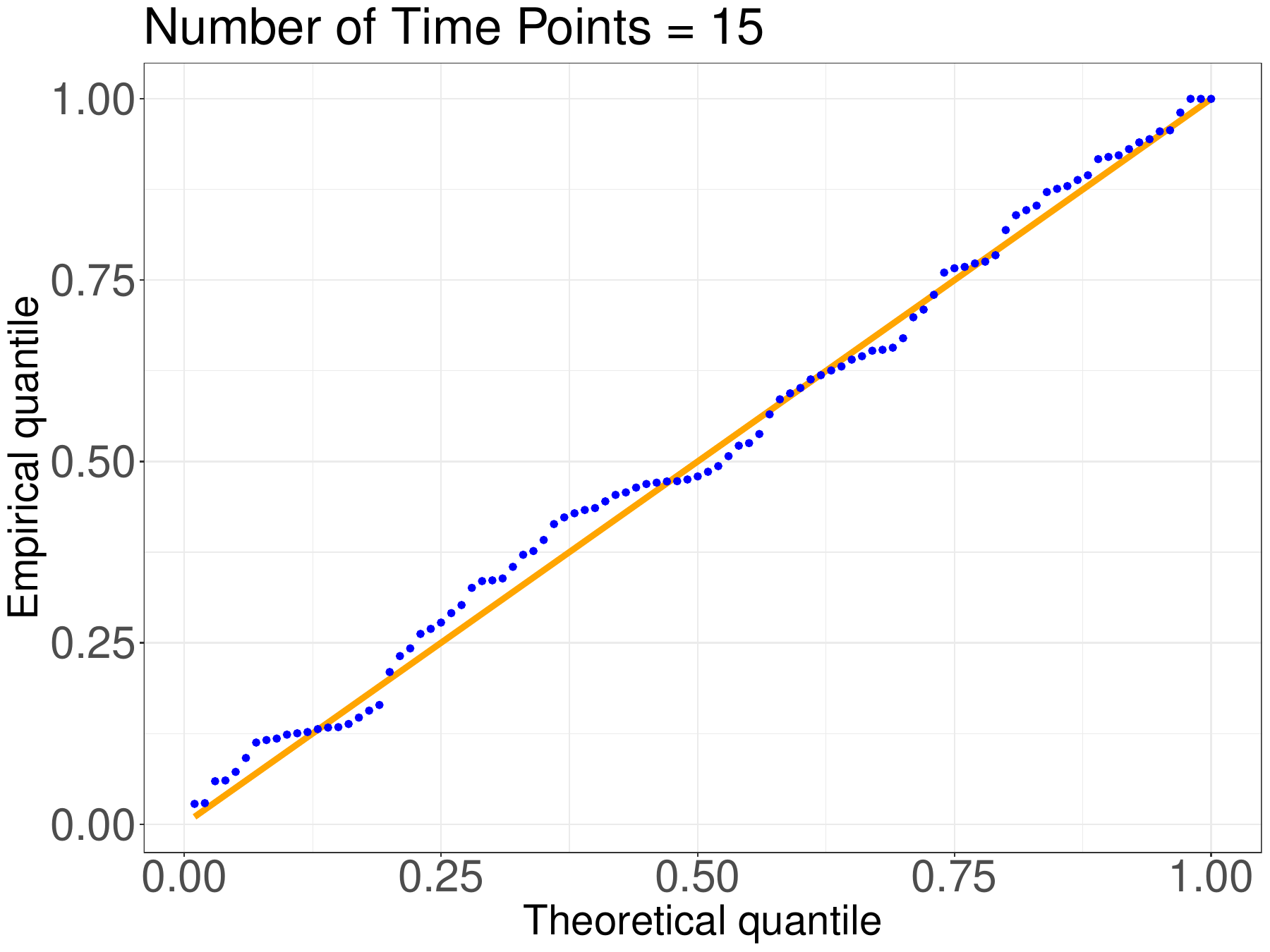} 
		\end{minipage}
	}
	\subfigure[]{
		\begin{minipage}[h]{0.46\textwidth}
			\centering
			\includegraphics[width=0.9\textwidth]{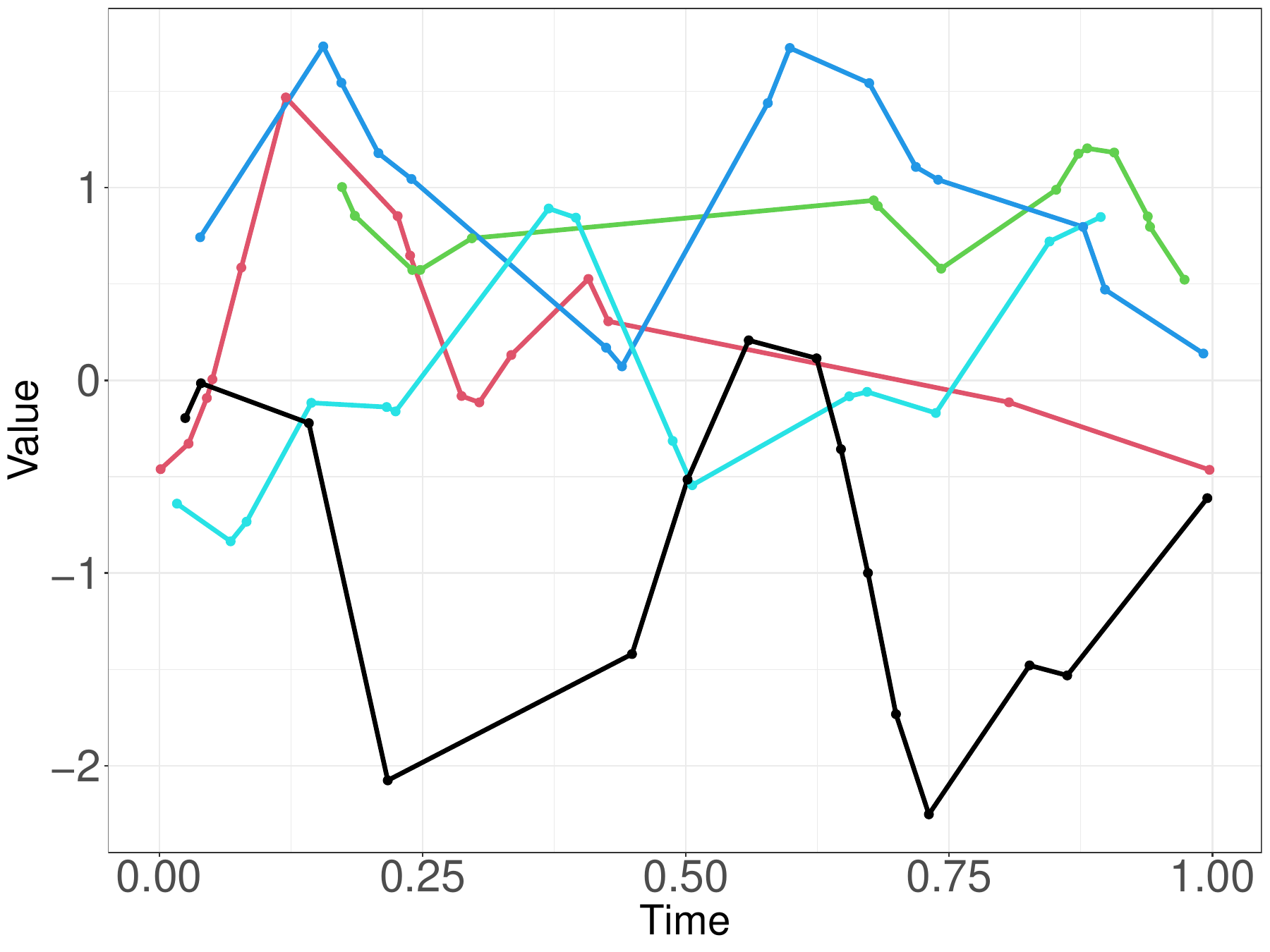} 
		\end{minipage}
	}
	\subfigure[]{
		\begin{minipage}[h]{0.46\textwidth}
			\centering
			\includegraphics[width=0.9\textwidth]{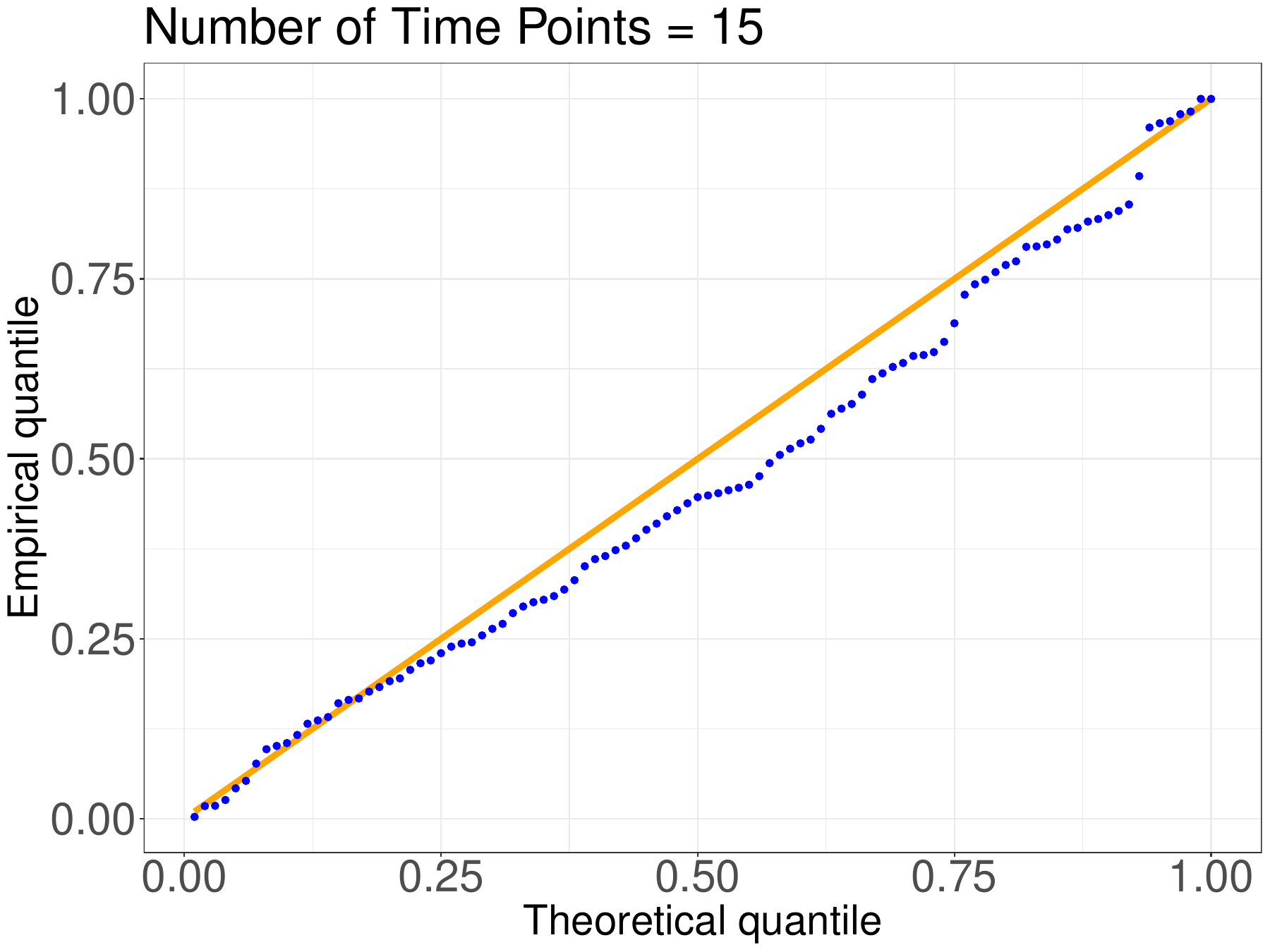} 
		\end{minipage}
	}
	\subfigure[]{
		\begin{minipage}[h]{0.46\textwidth}
			\centering
			\includegraphics[width=0.9\textwidth]{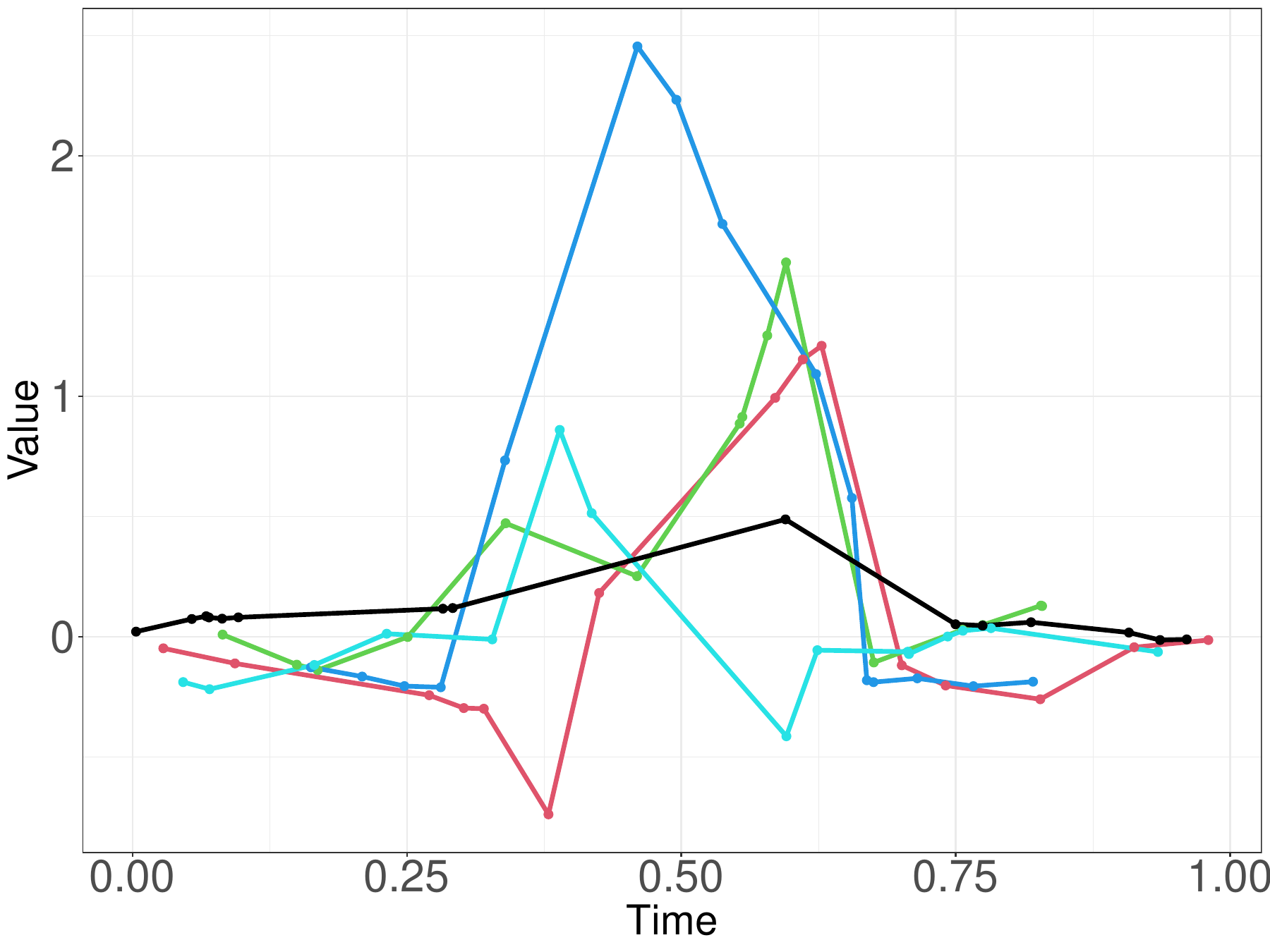} 
		\end{minipage}
	}
	\subfigure[]{
		\begin{minipage}[h]{0.46\textwidth}
			\centering
			\includegraphics[width=0.9\textwidth]{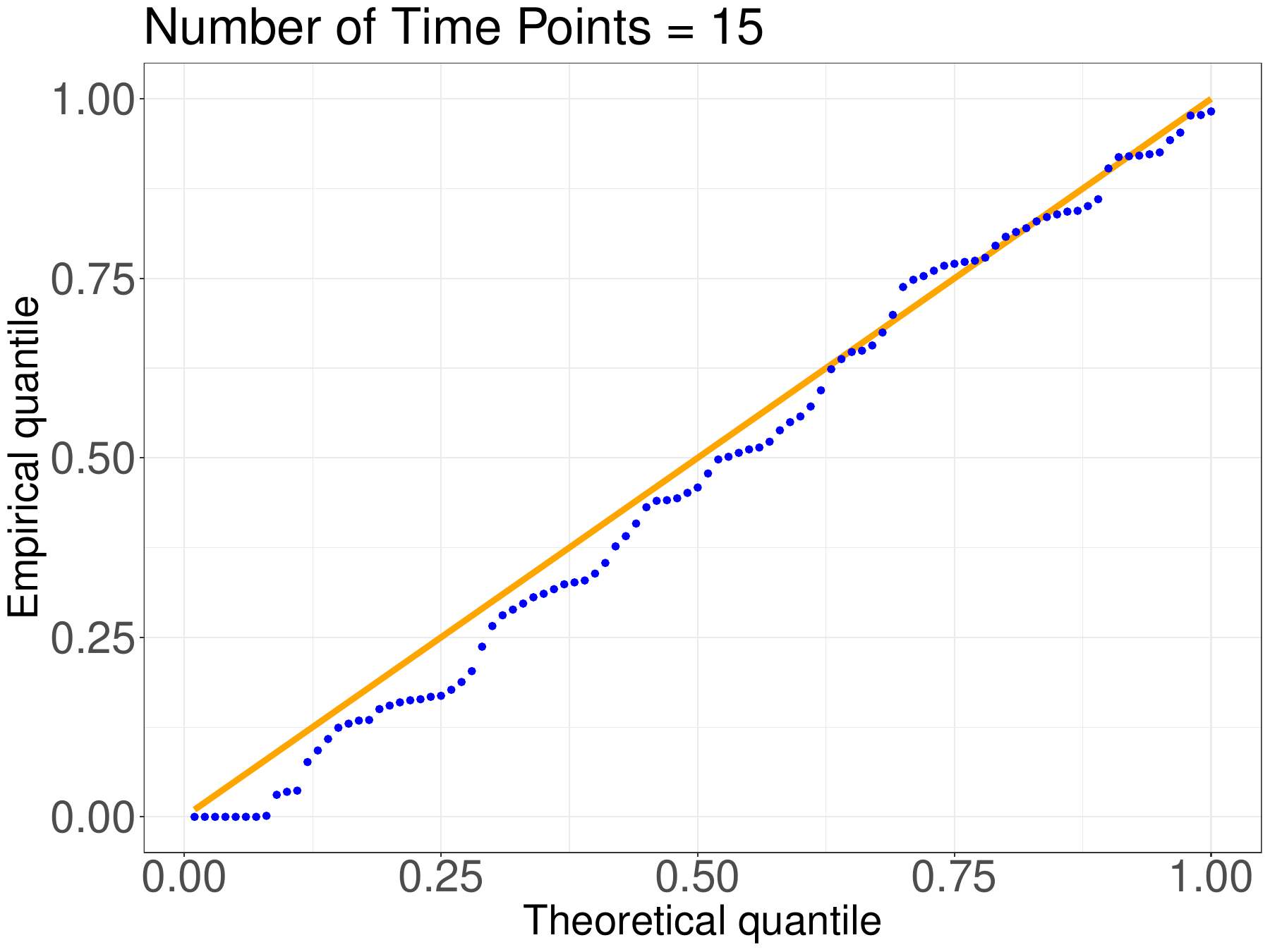} 
		\end{minipage}
	}
	\caption{\textbf{Left column}: Records of the first feature and the first 5 subjects for the dataset generated with $15$ time points. \textbf{Right column}: Quantile plots of the selective $p$-value for the corresponding kernel with $100$ generated datasets, where \textbf{(b)} is the result of \textbf{RQ Kernel}, \textbf{(d)} is the result of \textbf{PE Kernel}, and \textbf{(f)} is the result of \textbf{LPE Kernel}.}\label{f4}
\end{figure}

\begin{figure}[htbp]
	\centering
	\textbf{Statistical Power}\par\medskip\par\medskip
	\subfigure[]{
		\begin{minipage}[h]{0.6\textwidth}
			\centering
			\includegraphics[width=0.9\textwidth]{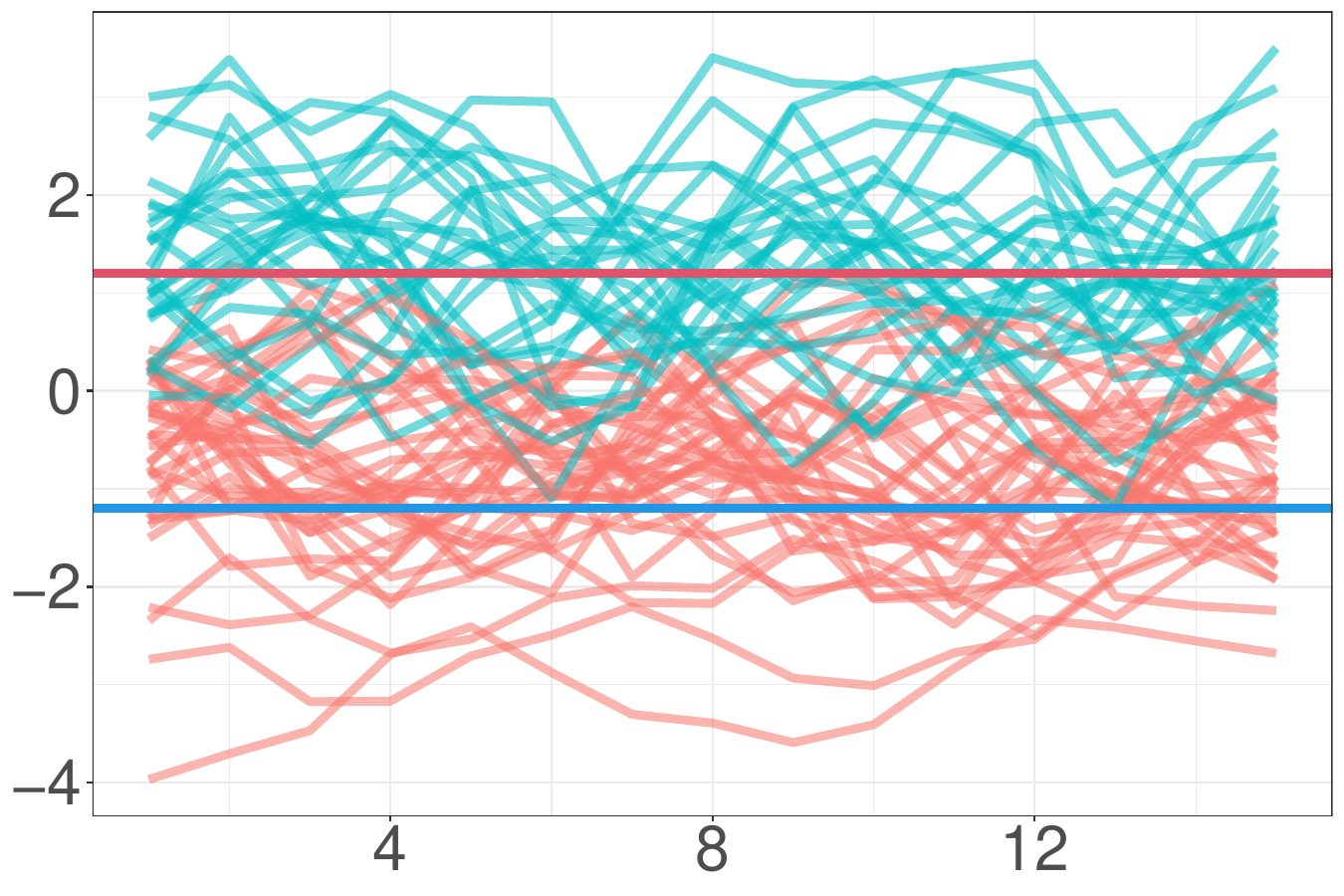}
		\end{minipage}
	}
	\quad
	\subfigure[]{
		\begin{minipage}[h]{0.46\textwidth}
			\centering
			\includegraphics[width=0.9\textwidth]{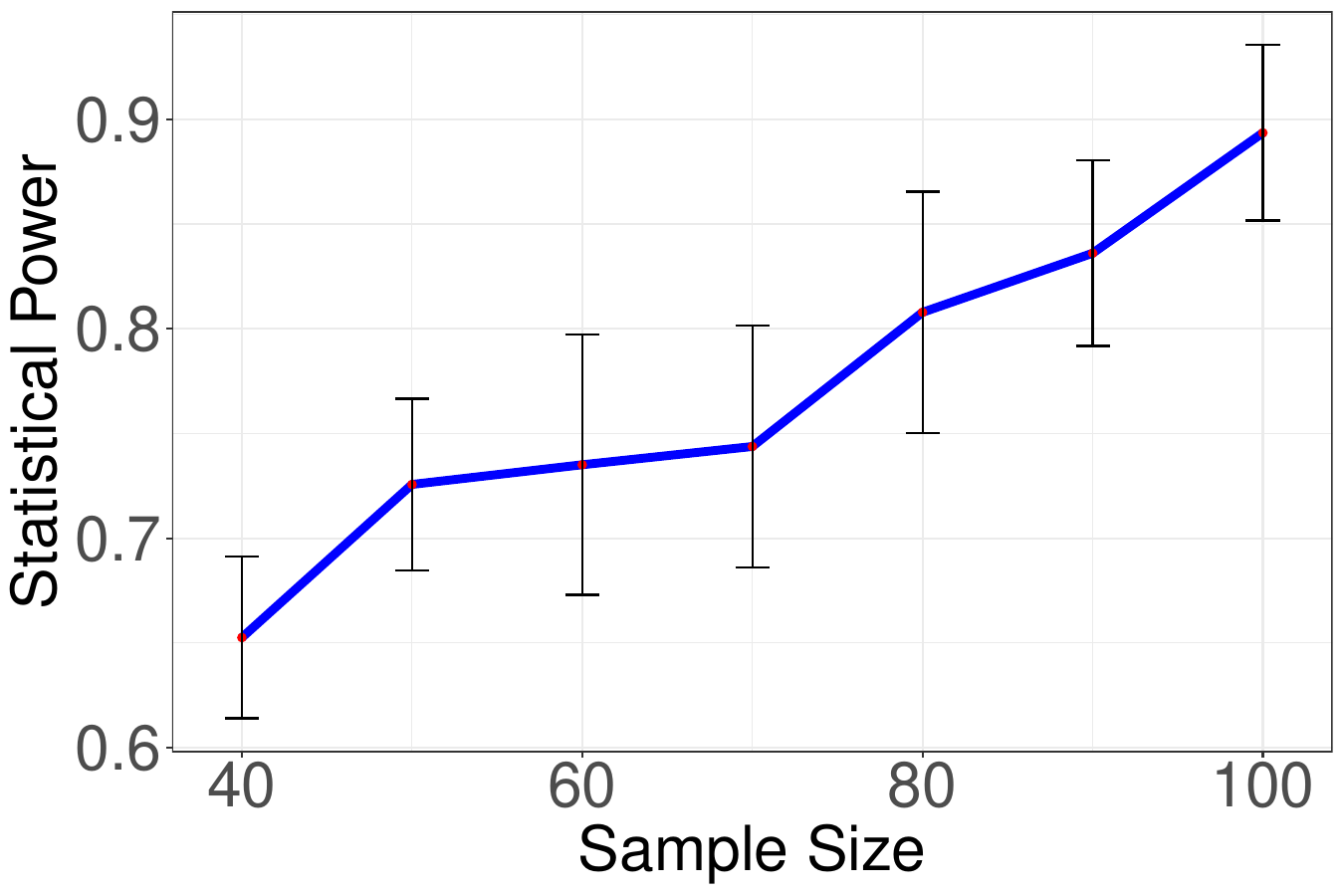} 
			\label{f5b}
		\end{minipage}
	}
	\subfigure[]{
		\begin{minipage}[h]{0.46\textwidth}
			\centering
			\includegraphics[width=0.9\textwidth]{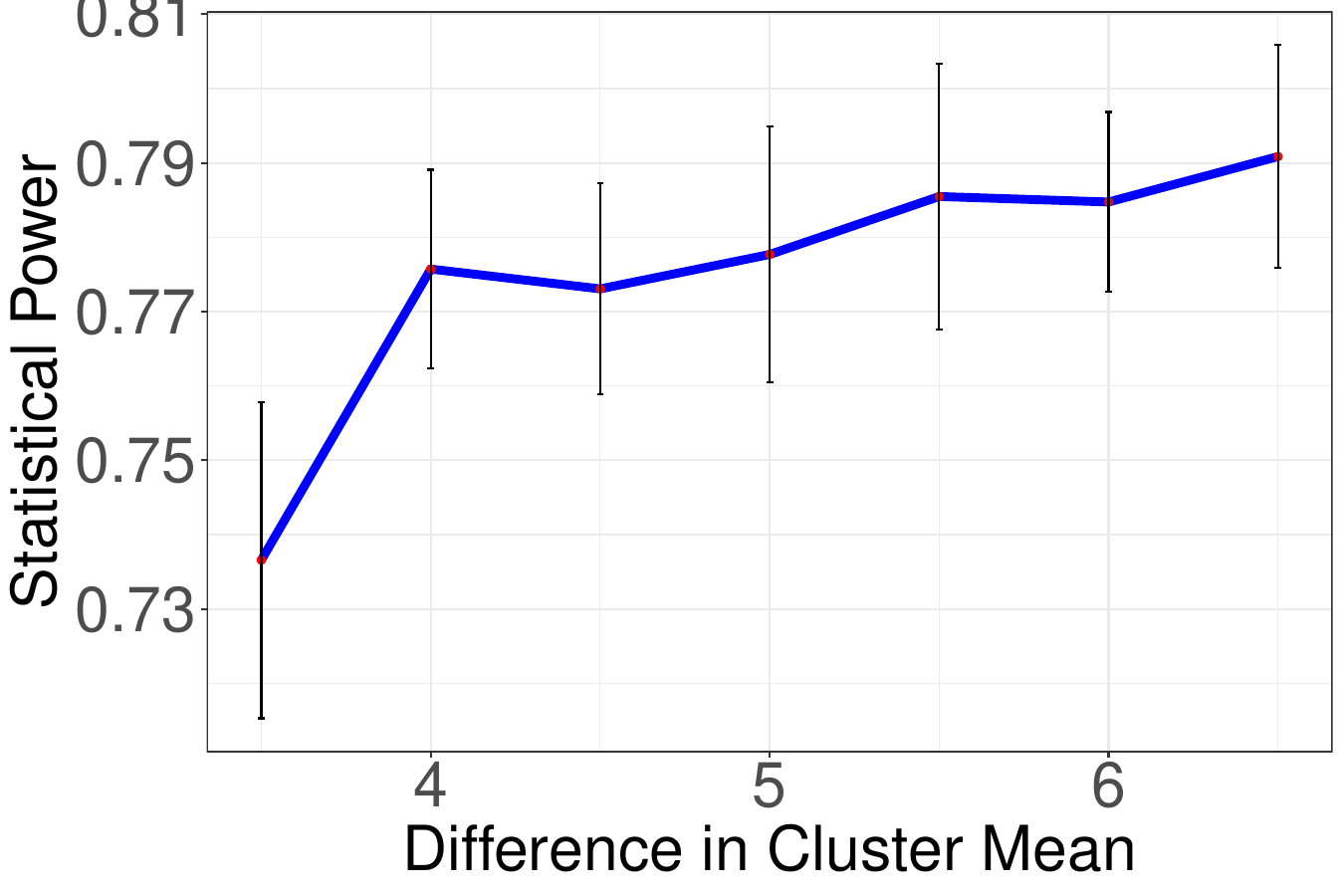} 
			\label{f5c}
		\end{minipage}
	}
	\caption{\textbf{(a)}: Example of the dataset generated under $H_1^{\{\mathcal{C}_1,\mathcal{C}_2\}}$ with $15$ time points and sample size $m = 80$, where population means are $\mu_i(\cdot) = 1.1$ for $i\leq 50$ and $\mu_i(\cdot) = -1.1$ for $i> 50$. \textbf{(b)}: Statistical power with sample size $m\in\{40,50,\cdots,100\}$, where population means are $\mu_i(\cdot) = 10$ for $i\leq m/2$ and $\mu_i(\cdot) = -10$ for $i> m/2$.  \textbf{(c)}: Statistical power with sample size $m=80$ and population means $\mu_i(\cdot) = k$ for $i\leq m/2$ and $\mu_i(\cdot) = -k$ for $i> m/2$, where $k\in\{3.5,4,\cdots,6.5\}$.}\label{f5}
\end{figure}

\begin{figure}[p]
	\centering
	\textbf{Q-Q plot of the selective $p$-value under global null (misspecification cases)}\par\medskip
	\subfigure[]{
		\begin{minipage}[h]{0.46\textwidth}
			\centering
			\includegraphics[width=0.9\textwidth]{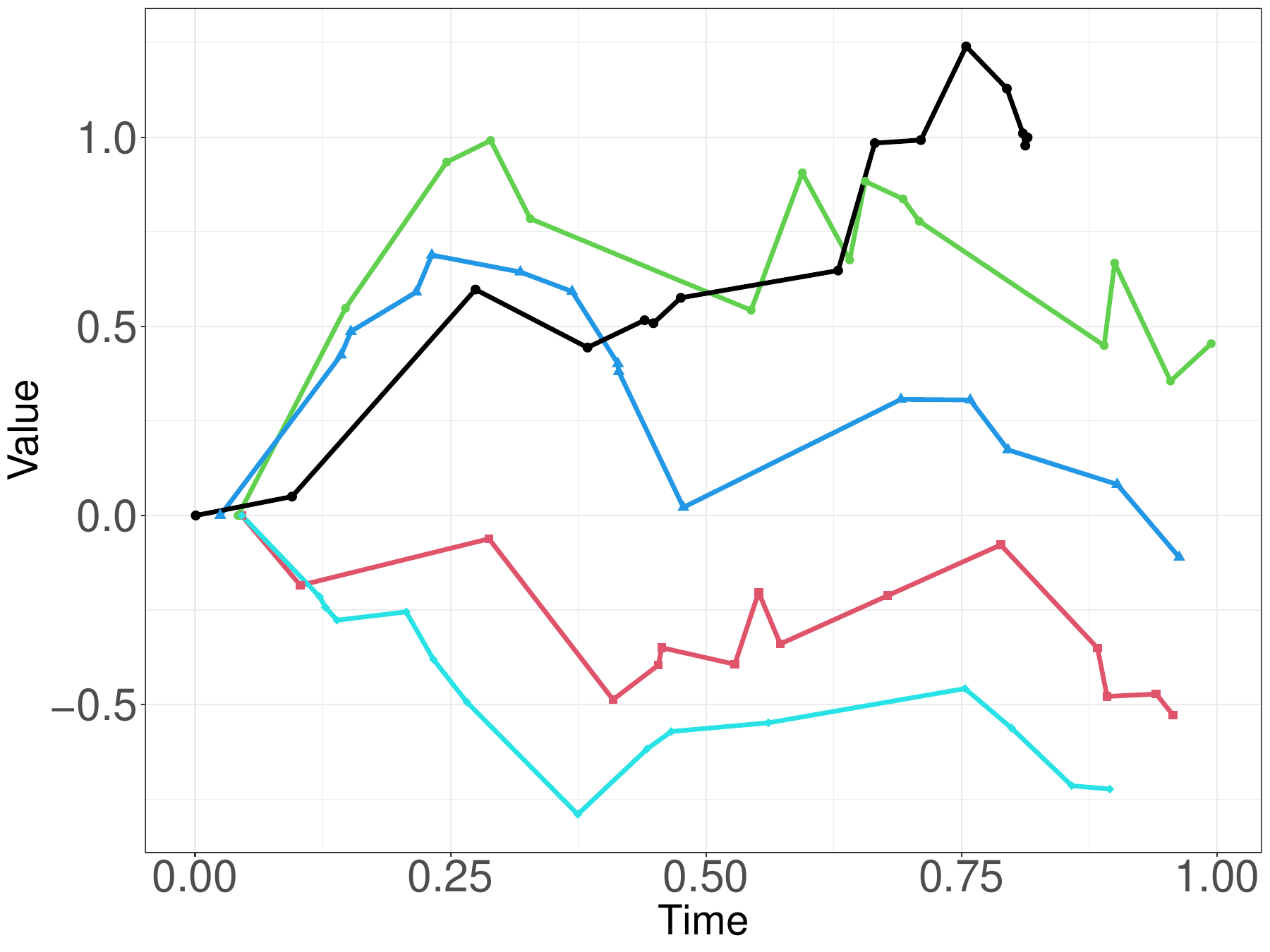}
			\label{f7a}
		\end{minipage}
	}
	\subfigure[]{
		\begin{minipage}[h]{0.46\textwidth}
			\centering
			\includegraphics[width=0.9\textwidth]{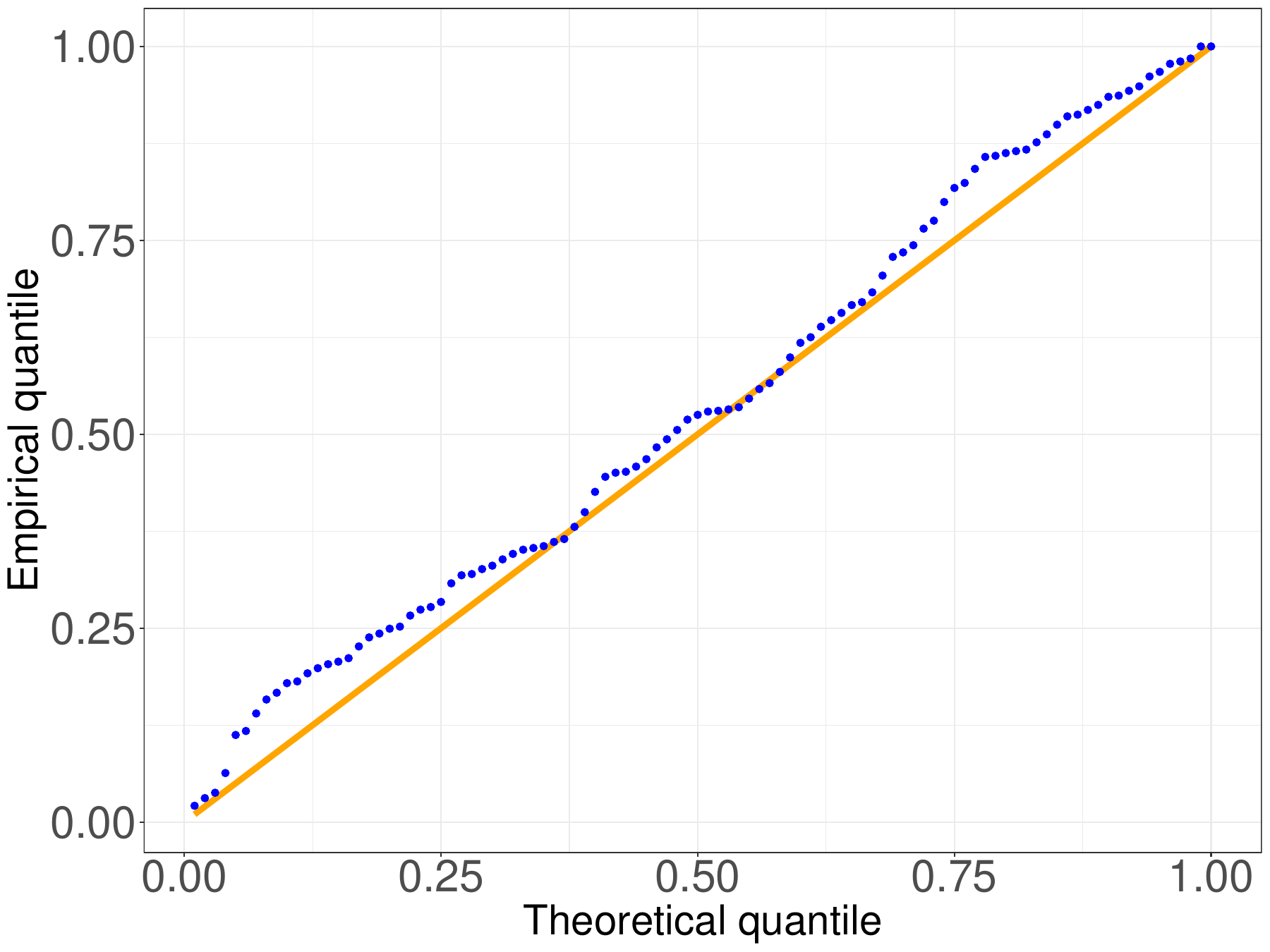}
		\end{minipage}
	}
	\subfigure[]{
		\begin{minipage}[h]{0.46\textwidth}
			\centering
			\includegraphics[width=0.9\textwidth]{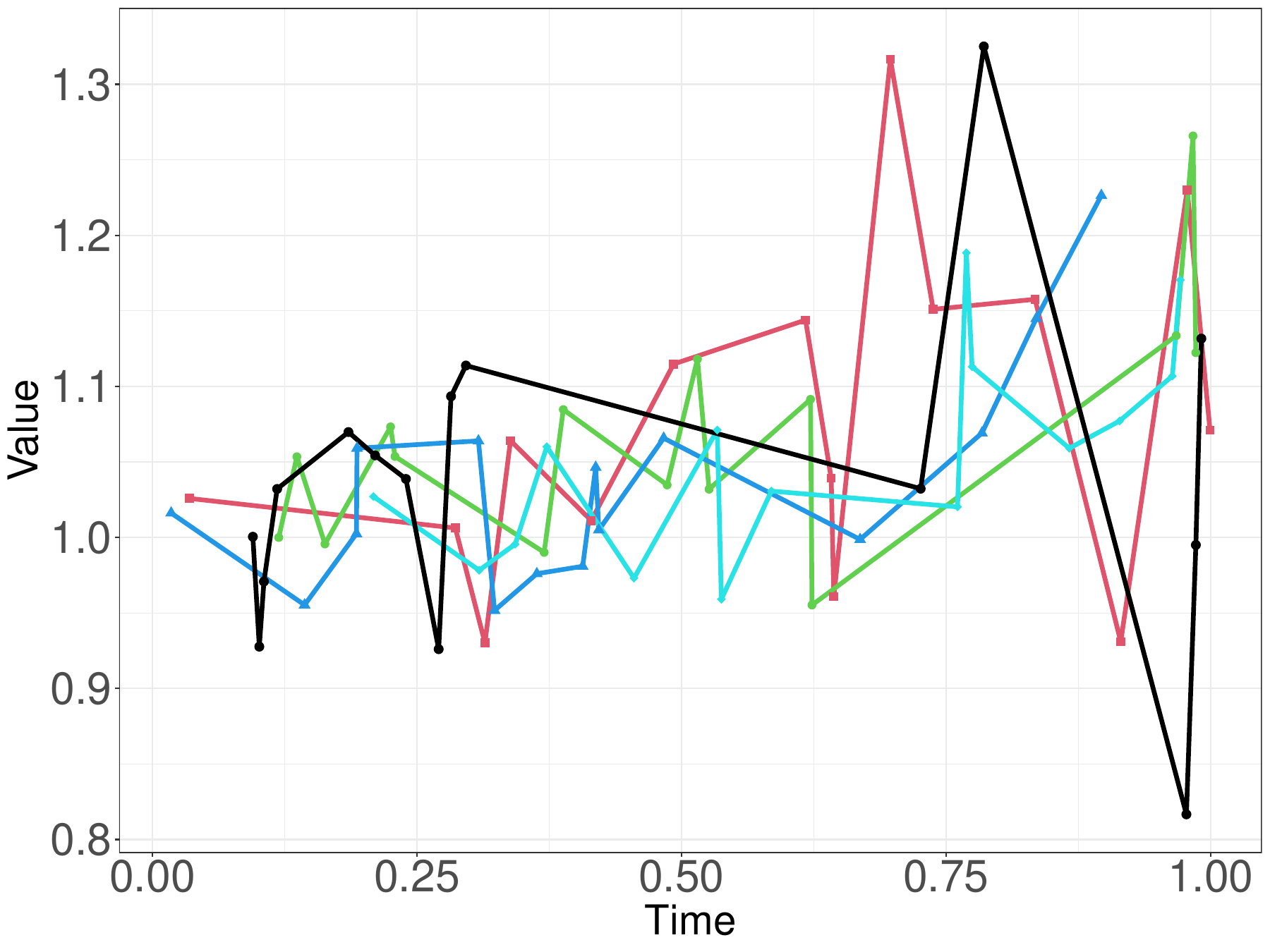}
			\label{f7c}
		\end{minipage}
	}
	\subfigure[]{
		\begin{minipage}[h]{0.46\textwidth}
			\centering
			\includegraphics[width=0.9\textwidth]{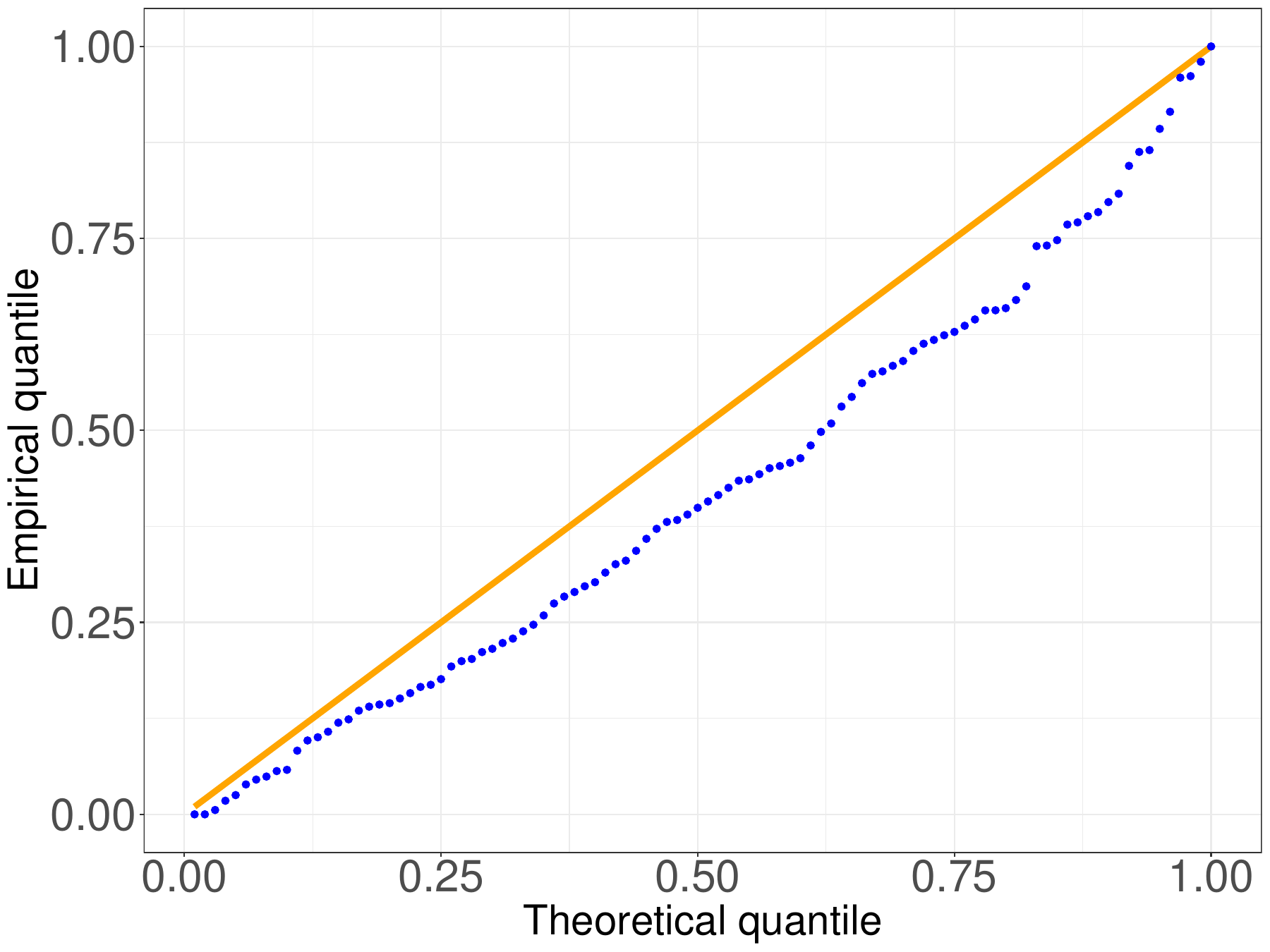} 
		\end{minipage}
	}
	\subfigure[]{
		\begin{minipage}[h]{0.46\textwidth}
			\centering
			\includegraphics[width=0.9\textwidth]{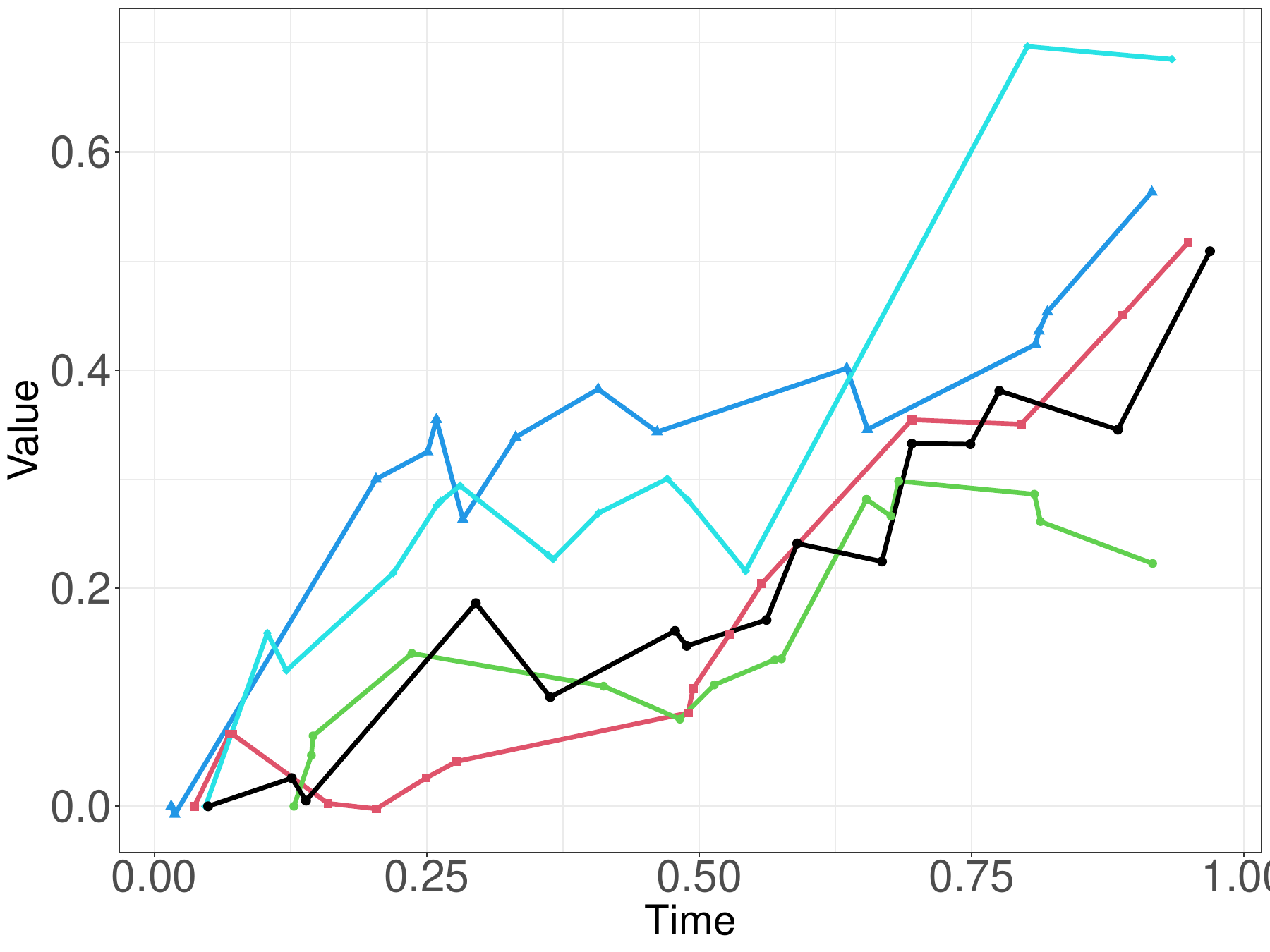}
			\label{f7e}
		\end{minipage}
	}
	\subfigure[]{
		\begin{minipage}[h]{0.46\textwidth}
			\centering
			\includegraphics[width=0.9\textwidth]{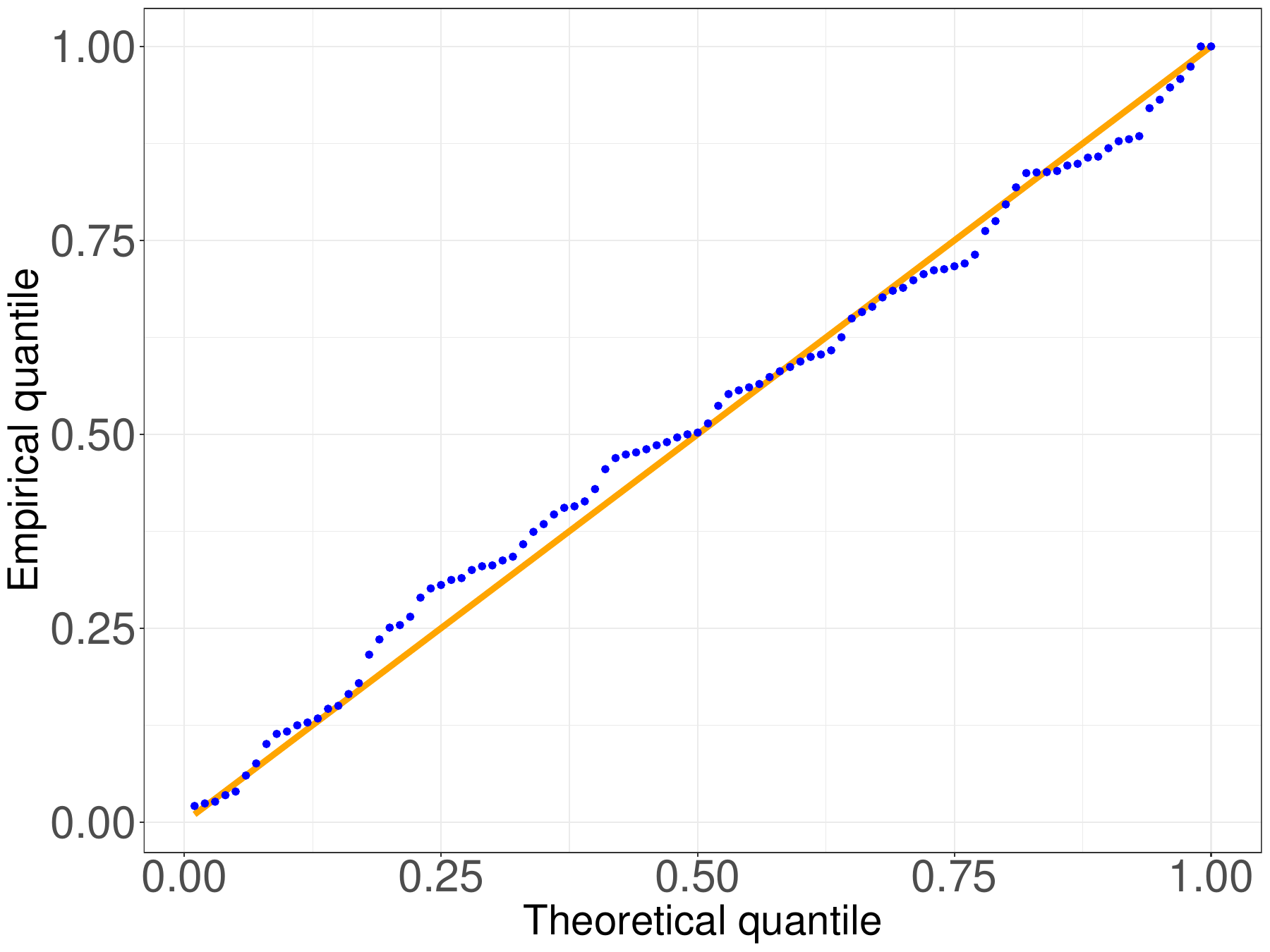} 
		\end{minipage}
	}
	\caption{\textbf{Left column}: Records of the first feature of the first 5 subjects for the dataset generated with $15$. \textbf{Right column}: quantile plots of the selective $p$-value. \textbf{(a)}: Each record is generated independently under the Wiener process. \textbf{(c)}: Each record is generated independently under the exponential Brownian motion. \textbf{(e)}: Each record is generated independently under the OU process.}\label{f2}
\end{figure}

\begin{figure}[p]
	\centering
	\subfigure[]{\begin{minipage}[h]{0.48\textwidth}\centering\includegraphics[width=0.95\textwidth]{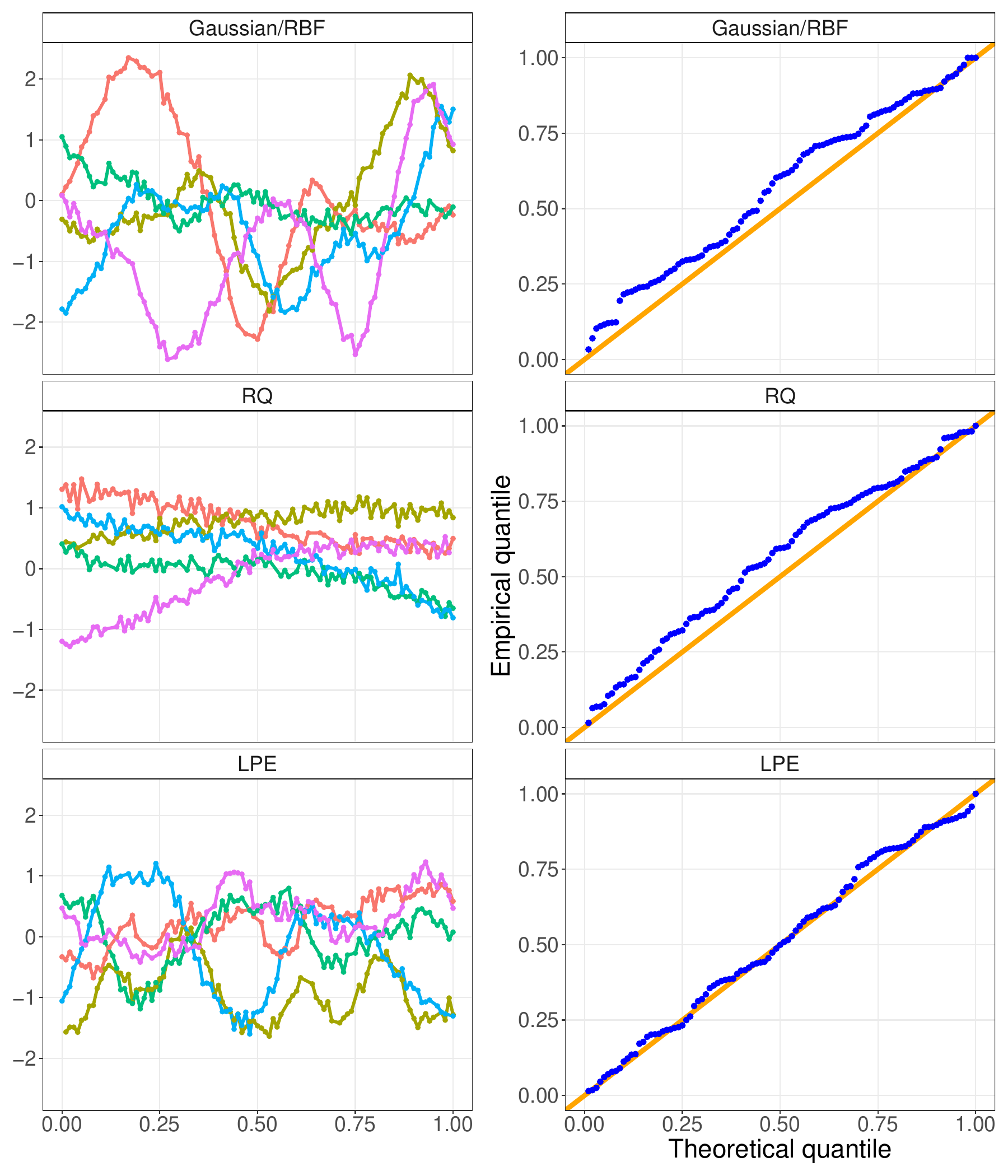}\end{minipage}}
	\subfigure[]{\begin{minipage}[h]{0.48\textwidth}\centering\includegraphics[width=0.95\textwidth]{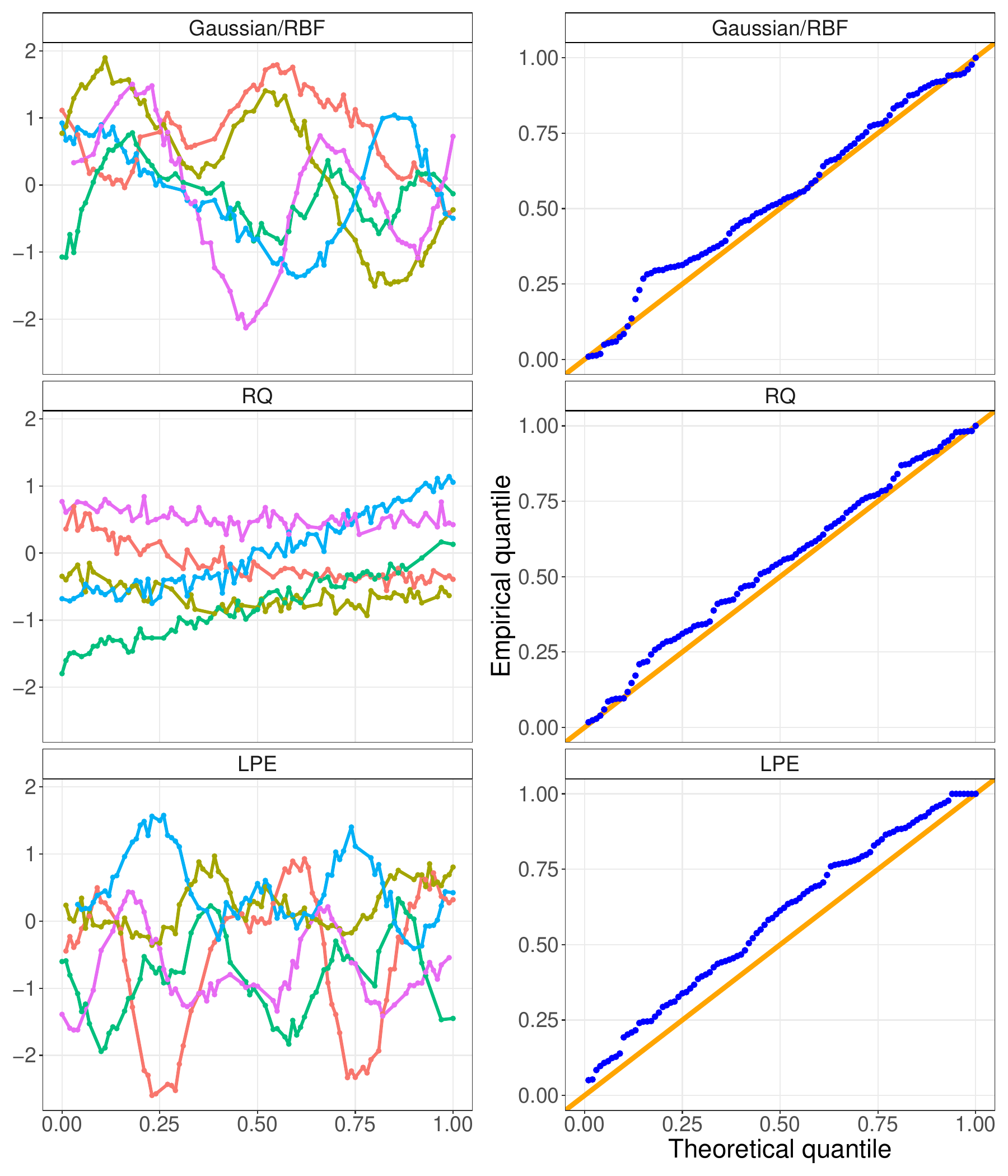}\end{minipage}}
	\caption{\textbf{Unequal observation times across individuals (I).} Each panel shows the first-feature trajectories of the first five subjects (left side within each panel) and the QQ plot of the selective $p$-values (right side within each panel) when observation times are removed independently for each subject. The two panels correspond to 10\% and 30\% loss rates. Within each panel, the three rows correspond to Gaussian/RBF, RQ, and LPE embeddings.}\label{fsparse1}
\end{figure}

\begin{figure}[p]
	\centering
	\subfigure[]{\begin{minipage}[h]{0.48\textwidth}\centering\includegraphics[width=0.95\textwidth]{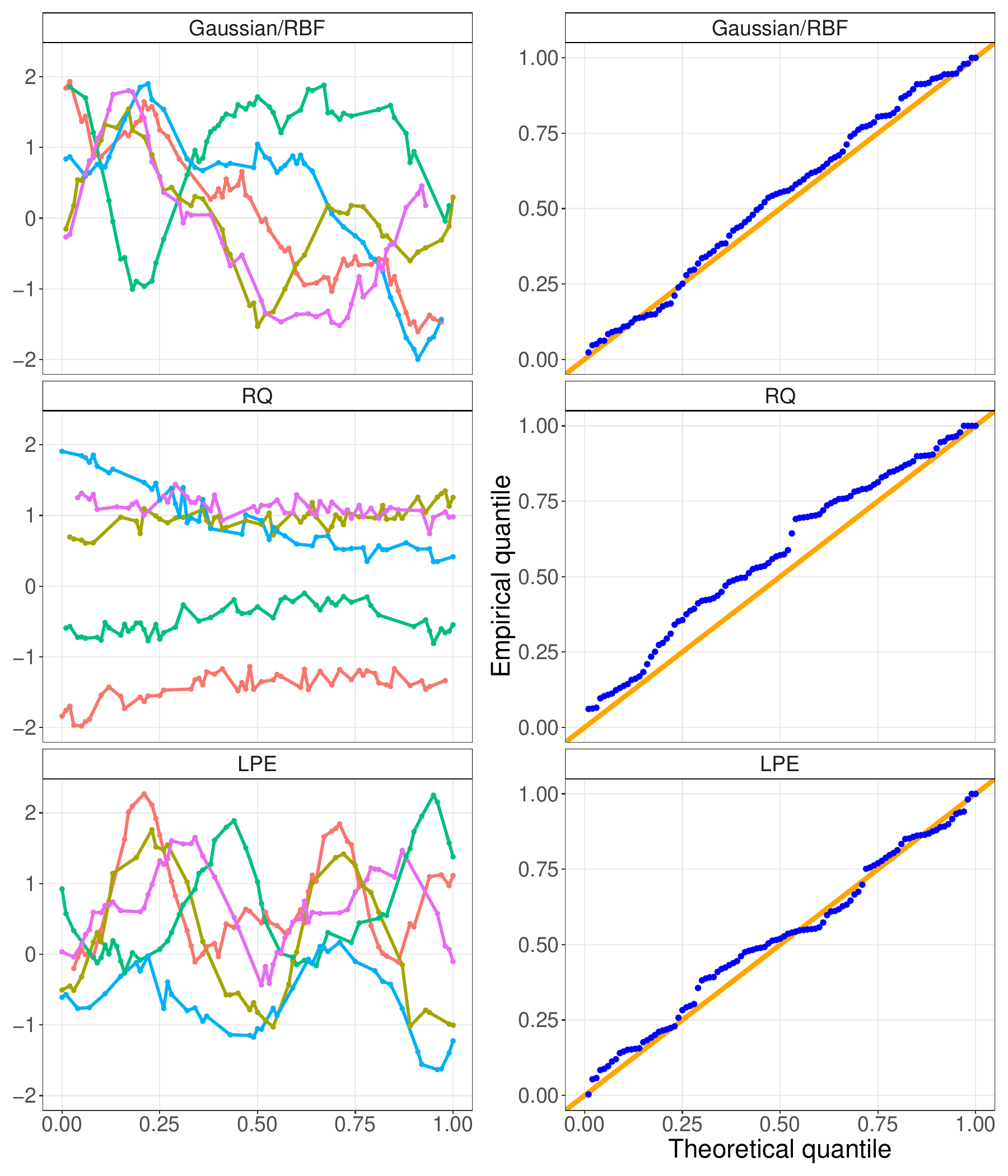}\end{minipage}}
	\subfigure[]{\begin{minipage}[h]{0.48\textwidth}\centering\includegraphics[width=0.95\textwidth]{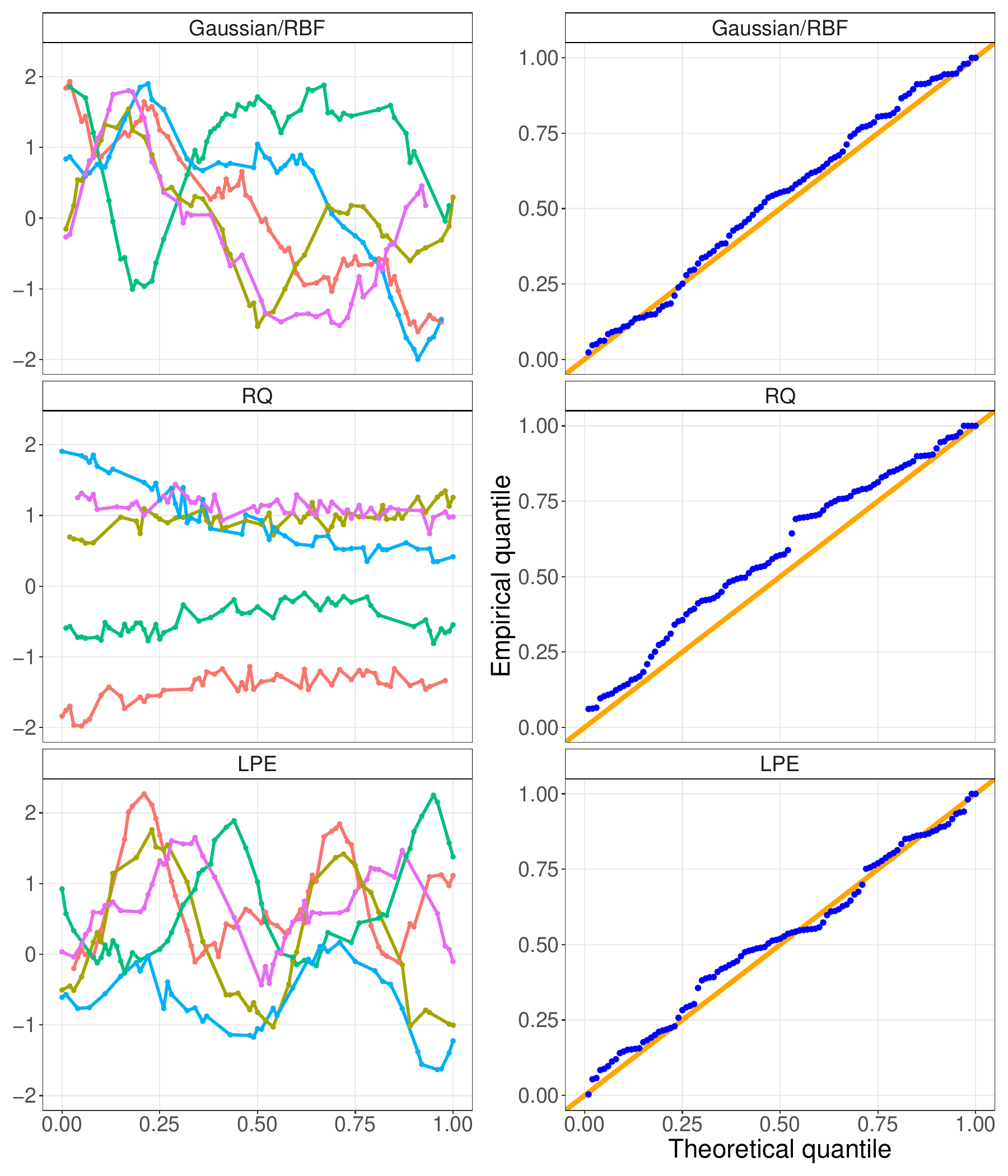}\end{minipage}}
	\caption{\textbf{Unequal observation times across individuals (II).} Same setup as Figure~\ref{fsparse1}, but with 50\% and 70\% loss rates. The sample-curve panels become visibly more fragmented as sparsity increases, while the QQ plots remain reasonably close to the diagonal across the three kernels.}\label{fsparse2}
\end{figure}

\begin{table}[p]
\centering
\footnotesize
\resizebox{0.95\linewidth}{!}{%
\begin{tabular}{lccccc}
\hline
Scenario & S1 corr. & S1 sel. & S2 att. & S2 corr. & S2 sel. \\
\hline
\texttt{coarse\_only}  & 1.00 & 0.93 & 0.93 & 0.00 & 0.043 \\
\texttt{null}          & 0.00 & 0.00 & 0.00 & --   & --    \\
\texttt{three\_cluster}& 0.46 & 0.49 & 0.23 & 1.00 & 0.957 \\
\hline
\end{tabular}}
\caption{Stagewise summary for the recursive three-cluster simulation.}
\label{tab:threecluster}

\vspace{0.3em}
\parbox{0.92\linewidth}{\footnotesize
\textit{Remark.} We note that S1 corr. = Stage~1 correct-split rate; 
S1 sel. = Stage~1 selective rejection rate; 
S2 att. = Stage~2 attempt rate; 
S2 corr. = conditional Stage~2 correct-split rate; 
S2 sel. = conditional Stage~2 selective rejection rate. 
Here, ``correct'' means that the estimated split matches the intended recursive split defined by the true latent labels: at Stage~1, the coarse split separates the first latent group from the other two, and at Stage~2, the finer split separates the remaining two latent groups. The table suggests that the recursive procedure can recover finer structure once the first-stage split reaches the appropriate branch. Under the \texttt{null} setting, the Stage~1 selective rejection rate is $0.00$ and Stage~2 is never attempted, which is consistent with the absence of any true split. Under the \texttt{coarse\_only} setting, the Stage~1 correct-split rate and selective rejection rate are both high ($1.00$ and $0.93$), while the conditional Stage~2 selective rejection rate remains low ($0.043$), as expected when only the first split is truly present. Under the \texttt{three\_cluster} setting, the Stage~1 split is recovered in $46\%$ of replicates, and conditional on proceeding to Stage~2, the finer split is recovered almost perfectly, with conditional Stage~2 correct-split rate $1.00$ and conditional selective rejection rate $0.957$. Thus, the table shows the expected recursive pattern: no false splitting under the null, strong detection of the first split when only the coarse split is present, and accurate recovery of the second split once the recursive procedure reaches the appropriate branch in the genuine three-cluster setting.
}
\end{table}

 {\begin{table}[p]
\centering
\footnotesize
\resizebox{0.98\linewidth}{!}{%
\begin{tabular}{cccc}
\hline
$\eta$ & Mean coarse-label accuracy & $\Pr(\mathrm{accuracy}\geq 0.90)$ & Rejection given accuracy $\geq 0.90$ \\
\hline
$0.0$ & $0.966$ & $0.940\ [0.874,\ 0.978]$ & $0.713\ [0.610,\ 0.801]$ \\
$0.1$ & $0.980$ & $1.000\ [0.964,\ 1.000]$ & $0.840\ [0.753,\ 0.906]$ \\
$0.2$ & $0.984$ & $1.000\ [0.964,\ 1.000]$ & $0.770\ [0.675,\ 0.848]$ \\
$0.3$ & $0.983$ & $0.970\ [0.915,\ 0.994]$ & $0.814\ [0.723,\ 0.886]$ \\
$0.4$ & $0.983$ & $0.960\ [0.901,\ 0.989]$ & $0.771\ [0.674,\ 0.850]$ \\
$0.5$ & $0.978$ & $0.950\ [0.887,\ 0.984]$ & $0.863\ [0.777,\ 0.925]$ \\
\hline
\end{tabular}}
\caption{Sensitivity to within-cluster mean-function heterogeneity. Brackets give exact 95\% binomial confidence intervals.}
\label{tab:heterogeneity}

\vspace{0.3em}
\parbox{0.94\linewidth}{\footnotesize
\textit{Remark.} Each heterogeneity level uses 100 replicates. Coarse-label accuracy is computed after optimal label switching, and the final column conditions on accuracy at least $0.90$. For $\eta>0$, the common within-cluster mean model is misspecified, so the reported rejection frequencies are sensitivity summaries rather than selective Type~I error estimates.
}
\end{table}}

 {\begin{figure}[p]
    \centering
    \includegraphics[width=0.95\linewidth]{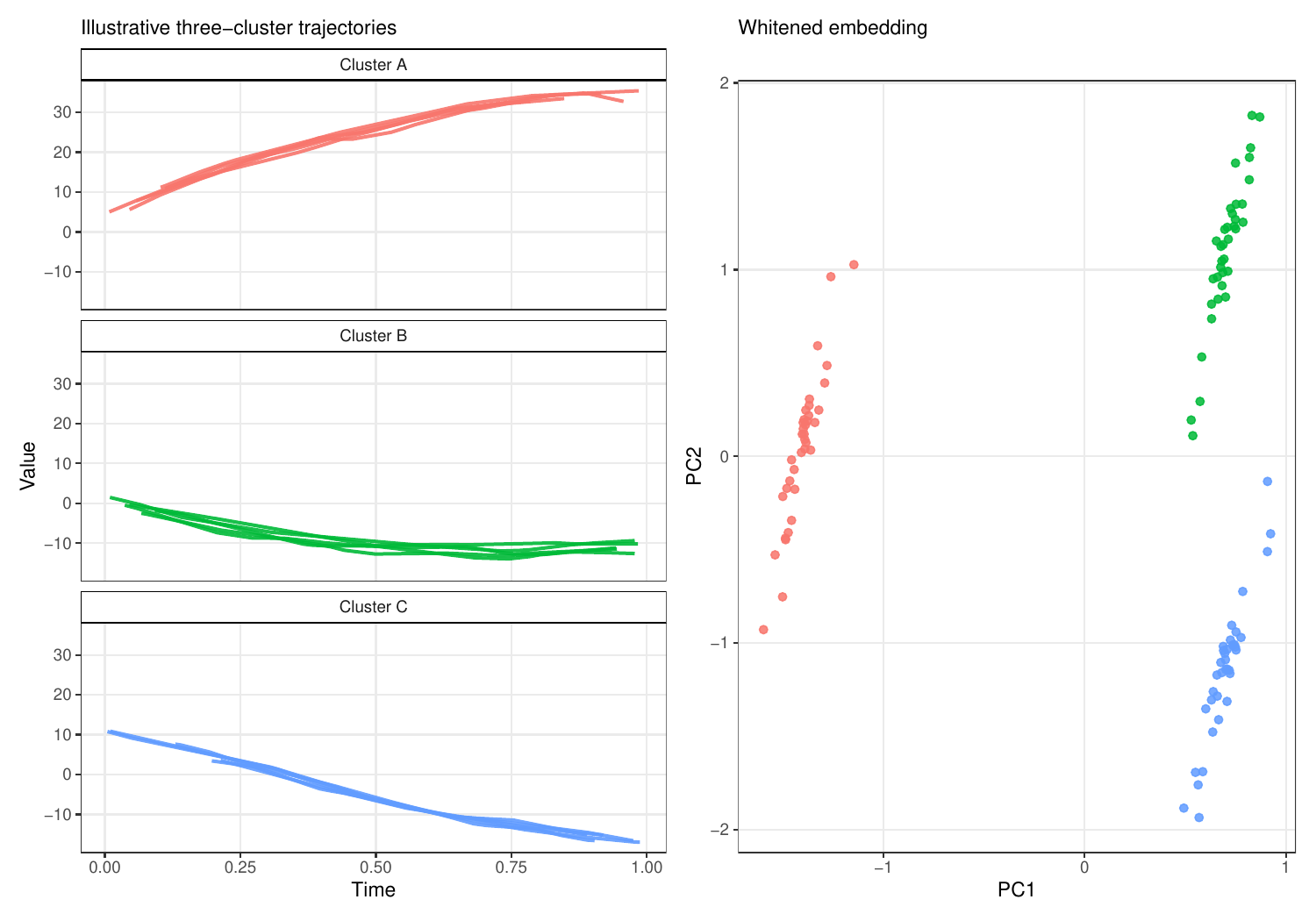}
    \caption{Illustrative dataset from the recursive three-cluster simulation. The figure shows sample longitudinal curves together with the corresponding scatter plot in the whitened low-dimensional embedding.}
    \label{fig:threecluster_illustrative}
\end{figure}}

 {\begin{figure}[p]
    \centering
    \includegraphics[width=0.95\linewidth]{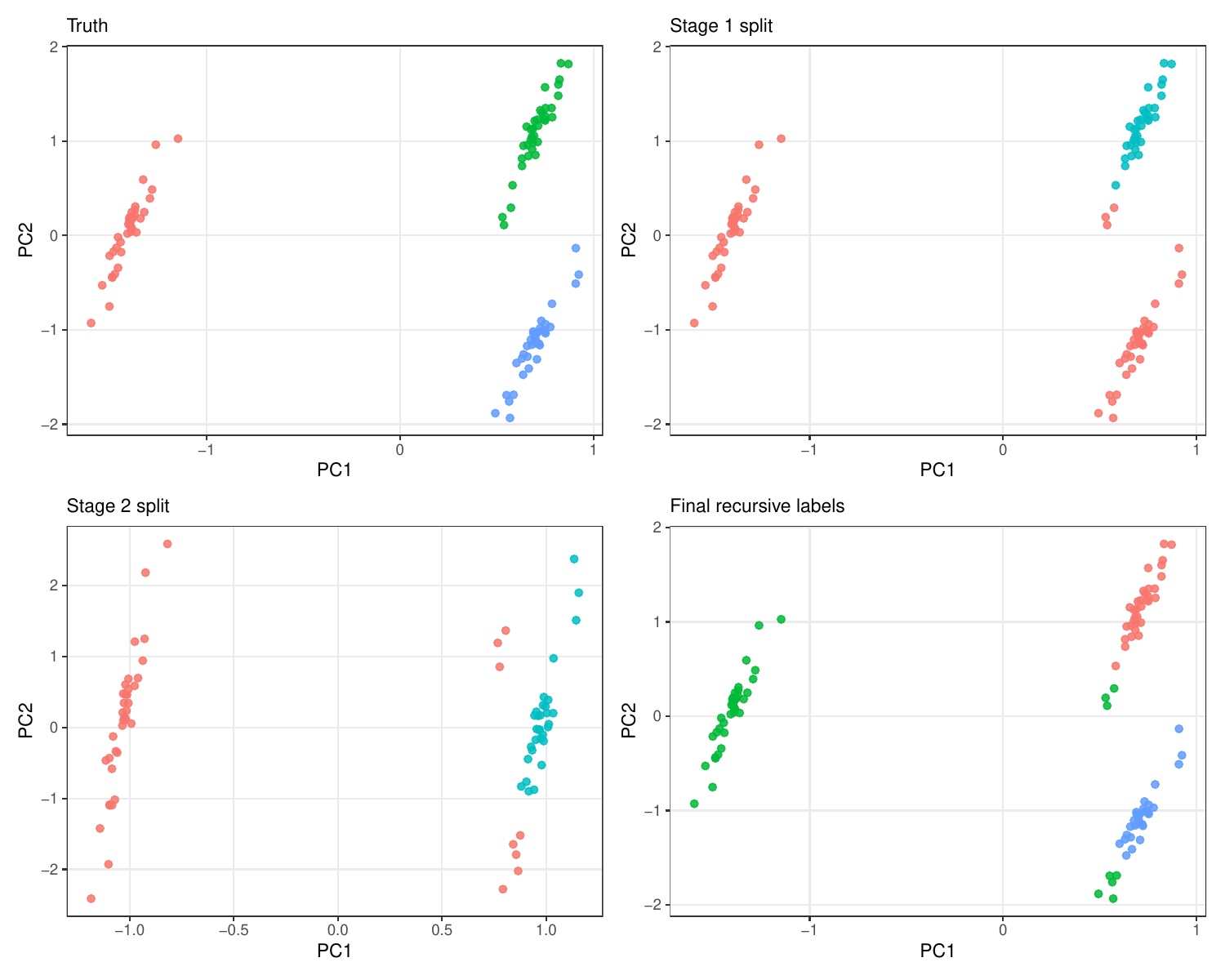}
    \caption{Recursive splitting in the whitened embedding for the three-cluster simulation. From left to right, the panels show the truth, the Stage~1 split on the full sample, the Stage~2 split on the selected branch after re-applying the full pipeline, and the final recursive labels.}
    \label{fig:threecluster_scatter}
\end{figure}}

 {\begin{figure}[p]
    \centering
    \includegraphics[width=0.95\linewidth]{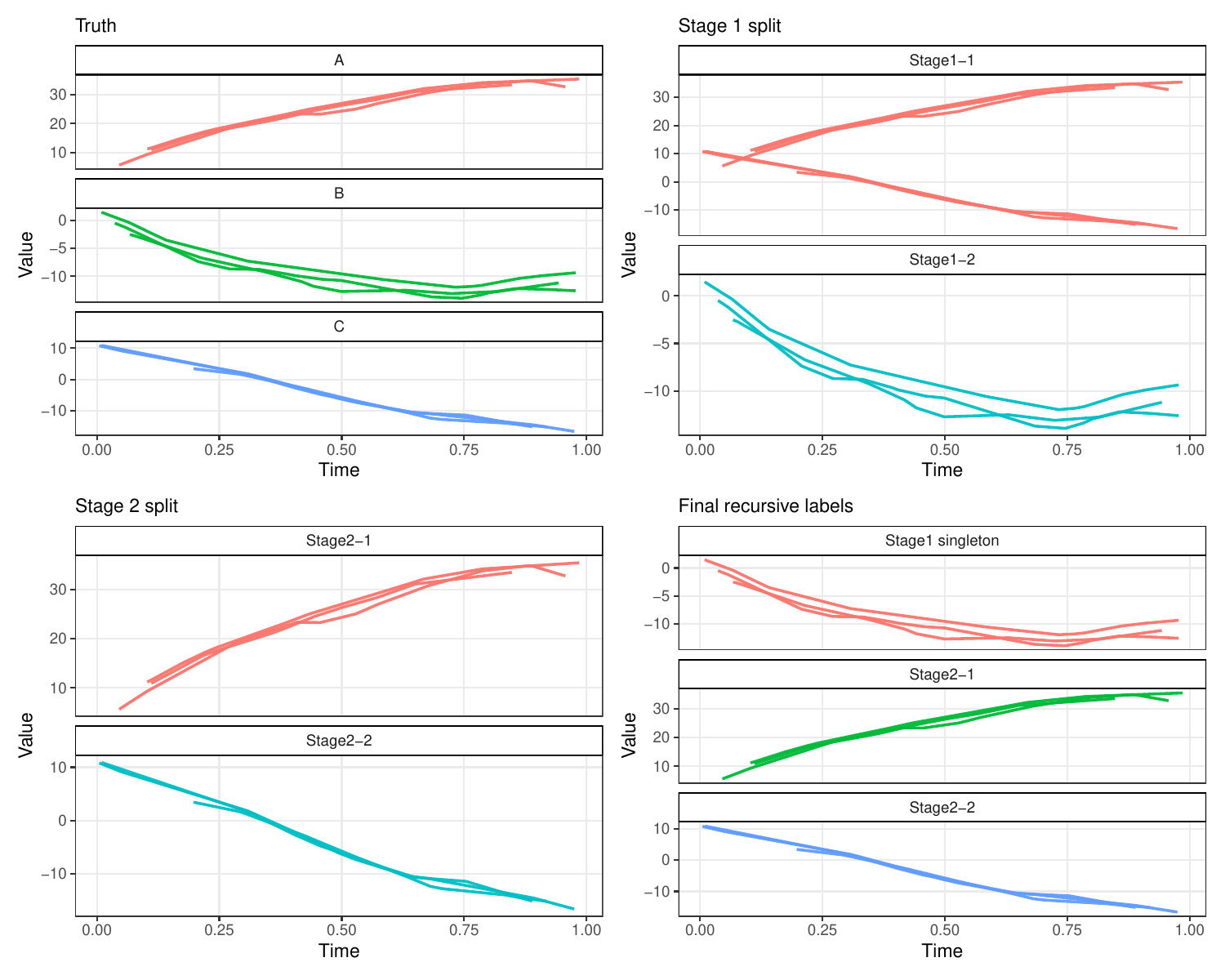}
    \caption{Recursive splitting shown in the longitudinal curves for the same representative three-cluster dataset. The panels are labeled as truth, Stage~1 split, Stage~2 split, and final recursive labels.}
    \label{fig:threecluster_curves}
\end{figure}}

\clearpage

\vskip 0.2in
\bibliographystyle{ims}
\bibliography{ref}

\end{sloppypar}

\end{document}